%% file: main.tex
\documentclass[a4paper,twocolumn,accepted=2025-02-26]{quantumarticle}
\pdfoutput=1

\usepackage[english]{babel}
\usepackage{graphicx}
\usepackage{dcolumn}
\usepackage{bm}
\usepackage{comment}
\usepackage{import}
\usepackage[normalem]{ulem} 
\usepackage{ mathrsfs }
\usepackage{ bbold }
\usepackage{ textcomp }
\usepackage{ stmaryrd }
\usepackage{float}
\usepackage{physics}
\usepackage{yhmath}
\usepackage{bm}
\usepackage{xfrac}
\usepackage{tabularx}
\usepackage{multirow}
\usepackage[dvipsnames, table]{xcolor}
\usepackage{algorithm}
\usepackage[noend]{algpseudocode}
\usepackage{amsmath}
\usepackage{amsthm}
\usepackage{amssymb}
\usepackage{hyperref}
\usepackage[dvipsnames]{xcolor}
\usepackage[capitalise]{cleveref}
\usepackage{caption}
\usepackage{subcaption}
\usepackage{ragged2e}
\usepackage{textgreek}

\crefname{section}{Sec.}{Secs.}
\Crefname{section}{Section}{Sections}

\newtheorem{theorem}{Theorem}[section]
\newtheorem{lemma}{Lemma}[section]

\begin{document}


\import{sections/}{titlepage.tex} 

\import{sections/}{abstract.tex} 

\maketitle 


\import{sections/}{section1.tex} 

\import{sections/}{section2.tex} 

\import{sections/}{section3.tex} 


\import{sections/}{acknowledgements.tex} 

\import{sections/}{appendix.tex} 

\clearpage

\bibliography{files/references.bib} 

\end{document}

%% file: sections/titlepage.tex



\title{Nonlinearity of the Fidelity in Open Qudit Systems: Gate and Noise Dependence in High-dimensional Quantum Computing}


\author{Jean-Gabriel Hartmann}
\email{jeangabriel.hartmann@ipcms.unistra.fr}
\affiliation{Université de Strasbourg, CNRS, Institut de Physique et Chimie des Matériaux de Strasbourg, UMR7504, F-67000 Strasbourg, France}
\orcid{0000-0002-1992-9675}

\author{Denis Jankovi\'c}
\email{denis.jankovic@ipcms.unistra.fr}
\affiliation{Université de Strasbourg, CNRS, Institut de Physique et Chimie des Matériaux de Strasbourg, UMR7504, F-67000 Strasbourg, France}
\affiliation{Institute of Nanotechnology (INT), Karlsruhe Institute of Technology, P.O. Box 3640, 76021 Karlsruhe, Germany}
\orcid{0000-0002-9550-6412}

\author{Rémi Pasquier}
\email{rpradeep@outlook.fr}
\affiliation{Université de Strasbourg, CNRS, Institut de Physique et Chimie des Matériaux de Strasbourg, UMR7504, F-67000 Strasbourg, France}
\orcid{0000-0003-0058-4125}

\author{Mario Ruben}
\email{mario.ruben@kit.edu}
\affiliation{Institute of Nanotechnology (INT), Karlsruhe Institute of Technology, P.O. Box 3640, 76021 Karlsruhe, Germany}
\affiliation{Institute for Quantum Materials and Technologies (IQMT), Karlsruhe Institute of Technology, P.O. Box 3640, 76021 Karlsruhe, Germany}
\affiliation{Université de Strasbourg, CNRS, Centre Européen de Science Quantique - Institut de Science et d'Ingénierie Supramoléculaires, UMR 7006, F-67000 Strasbourg, France}
\orcid{0000-0002-7718-7016}

\author{Paul-Antoine Hervieux}
\email{hervieux@unistra.fr}
\affiliation{Université de Strasbourg, CNRS, Institut de Physique et Chimie des Matériaux de Strasbourg, UMR7504, F-67000 Strasbourg, France}
\orcid{0000-0002-4965-9709}


             

%% file: sections/abstract.tex

\begin{abstract}
High-dimensional quantum computing has generated significant interest due to its potential to address scalability and error correction challenges faced by traditional qubit-based systems. This paper investigates the Average Gate Fidelity (AGF) of single qudit systems under Markovian noise in the Lindblad formalism, extending previous work by developing a comprehensive theoretical framework for the calculation of higher-order correction terms. We derive general expressions for the perturbative expansion of the Average Gate Infidelity (AGI) in terms of the environmental coupling coefficient and validate these with extensive numerical simulations, emphasizing the transition from linear to nonlinear behaviour in the strong coupling regime. Our findings highlight the dependence of AGI on qudit dimensionality, quantum gate choice, and noise strength, providing critical insights for optimising quantum gate design and error correction protocols. Additionally, we utilise our framework to identify universal bounds for the AGI in the strong coupling regime and explore the practical implications for enhancing the performance of near-term qudit architectures. This study offers a robust foundation for future research and development in high-dimensional quantum computing, contributing to the advancement of robust, high-fidelity quantum operations.
\end{abstract}


%% file: sections/section1.tex

\section{\label{sec:intro} Introduction}
High-dimensional quantum computing (QC) has generated remarkable scientific interest of late, introducing a shift from traditional computing paradigms. While qubit-based quantum information processing (QIP) platforms, particularly superconducting qubits, have the highest technological maturity, they are faced with near-term technical challenges of scalability and error correction \cite{ezratty_perspective_2023, de_leon_materials_2021}. On the other hand, recent advancements have highlighted the potential of qudits — quantum systems with $d$ levels — as powerful alternatives for novel QC architectures \cite{moreno-pineda_molecular_2018, moreno-pineda_measuring_2021, chi_programmable_2022, ringbauer_universal_2022}.

Indeed, qudits offer several advantages, including (i) lower decoherence rates in certain physical systems \cite{wang_qudits_2020}, (ii) enhanced quantum error correction through additional levels \cite{chiesa_embedded_2021, petiziol_counteracting_2021, miyahara_decoherence_2023} and stabiliser codes \cite{gunderman_local-dimension-invariant_2020}, as well as (iii) higher information density for reducing circuit complexity and enabling novel algorithm design \cite{wernsdorfer_synthetic_2019}. They also promise more robust flying quantum memories \cite{zheng_entanglement_2022, bouchard_high-dimensional_2017}. 

While classical computing ultimately settled on bits once sufficient scalability and fault-tolerance were achieved, early platforms did experiment with multi-level systems \cite{impagliazzo_ternary_2011}. Analogously, it can be argued that quantum computing's infancy stage could benefit from exploring higher-dimensional bases to address these current issues. In the push towards universal QC, increasing the total Hilbert space dimension of quantum systems is critical \cite{boixo_characterizing_2018}. This dimension is determined by $d^n$, where $d$ is the dimensionality and $n$ the number of qudits. Despite impressive advancements in superconducting platforms \cite{arute_quantum_2019}, scaling the number of qubits continues to pose significant challenges. Thus, qudit-based approaches may provide an avenue for increasing the Hilbert space with fewer physical units \cite{arute_quantum_2019, ringbauer_universal_2022, godfrin_operating_2017, thiele_electrically_2014}.

On the other hand, with the increased number of excited states utilised to implement higher-dimensional states in physical systems, qudits may introduce a greater number of error channels compared to qubits. This could lead to increased sensitivity to environmental noise, affecting coherence times and complicating error correction processes \cite{otten_impacts_2021}. In a previous paper \cite{jankovic_noisy_2024}, we studied this scenario in detail by comparing the effects of Markovian noise on multi-qubit and single-qudit systems of equal Hilbert space dimension. We developed a theoretical model of the first-order effects of the noise in the Lindblad formalism, demonstrating how the noise impacts the performance of quantum gates in these different systems, and supported these results with numerical simulations.

The fundamental quantities investigated were the average gate fidelity (AGF, $\bar{\mathcal{F}}$) - equivalently infidelity (AGI, $\bar{\mathcal{I}}$) - and the figure-of-merit ($\tau:=t_{\mathrm{gate}}/T_2$). As the name suggests, the AGF is useful in that it integrates out any specific features due to a particular choice of initial state. Therefore, unlike the state fidelity or process fidelity, which compute the target-to-output overlaps of quantum state transfers and unitary gates, respectively, the AGF is fundamental as a platform-agnostic measure of the quality of a system's interaction with its environment \cite{nielsen_quantum_2000, nielsen_simple_2002}. Similarly, the figure-of-merit is important for quantifying realistically the circuit depth that can be achieved in a system, based on the gate time, $t_{\mathrm{gate}}$, and decoherence time, $T_2$.


Those results showed that in the quasi-error-free - weak coupling - regime, the first-order response of the AGI (a linear function of the dimensionless coupling strength $\gamma t_{\mathrm{gate}}$) was sufficient to characterise the behaviour of the qudit or multi-qubit system. Specifically, a scaling law was derived, showing how the AGI is affected by qudit dimension and the conditions on the figure-of-merit under which qudits may be competitive with multi-qubit systems. These conditions lead to stricter requirements on the figure-of-merit as the qudit dimension increases. However, first-order approximations are insufficient for fully understanding the behavior of qudits under realistic, near-term, conditions where higher-order noise effects are expected to become significant, either through stronger coupling or longer gate times (or circuit depths) approaching the decoherence limit.

Thus, this paper aims to extend the understanding of the AGF of single qudit systems under Markovian noise conditions through a general perturbative expansion. Specifically, we focus on higher-order correction terms and their implications on quantum gate performance in the strong coupling regime. Building upon our previous work, we develop a comprehensive theoretical framework that includes these nonlinear effects, with a constructive method generalising the AGF to arbitrary order in $\gamma t$. This is supported by detailed numerical simulations, with an emphasis on the differences between the first and second order terms in the case of pure dephasing. By doing so, we aim to capture the nuanced impact of noise on the fidelity of quantum gates more precisely.

The key research question addressed in our study is: How do higher-order correction terms and noise coupling strength influence the AGI of single qudit systems, and what implications do these have for the selection and design of quantum gates? Our findings provide important insights on: (i) setting benchmarks, as well as limits, for the performance of noisy quantum systems of arbitrary dimension by identifying gate-dependent effects, (ii) methodologies for improving performance through optimising basis gates that have favourable decoherence characteristics, and (iii) advancing error correction protocols by enabling the cancellation of higher-order noise effects.

The paper is structured as follows: We begin in \cref{sec:2:subsec:1} with a review of the relevant theoretical background concerning the Lindblad formalism for qudit open quantum systems and the superoperator representation of noisy quantum channels, as well as generalised quantum gates on qudits. In \cref{sec:2:subsec:2}, we present our general result for the perturbative expansion of the AGI in noise strength, deriving the general expressions to arbitrary order in $\gamma t$ and discussing the implications of its structure. \Cref{sec:2:subsec:3} examines the nonlinearity of the AGI in the strong coupling regime of pure dephasing through numerical simulations that elucidate key characteristics of this behaviour and motivate the proceeding investigations that make the link with the theoretical results. Subsequently, in \cref{sec:2:subsec:4} we explore the gate- and noise-dependence of the AGI, with particular emphasis on how the choice of quantum gate can influence the performance of the quantum channel. Finally, \cref{sec:2:subsec:5} presents a detailed analytical and numerical study of the first- and second-order correction terms to the AGI, extending the results of the previous paper and placing them on a more rigorous theoretical foundation that serves future investigations.


%% file: sections/section2.tex

\section{\label{sec:2} Results and Discussion}

\subsection{\label{sec:2:subsec:1} Mathematical Foundations of the Average Gate Fidelity}

We begin with the study of a single qudit, of arbitrary dimension $d$, whose state is represented by a $d \times d$ density matrix $\rho (t)$, and evolves according to the Gorini-Kossakowski-Sudarshan-Lindblad (GKSL) master equation \cite{lindblad_generators_1976, gorini_completely_1976}, which may be expressed in superoperator form as \cite{breuer_theory_2007}

\begin{align} \label{eq:GKSL}
    \partial_t \rho &= \mathcal{S} \left[ \rho \right] + \mathcal{D} \left[ \rho \right] . 
\end{align}
Here, setting $\hbar = 1$, the Liouvillian superoperator $\mathcal{S}$ represents the unitary evolution operator of the von Neumann equation, and $\mathcal{D}$ the dissipation superoperator coupling the qudit to the environment,
\begin{align}\label{eq:super}
    \mathcal{S} \left[ \rho \right] &:= -i \left[ H, \rho \right] , \\
    \mathcal{D} \left[ \rho \right] &:= \sum_{k=1}^K \gamma_k \mathcal{L}_k[\rho] \\
    &= \sum_{k=1}^K\gamma_k\left(L_k\rho L_k^\dag - \frac{1}{2}\left\{L_k^\dag L_k,\rho\right\}\right) ,
\end{align}
where $H$ is the time-independent interaction Hamiltonian, and $L_k$ one of $K$ possible collapse operators characterising the Markovian noise with coupling coefficient $\gamma_k$ and $\mathcal{L}_k$ the matrix superoperator form \cite{manzano_short_2020}.

The solution to the master equation for $\rho(0) = \rho_0$ is therefore given by the completely-positive and trace-preserving (CPTP) quantum channel $\mathcal{E}$ as \cite{watrous_theory_2018} 
\begin{align}\label{eq:mesolution1}
    \rho(t) &= \mathcal{E}\left[ \rho_0 \right] \\
    &= e^{(\mathcal{S} + \mathcal{D})t}\rho_0 . \label{eq:mesolution2}
\end{align}

Considering the application of a quantum gate $\mathcal{U} \left[ \rho \right] = U \rho U^{\dag}$ to the system then corresponds to a composition operation, that in the matrix superoperator representation reduces to the matrix product, from which we may express the AGF, $\bar{\mathcal{F}}(\mathcal{E}, \mathcal{U})$, of implementing a unitary operator over the quantum channel in the presence of a noisy environment \cite{nielsen_simple_2002}:
\begin{align}
    \bar{\mathcal{F}}(\mathcal{E}, \mathcal{U}) &:= 1 - \bar{\mathcal{I}}(\mathcal{E}, \mathcal{U}) \\ &= \int_{\mathscr{H}}\, \left\langle U^{\dagger} \mathcal{E}[\rho_0] U\right\rangle_0 d \rho_0 \\
    &= \int_{\mathscr{H}}  \,\left\langle\left(\mathcal{U^\dag \circ E}\right)[\rho_0]\right\rangle_0 d \rho_0 , \label{eq:AGF}
\end{align}
where the matrix product has been rewritten in the superoperator representation. The integral is taken uniformly over the Haar measure $\mathscr{H}$ of the state space, with the expectation value $\ev{\mathcal{A}}_0 = \Tr{\mathcal{A} \rho_0}$ representing the average over all initial states. For convenience, we shall use $\bar{\mathcal{F}}$ and $\mathcal{F}$ interchangeably for the AGF, and likewise for the AGI.

In this study we shall consider pure initial states, $\Tr{\rho^2_0}=1$. Then the state space is reduced to the complex-projective space $\mathbb{C}\mathbf{P}^{d-1}$ such that the Haar measure (over all $\rho_0$) then induces the Fubini-Study measure, the integral of which is normalised, $\int_{\mathscr{H}} d \rho_0 = 1$, \cite{bengtsson_geometry_2006} and may be calculated analytically (see \cref{app:A:subsec:3}). Furthermore, unless specified, we shall also assume that: (i) the dissipator term is dominated by a single noise channel, such that $\mathcal{D} = \gamma \mathcal{L}$, (ii) the associated superoperators $\mathcal{S}$ and $\mathcal{L}$ are time-independent and (iii) we are working in the interaction picture, such that the free evolution Hamiltonian $H_0=\mathbb{0}_d$, and therefore $\mathcal{S}_0=\mathbb{0}_{d\times d}$ such that the unitary evolution simplifies to $\mathcal{U} = e^{(\mathcal{S}_0 + \mathcal{S}_c)t}=e^{\mathcal{S}t}$ with $\mathcal{S} = \mathcal{S}_c$ corresponding to only the control (driving) Hamiltonian $H=H_c$ implementing the quantum gate. This allows us to rewrite \cref{eq:mesolution1,eq:mesolution2} as
\begin{align} \label{eq:supersol}
    \mathcal{E}\left[ \rho_0 \right] = e^{(\mathcal{S} + \gamma \mathcal{L})t} \rho_0 .
\end{align}

We note that the $(d \times d)\times (d \times d)$ superoperators acting on the $\mathcal{H} \otimes \mathcal{H}$ Hilbert space of the quantum dynamical semigroup (QDS) may be written explicitly in Liouville matrix form in terms of the standard operator form as \cite{havel_robust_2003}:
\begin{align}
    &\mathcal{U}\left[ \cdot \right] := \left( U^* \otimes U \right) \mathrm{vec}\left( \cdot \right) , \label{eq:Uprepost} \\
    &\mathcal{S}\left[ \cdot \right] := -i \left( \mathbb{1}_d \otimes H  - H^T\otimes \mathbb{1}_d  \right) \mathrm{vec}\left( \cdot \right) , \label{eq:Sprepost} \\
    &\mathcal{L} \left[ \cdot \right] \notag \\
    &:= \left( L^* \otimes L - \frac{1}{2}\left(L^{\dag} L \otimes \mathbb{1}_d + \left( \mathbb{1}_d \otimes L^{\dag} L \right)^T \right) \right) \mathrm{vec}\left( \cdot \right) , \label{eq:Lprepost} 
\end{align}
where we identify the adjoint ($\dag$), complex-conjugate ($*$) and transpose ($T$) operations, $\mathbb{1}_d$ the $d \times d$ identity matrix, $\otimes$ the tensor, or equivalently here for linear maps, the Kronecker product, and $\mathrm{vec}(\cdot)$ the vectorisation operation that, when applied to a $d \times d$ density matrix $\rho$, produces a $1\times d^2$ column vector by stacking vertically each column of the matrix from left to right. This corresponds to the explicit mapping
\begin{align}
    \mathrm{vec}(\rho) \; : \; \sum_{i,j}\rho_{ij}\ketbra{i}{j} \; \rightarrow \; \sum_{i,j}\rho_{ij}\ket{j} \otimes \ket{i} ,
\end{align}
of the quantum state from the Hilbert to the Fock-Liouville space \cite{manzano_short_2020}. Furthermore, the quantum gate $U_g$ and associated control Hamiltonian $H_c$ (and their respective superoperator representations $\mathcal{U}$ and $\mathcal{S}$) are related by the matrix exponent (and, inversely) logarithm operations. Therefore, given some quantum gate, it may be implemented by the control term,
\begin{align}
    H_c t_g &= i \log U_g ,
\end{align}
where the now time-independent $H_c$ represents a single, ideal control pulse that implements the gate in time $t_g$, which can be modulated by the absolute amplitude of the pulse $\norm{H_c}$. Existence and uniqueness of the matrix logarithm map are well established under the conditions that $U_g$ is unitary and $-\pi < H_c t_g \leq \pi$ \cite{loring_computing_2014}. This is an idealisation that necessitates simultaneous control over all pulse amplitudes, phases and detunings on each possible transition between all $d$ states. Platform-dependent physical constraints on the control pulses may preclude experimental realisation of such a control term. Nevertheless it allows for a time-optimal multi-chromatic pulse for studying the robustness of quantum gates that is independent of constraints imposed by the physical platform or pulse control technique \cite{aifer_quantum_2022}.

We now precisely define the gate and collapse operators that will be of interest in this study. We begin with the following set of single-qudit quantum gates $U_g \in \lbrace \mathbb{1}, X, Z, F \rbrace$ that generalise the single-qubit identity, Pauli-$x$, Pauli-$z$ and Hadamard gates to $d$ dimensions, and generate the $d$-dimensional generalised Clifford algebra (GCA) \cite{vourdas_quantum_2004}
\begin{align}
    \mathbb{1}_d &:= \sum_{j=0}^{d-1} \ketbra{j}{j} , \label{eq:identity} \\
    X_d :&= \sum_{j=0}^{d-1} \ketbra{(j+1)\,\mathrm{mod}\,d}{j} , \label{eq:X} \\
    Z_d &:= \sum_{j=0}^{d-1} \omega^j  \ketbra{j}{j} \label{eq:Z} , \\
    F_d &:= \frac{1}{\sqrt{d}} \sum_{j=0}^{d-1}\sum_{k=0}^{d-1} \omega^{jk}  \ketbra{j}{k} , \label{eq:F} 
\end{align}
where $\omega := e^{\frac{2 \pi i}{d}}$ is the $d$-th root of unity. The subscript $d$ is used to specify the dimension of the operator, and may be omitted in cases where it is clear from context or when referring to the general operator. The $X$, $Z$ and $F$ gates retain unitarity (and tracelessness for $X$ and $Z$) as well as the relations $X^d=Z^d = \mathbb{1}_d$ and $X = F Z F^{\dag}$ of involution and change-of-basis, respectively. However, it is worth noting that they are no longer Hermitian for $d>2$. Furthermore, in the generalised forms, it can be observed that (i) $X$ is extended from a bit-flip (NOT) gate to a cyclic permutation (INC/ SHIFT) gate that increments the qudit state, (ii) $Z$ is extended from a phase-flip to a phase-shift (CLOCK) gate of each state over the $d$ roots of unity, and (iii) the Walsh-Hadamard gate is extended to the matrix form of the Quantum Fourier Transform (QFT) (equivalently referred to as the Discrete Fourier Transform, Sylvester or Chrestenson matrix).

In addition to this discrete set of Clifford gates, we can also consider non-Clifford gates that may be useful for quantum error correction (QEC) protocols for universal QC. In general, since the unitary group $\mathbf{U}$ is not finite, any unitary matrix sampled randomly from, for example, the Haar measure over $\mathbf{U}$ will almost always be non-Clifford. Therefore, by including in our investigations the Haar-random gates enables comparison of the differences in behaviour of the Clifford and non-Clifford gates. More specifically, one noteworthy non-Clifford gate is the (generalised) $\pi/8$ gate \cite{howard_qudit_2012},
\begin{align}
    T_d := \sum_{j=0}^{d-1} e^{i \frac{j\pi}{8}} \ketbra{j}{j} ,\label{eq:T_gate}
\end{align}
which when considered together with the $Z_d$ gate, we may extend to a generalised-PHASE gate
\begin{align}
    \mathrm{PHASE}_d (\phi) := \sum_{j=0}^{d-1} e^{i j \phi} \ketbra{j}{j} .\label{eq:PHASE_gate}
\end{align}
Setting $\phi$ to $\frac{\pi}{8}$, $\frac{2\pi}{d}$ or $\frac{\pi}{4}$ recovers $T_d$, $Z_d$ or the $S_d=\sqrt{Z_d}$ gates, respectively, where we may note that special cases including $Z_d$ and $S_d$ are Clifford, while in general for arbitrary choices of phase, including $\phi=\frac{\pi}{8}$, will likely be non-Clifford (see \cref{app:B:subsec:1} for further details on the gate definitions).


Regarding the choice of Markovian collapse operators $L \in \lbrace J_z, J_x, J_- \rbrace$ representing the coupling of the qudit to the environment in the master equation, we shall primarily consider the effect of pure dephasing by the operator $J_z$, and of secondary consideration are the bit-flip error and spin relaxation effected by the operators $J_x$ and $J_-$, respectively. These higher-order spin operators from spin-$\frac{1}{2}$ to spin-$\frac{d - 1}{2}$ can be obtained through the generalised Gell-Mann matrices, and written as
\begin{align}
    J_z &:= \sum_{j = 1}^{d} \frac{1}{2} (d + 1 - 2j) \ketbra{j}{j} ,\label{eq:Jz} \\
    J_x &:= \sum_{j = 1}^{d-1} \frac{1}{2} \sqrt{j(d-j)} \left( \ketbra{j+1}{j} + \ketbra{j}{j+1}  \right) ,\label{eq:Jx} \\
    J_- &:= \sum_{j = 1}^{d-1} \sqrt{j(d-j)} \left( \ketbra{j+1}{j}  \right) ,\label{eq:Jm}
\end{align}
where $J_-=J_x-iJ_y$. It is also possible to consider configurations of linear combinations of these operators, which can be used to model more complex and realistic environmental interactions. However, for the purposes of this study, we are interested in understanding their individual interaction with the control Hamiltonian and consequent effect on the AGI in a platform-agnostic way and thus shall only consider each of the operators separately.

\subsection{\label{sec:2:subsec:2} Perturbative Expansion of the AGI}

Given the simplified expression for the time-evolution of the quantum channel $\mathcal{E}$ in (\ref{eq:supersol}), we want to study the effects on the AGF, not only from the choice of $\mathcal{S}$ or $\mathcal{L}$, but also from the coupling strength $\gamma$. Therefore, performing a Taylor series expansion of the AGF in powers of $\gamma$,
\begin{align}\label{eq:agfexpansion}
    \bar{\mathcal{F}}(\mathcal{E}, \mathcal{U}) &= 1 - \sum_{m=1}^{\infty} \gamma^m F^{(m)}(t) .
\end{align}
To calculate explicitly the correction terms $F^{(m)}$, we begin by expanding the solution to the master equation,
\begin{align} \label{eq:rhoexpansion}
    \rho (t) = \rho^{(0)} + \sum_{m=1}^{\infty}\gamma^m \rho^{(m)},
\end{align}
where the $m$-th order correction to the non-perturbed solution of the master equation can be written as \cite{martinez-carranza_alternative_2012, villegas-martinez_application_2016}

\begin{widetext}
    \begin{align} \label{eq:rhocorrection}
    \rho^{(m)} &= e^{\mathcal{S}t} \left( \int_0^t \int_0^{t_1} \cdots \int_0^{t_{m - 1}} \left( \prod_{i=1}^m  e^{-\mathcal{S}t_i} \mathcal{L} e^{\mathcal{S}t_i} \right) dt_{m} \cdots dt_2 dt_1 \right) \rho_0 ,
\end{align}
\end{widetext}
and we separate the zero order term $\rho^{(0)}$ from the summation, since it corresponds to only unitary evolution through the von Neumann equation in the abscence of environmental coupling.

We may then define a new quantity $\Tilde{M}^{(m)}(t)$ containing the integral terms, such that $\rho^{(m)} := e^{\mathcal{S}t} \left( \Tilde{M}^{(m)}(t) \right) \rho_0$, and therefore the $m$-th order of the solution to the master equation in terms of the quantum channel reduces to $\mathcal{E}^{(m)} = \mathcal{U} \circ M^{(m)}(t)$. Now, in order to render this term $\Tilde{M}^{(m)}(t)$ into a more computationally tractable form, we make use of the eponymous lemma of Campbell resulting from the Baker-Campbell-Hausdorff (BCH) Formula \cite{campbell_law_1896, hall_lie_2015},
\begin{align}\label{eq:BCH}
    e^X Y e^{-X} = \sum_{n=0}^{\infty} \frac{\comm{(X)^n}{Y}}{n!} ,
\end{align}
utilising the iterated commutator defined by the recursion relation $\comm{(X)^n}{Y} = \comm{X}{\comm{(X)^{n-1}}{Y}}$ with the halting condition $\comm{(X)^0}{Y} = Y$. With $\mathcal{S}$ and $\mathcal{L}$ time-independent, $\Tilde{M}^{(m)}(t)$ simplifies to
\begin{widetext}
    \begin{align} 
    \Tilde{M}^{(m)}(t) &= t^m \sum_{n_1 = 0}^{\infty} \sum_{n_2 = 0}^{\infty} \cdots \sum_{n_m = 0}^{\infty}  \left( \prod_{i = 1}^{m} \frac{(-t)^{n_i}\comm{(\mathcal{S})^{n_i}}{\mathcal{L}}}{n_i! \sum_{j=i}^{m}\left( n_j + 1 \right)} \right) \label{eq:Mexpansion} \\
    &= t^m \frac{\mathcal{L}^m}{m!} +  t^m\sum_{n_1 = 1}^{\infty} \sum_{n_2 = 1}^{\infty} \cdots \sum_{n_m = 1}^{\infty}  \left( \prod_{i = 1}^{m} \frac{(-t)^{n_i}\comm{(\mathcal{S})^{n_i}}{\mathcal{L}}}{n_i! \sum_{j=i}^{m}\left( n_j + 1 \right)} \right) . \label{eq:Mseparated}
\end{align}
\end{widetext}
For the sake of future convenience, we define the new quantity $\Tilde{M}^{(m)}(t) := t^m M^{(m)}(t)$, from which we obtain the following general form for the correction terms of the AGF in \cref{eq:agfexpansion} in terms of \cref{eq:supersol}:
\begin{align}
    &F^{(m)}(t) \notag \\
    &= -\int_{\mathscr{H}}  \ev{\mathcal{U}^{\dag} \circ  \mathcal{U} \circ \left(t^m M^{(m)}(t) \right) \left[ \rho_0 \right]}_0 d\rho_0 \\
    &= -t^m \int_{\mathscr{H}} \Tr{M^{(m)} \left[ \rho_0 \right] \rho_0} d\rho_0 . \label{eq:channelcorrection}
\end{align}
This integral over the Fubini-Study measure of pure initial states may now be evaluated using results from Weingarten calculus \cite{collins_weingarten_2022} (see \cref{app:A:subsec:3}). From here we obtain our main result for the general perturbative expansion of the AGI, $\mathcal{I} = 1 - \mathcal{F} = \sum_{m=1}^{\infty} \mathcal{I}^{(m)}$:
\begin{align}
   \mathcal{I} = \frac{-1}{d(d+1)}\sum_{m=1}^{\infty} (\gamma t)^m \Tr{M^{(m)}(t)}, \label{eq:AGFcorrection}
\end{align}
where $d$ is the dimensionality of the qudit system, and the powers of $(\gamma t)$ have been grouped to form a single, dimensionless quantity.

\begin{figure}[t]
    \centering
    \includegraphics[width=\columnwidth]{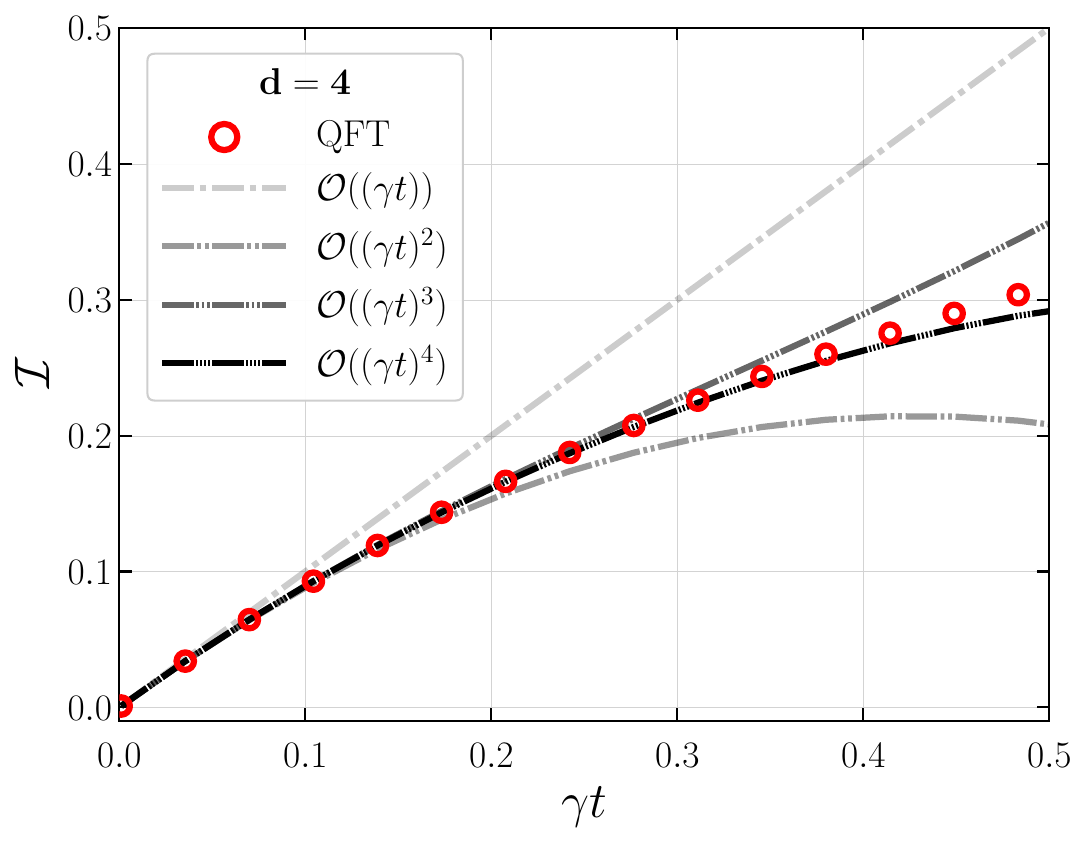}
    \caption{\justifying \textbf{Comparison of the first four correction terms of the AGI to numerical simulations.} AGIs for the QFT gate applied to a $d=4$ qudit undergoing pure dephasing ($L = J_z$) at couplings $\gamma t \in \left[ 0, 0.5 \right]$. The discrete points (red) represent the simulated AGI values, while the successive dashed-dotted lines represent the correction terms from first to fourth order.}
    \label{fig:01}
\end{figure}

As an illustrative example of the utility of this expression, in \cref{fig:01} we present the cumulative correction terms up to fourth order in $\gamma t$ for the QFT gate $F_4$ acting on a qudit of $d=4$. With $t=1$ normalised by the amplitude of the control Hamiltonian $H_c$, the pure dephasing operator $L=J_z$ was applied over the couplings $\gamma t \in \left[ 0, 0.5 \right]$. While the first-order approximation shows good agreement with the simulated AGI for small $\gamma t \ll 1$, the deviation becomes significant as the coupling grows large, necessitating the addition of the higher-order corrections.

Looking closer at this general result for $\mathcal{I}$, let us consider the case of no coherent driving. This is implemented through a zero control Hamiltonian, $H = \mathbb{0}_d$, giving $\mathcal{S}=\mathbb{0}_{d\times d}$ and corresponding to evolution by the identity gate $U_g = \mathbb{1}_d$. Then, the system evolves only through decay due to the collapse operator $\mathcal{L}$, since only the iterated commutator terms $\comm{(\mathcal{S})^{n_i}}{\mathcal{L}}$ for $n_i=0$ in $M^{(m)}(t)$ survive, resulting in 
\begin{align}\label{eq:mthorderAGIcorrection}
    \mathcal{I}^{(m)}_{0} = \frac{-(\gamma t)^m}{d(d+1)} \frac{\Tr{\mathcal{L}^m}}{m!},
\end{align}
which is precisely the first term in \cref{eq:Mseparated}, with the subscript $0$ used to refer to the gate-independent correction term. Therefore, generalising to nonzero $\mathcal{S}$, it is clear that the $m$-th order correction can always be separated into the sum of two parts as in \cref{eq:Mseparated}: These gate-independent terms $\mathcal{O}((\gamma t)^m)$ (when all $n_i=0$) and the product of the gate-dependent iterated commutators $\mathcal{O}((\gamma t)^m \prod_{i=1}^m t^{n_i})$.

Combining the summation over $m$ in \cref{eq:AGFcorrection} and the $m$-th order expression in \cref{eq:mthorderAGIcorrection} allows us to rewrite the gate-independent part as a Taylor expansion,
\begin{align}
    \mathcal{I}_{0} = \frac{\Tr{ \mathbb{1}_{d\times d} - e^{\gamma t \mathcal{L}}}}{d(d+1)}.
\end{align}

Furthermore, as long as the collapse operator is real and symmetric, obeying $L=L^*=L^T=L^{\dag}$, (as for $J_z$, for example), it is possible to use a multinomial expansion on the operator form of $\mathcal{L}$ in \cref{eq:Lprepost} to obtain explicit expressions to arbitrary order $(\gamma t)^m$ for the gate-independent correction in \cref{eq:mthorderAGIcorrection}:
\begin{align}
    &\Tr{\mathcal{L}^m} \notag \\
    &= \Tr{\left( L^* \otimes L - \frac{1}{2}L^{\dag}L \otimes \mathbb{1} - \frac{1}{2} \mathbb{1}\otimes L^{\dag}L \right)^m} \\
    &= m!\sum_{k_1+k_2+k_3=m}\frac{\Tr{L^{k_1+2k_2}}\Tr{L^{k_1+2k_3}}}{k_1!k_2!k_3!(-2)^{k_2+k_3}}.
\end{align}

In the case of a pure (spin) dephasing channel $L = J_z$, considering that $\Tr{J_z^m} \sim d^{m+1}$, it can be easily seen that this expression for the $m$-th-order correction scales as
\begin{align}
\mathcal{O}\left( \Tr{\mathcal{L}^m} \right) &\sim d^{2(m+1)} \\
    \implies \mathcal{O}\left( \mathcal{I}_0^{(m)} \right) &\sim (\gamma t)^m d^{2m} .
\end{align}
Now, given this scaling of the gate-independent AGI, let us refer to the work of Chiesa \textit{et al.} \cite{chiesa_molecular_2020}, where the authors proposed the use of a physical qudit to encode a logical qubit. Using their binomial encoding and error-correction protocol would result in a logical qubit robust to pure dephasing up to order $(\gamma t)^m$ where $2m+1=d$. Effectively, the AGI of such a logical qubit would be of the same order as $\mathcal{I}_0^{(m)}$. In turn, to maintain $\mathcal{I}_0 \ll 1$, this sets a limit on the maximum dimension (equivalently, order $m$ of noise correction), such that $d<\frac{1}{\sqrt{\gamma t}}$. Such investigations for protecting against decoherence have already been studied with alternative logical qubit encodings and a discrete error evolution protocol in the work of Miyahara \textit{et al.} \cite{miyahara_decoherence_2023}.

In what follows in the remainder of this study, we shall focus on the first- and second-order correction terms to the AGF. We note that the preceding analytical approach provides a framework for computing any correction term to arbitrary order. However, explicit calculations up to second-order serve as an extension of our prior work \cite{jankovic_noisy_2024}, in which the first-order correction to the AGF was investigated and the consequent conditions on gate efficiency derived in terms of the figure-of-merit as a function of qudit dimension. Thus, we study in detail the effects due to the second-order correction terms as they relate to our particular system of interest, particularly with regard to the choice of quantum gate and collapse operator.

\subsection{\label{sec:2:subsec:3} Nonlinearity of the AGI in the Strong Coupling Regime}
Let us now motivate our investigation into the AGI in the regime of strong coupling between the qudit and the environment. As identified in \cref{fig:01}, the linear first-order approximation may be sufficient in the weak coupling regime, but nonlinear effects become significant above $\gamma t \sim 10^{-1}$. It is therefore necessary to identify a clear transition or separation between these two regimes, and how the AGI behaves in the limit of strong coupling.

\begin{figure}[t]
    \centering
    \includegraphics[width=\columnwidth]{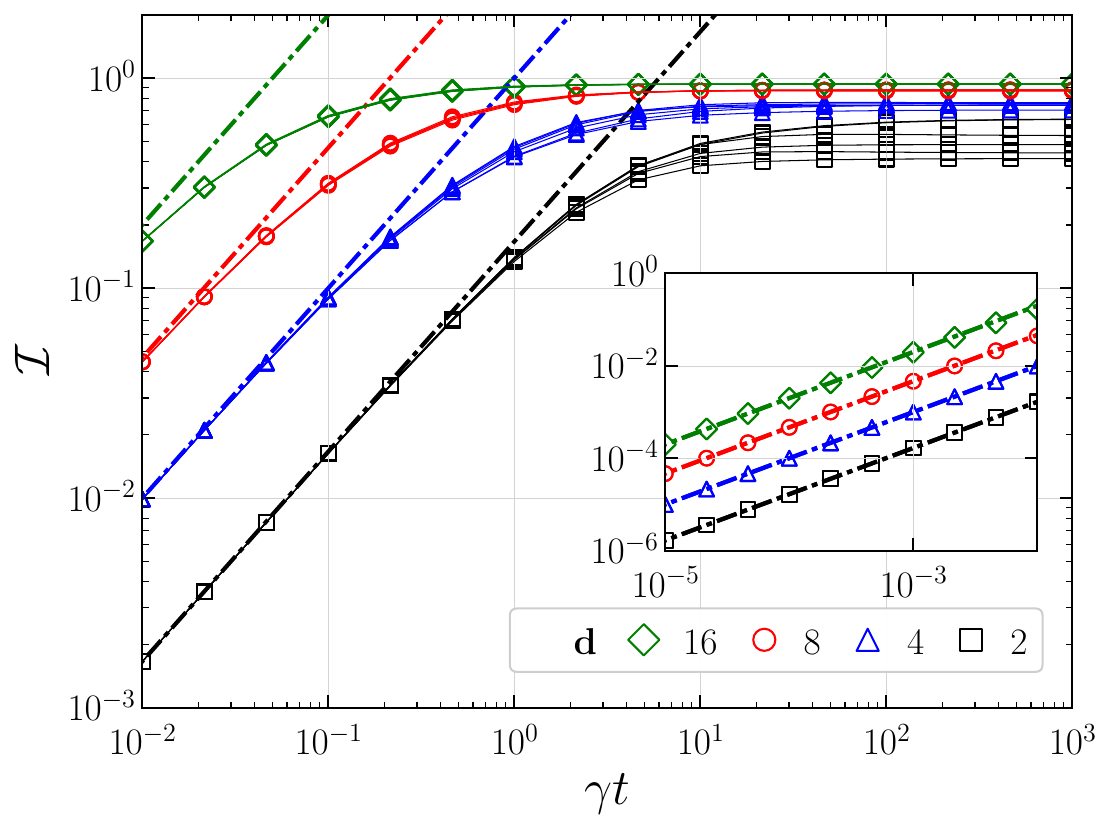}
    \caption{\justifying \textbf{Deviation from linearity of the AGIs for qudits under strong dephasing.} Simulations of the AGI against the strength of pure dephasing ($L = J_z$) are plotted on log-log axes over $\gamma t \in \left[10^{-2}, 10^{3}  \right]$. The simulations were performed on a set of qudit dimensions, $d \in \lbrace 2, 4, 8, 16 \rbrace$ each for 100 Haar-random quantum gates. The dashed lines represent the linear regime given by the first-order correction to the AGI at each dimension. The inset shows the linear regime for smaller values of $\gamma t \in \left[10^{-5}, 10^{-2}  \right]$. For stronger noise the AGIs exhibit a transition from the linear to nonlinear regime that then saturates at a stable plateau value. These plateau values vary for qudit dimension and also for different quantum gates, highlighting the gate-dependence. The saturation points $(\gamma t)^*$ of the stable regime are also dependent on both the qudit dimension and gate type, with higher-dimensional qudits deviating from linearity and saturating earlier.}
    \label{fig:02}
\end{figure}

To investigate, we perform calculations of the AGI over a large range of scales for $\gamma t$. Therefore, in \cref{fig:02}, we show numerical simulations of the AGIs for qudits under pure dephasing ($L = J_z$) where the coupling strengths varied logarithmically in $\gamma t \in \left[ 10^{-5}, 10^{3} \right]$. Evolutions were repeated on qudits with dimensions $d \in \lbrace 2, 4, 8, 16 \rbrace$ for 100 Haar-random quantum gates (see \cref{app:C:subsec:1} for a discussion on methods for generating random quantum gates), with the inset showing the linear regime for small $\gamma t \in \left[ 10^{-5}, 10^{-2} \right]$ and the main figure showing the transition region and strong-coupling regime. The dashed-dotted straight lines give the first-order correction terms at each qudit dimension, defined by the linear function \cite{jankovic_noisy_2024}
\begin{align}\label{eq:Jzfirstorderexplicit}
    \mathcal{I}^{(1)} = \frac{d (d - 1)}{12} \gamma t ,
\end{align} 
showing the $d$-dependence in the weak coupling regime.

Further effects of the system dimension on the AGI are notable for larger couplings: The start of deviation from linearity, the plateau values $\mathcal{I}^*$ that the AGI curves converge to in the large $\gamma t$ limit, and the associated saturation points $(\gamma t)^*$ (both denoted by the superscript $*$) at which the curves reach their respective plateau values. Each of these effects confirms an overall decrease in robustness as the system dimension increases: Compared to the behaviour for qubits, the AGIs (for arbitrary gates) of higher-dimensional qudits begin to deviate from the linear regime, and reach the higher plateau limits, at lower noise thresholds. These quantities shall be studied in more detail in \cref{sec:2:subsec:4}.

Finally, it is interesting to note the gate dependence of the $\mathcal{I}^*$ and $(\gamma t)^*$, which is more pronounced for smaller $d$. Given the sample size of 100 gates, it is apparent that, at each dimension, there exists an upper and lower bound to the AGI plateau values, with all the sampled gates converging within this range. It is also noticeable (particularly so for $d=2$) that the $(\gamma t)^*$ and $\mathcal{I}^*$ are positively correlated, as curves with lower plateau infidelity appear to reach their plateau values at lower $(\gamma t)^*$ than those with higher plateau infidelities. 

\begin{figure}[t]
    \centering
    \includegraphics[width=\columnwidth]{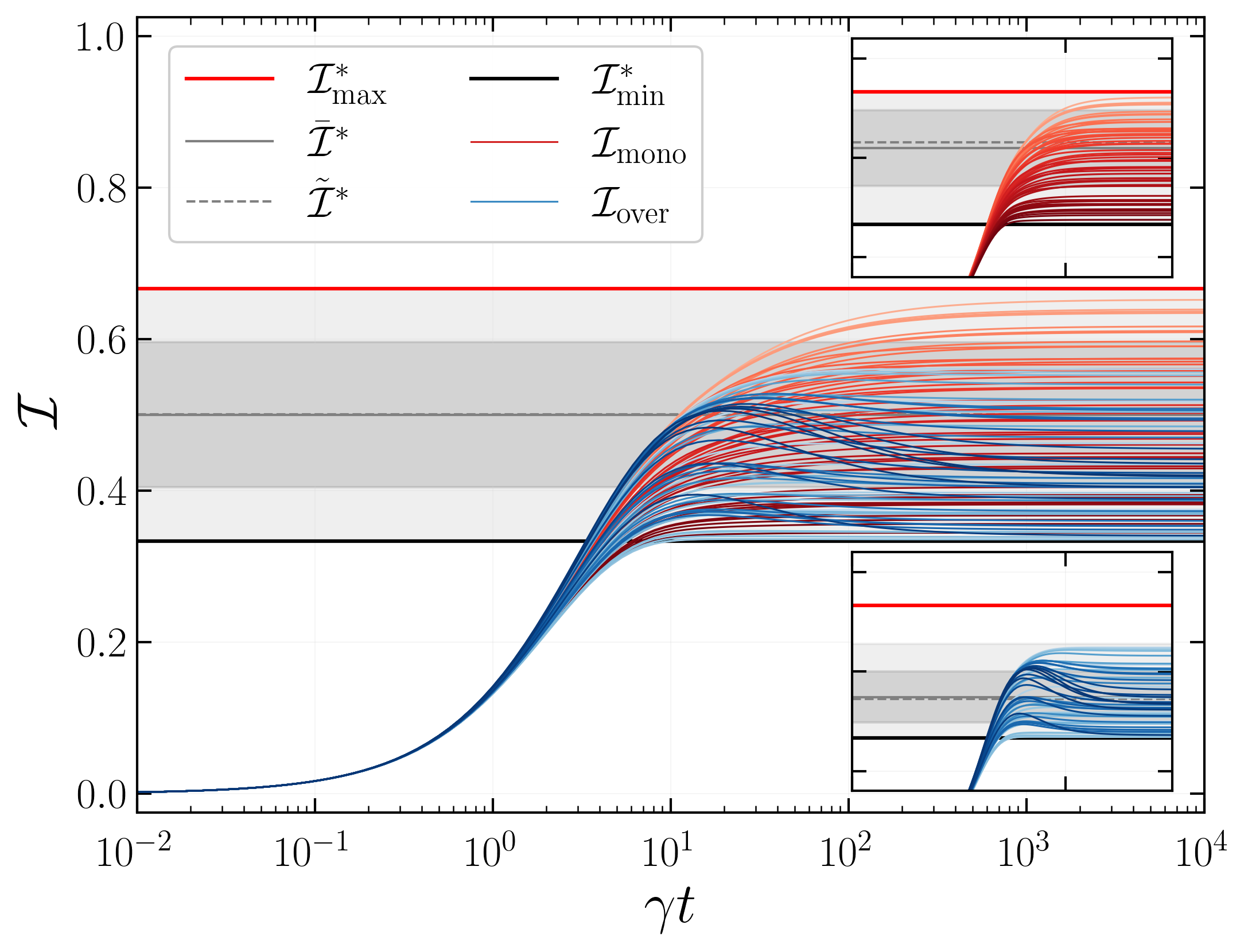}
    \caption{\justifying \textbf{Large-$\gamma t$ behaviour of the AGIs for Haar-random qubit gates under pure dephasing.} Simulations of the AGI were performed for a sample of one million (50 displayed) Haar-random qubit ($d=2$) gates under pure dephasing ($L = J_z$) for $\gamma t \in \left[ 10^{-2}, 10^{4} \right]$. For each gate, the AGI curves fall into one of two groups, exhibiting different behaviour near their respective saturation points ($(\gamma t)^*$). The curves in red ($\mathcal{I}_{\mathrm{mono}}$) approach their plateau values monotonically from below. The blue curves ($\mathcal{I}_{\mathrm{over}}$) rapidly approach their plateau values, before overshooting once with a single turning point, and then converging monotonically from above. The degree of blue shading for each curve indicates the degree of overshoot above the plateau value. All sampled gates converged to their $\mathcal{I}^*$ plateau values within the range $\left[\mathcal{I}^*_{\mathrm{min}}, \mathcal{I}^*_{\mathrm{max}}\right]$ indicated by the black and red horizontal lines. In the figure and insets, the lighter and darker shaded regions indicate the vertical extent of the sampled gates and their standard deviations, with the solid and dashed lines indicating the mean ($\bar{\mathcal{I}}^*$) and median ($\tilde{\mathcal{I}}^*$) AGI values, respectively. The inset figures show each group of gates in isolation. The red group span the full bounding range, but are weighted upwards by outliers near the upper limit. The blue group do not span the full range, but are weighted towards the lower limit. Only the average AGI taken over both groups converges towards the expected mean value of $0.5$.}
    \label{fig:03}
\end{figure}

Studying these effects of the large-$\gamma t$ behaviour in more detail, in \cref{fig:03} we present simulations of a sample of 1 million Haar-random qubit ($d=2$) gates over $\gamma t \in \left[ 10^{-2}, 10^{4} \right]$. This larger sample size provides stronger evidence that the $\mathcal{I}^*$ are indeed bounded above and below, as shown by the red and black horizontal lines, respectively. It was observed that the randomly generated quantum gates fall into one of two groups, indicated by the red and blue shaded curves. The red shaded curves ($\mathcal{I}_{\mathrm{mono}}$) approach their plateau values monotonically from below. The blue shaded curves ($\mathcal{I}_{\mathrm{over}}$) all experience an overshoot above their plateau value, with a single turning point, before settling to $\mathcal{I}^*$ monotonically from above. From the full sample, approximately $73\%$ of all gates fell in the red, monotonic group, and the remaining $27\%$ in the blue, overshoot group. The degree of blue shading represents the amount of overshoot, defined by the difference $\mathcal{I}_{\mathrm{max}} - \mathcal{I}^*$ between the height of the peak and the plateau value. Furthermore, there appears to be some correlation between this quantity and the saturation point; the larger the overshoot the longer the curve takes to settle. Gates that experience this overshoot may therefore perform worse at intermediate noise strengths over the strong coupling limit. 

The insets show each category plotted separately, from which it may be seen that, unlike the blue curves, the red curves span the range between the upper and lower bounds, indicated by the lightly-shaded regions. The darker shaded regions, show the standard deviation about the mean, $\bar{\mathcal{I}}^*$, with the dashed line being the median value, $\tilde{\mathcal{I}}^*$. We can see from these quantities that, despite spanning the range, the red curves are not distributed uniformly, but instead weighted upwards towards the upper bound, while the blue curves are weighted towards the lower bound. Both groups have significant numbers of outliers at either extreme, shifting the medians away from the means. In fact, it is only when averaging over the combined samples that the distribution becomes centered about the mean of $\bar{\mathcal{I}^*}=0.5$ (see \cref{app:C:subsec:2} for further investigations on the distributions of these groups of curves within the bounding region). 

\subsection{\label{sec:2:subsec:4} Gate- and Noise-Dependence of the Fidelity}
Having identified particular aspects of interest in the AGI behaviour in the strong coupling regime, let us investigate these characteristics in further detail.

Repeating the simulations presented in \cref{fig:03} for larger dimensions, we observed that the lower and upper bounds, $\mathcal{I}^*_{\mathrm{min}}$ and $\mathcal{I}^*_{\mathrm{max}}$ respectively, and the mean value, $\bar{\mathcal{I}}^*$, depend only on the dimension $d$ of the system, and are given precisely by the following expressions, for any $d$:
\begin{align}
    \mathcal{I}^*_{\mathrm{max}} &= 1 - \frac{1}{d + 1} , \label{eq:Imax} \\
    \bar{\mathcal{I}}^* &= 1 - \frac{1}{d} , \label{eq:Iavg} \\
    \mathcal{I}^*_{\mathrm{min}} &= 1 - \frac{2}{d + 1} . \label{eq:Imin}
\end{align}
In \cref{app:C:subsec:2}, we present these further examples of the distributions of $\mathcal{I}^*$ for ensembles of Haar random gates with dimensions $d=2,4,8,16$, which follow these expressions. It has previously been shown \cite{horodecki_general_1999} that, for the case of a qudit in the singlet state evolving in a noisy quantum channel, the infidelity is bounded by $1 - \frac{2}{d + 1} \leq \mathcal{I}^* \leq 1 - \frac{1}{d}$. This result matches with our case of a single qudit in the large-$d$ limit due to convergence of the mean and upper bound. However, for our single qudit system we shall use our existing framework to prove these limits for arbitrary $d$.

Writing the quantum channel as $\mathcal{E} = \mathcal{U} \circ M$, and substituting into the expression for the AGF in \cref{eq:AGF},
\begin{align}
    \mathcal{F} &= \int_{\mathscr{H}} \Tr{M\left[ \rho_0 \right] \rho_0} d\rho_0 \\
    &= \frac{\Tr{M} + \Tr{M\left[ \mathbb{1} \right]}}{d(d+1)} .
\end{align}
Given that the quantum channel is CPTP, and that $\mathcal{U}$ is a unitary operator, we know that the superoperator $M$ must also be trace-preserving and thus $\Tr{M\left[ \mathbb{1} \right]} = d$. Furthermore, in the strong coupling limit where the collapse operator $L=J_z$ describes pure dephasing, it is clear that the action of $M$ on a state $\rho$ is to transform it to a diagonal density matrix where the coherences (off-diagonal elements) have all decayed to zero. Expressing $M=\sum_{k=1}^K E_k^* \otimes E_k$ in the Kraus representation, the trace can be written as
\begin{align}
    \Tr{M} = \sum_{k=1}^K \sum_{i=0}^{d-1} \abs{E_{k, \, ii}}^2.
\end{align}
Furthermore, it can be shown that (see \cref{app:A:subsec:4} for a detailed derivation) each of the $d$ squared-elements are bounded by $0 \leq \abs{E_{k, \, ii}}^2 \leq 1$ and hence in the large noise regime the AGI is bounded by
\begin{align}\label{eq:upperlowerbounds}
    1 - \frac{2}{d+1} \leq \mathcal{I}^* \leq 1 - \frac{1}{d+1} .
\end{align}

Given the significance of this result, it is now of interest to consider how the behaviour is affected by the choice of quantum gate. In other words, we want to identify any particular unitary operators $U_g$, or properties thereof, whose AGIs $\mathcal{I}(\mathcal{E}, U_g)$ when $(\gamma t) \gg 1$ are precisely the $\mathcal{I}^*_{\mathrm{min}}$, $\mathcal{I}^*_{\mathrm{max}}$ or $\bar{\mathcal{I}}^*$.


\begin{figure}[t!]
    \centering
    \includegraphics[width=\columnwidth]{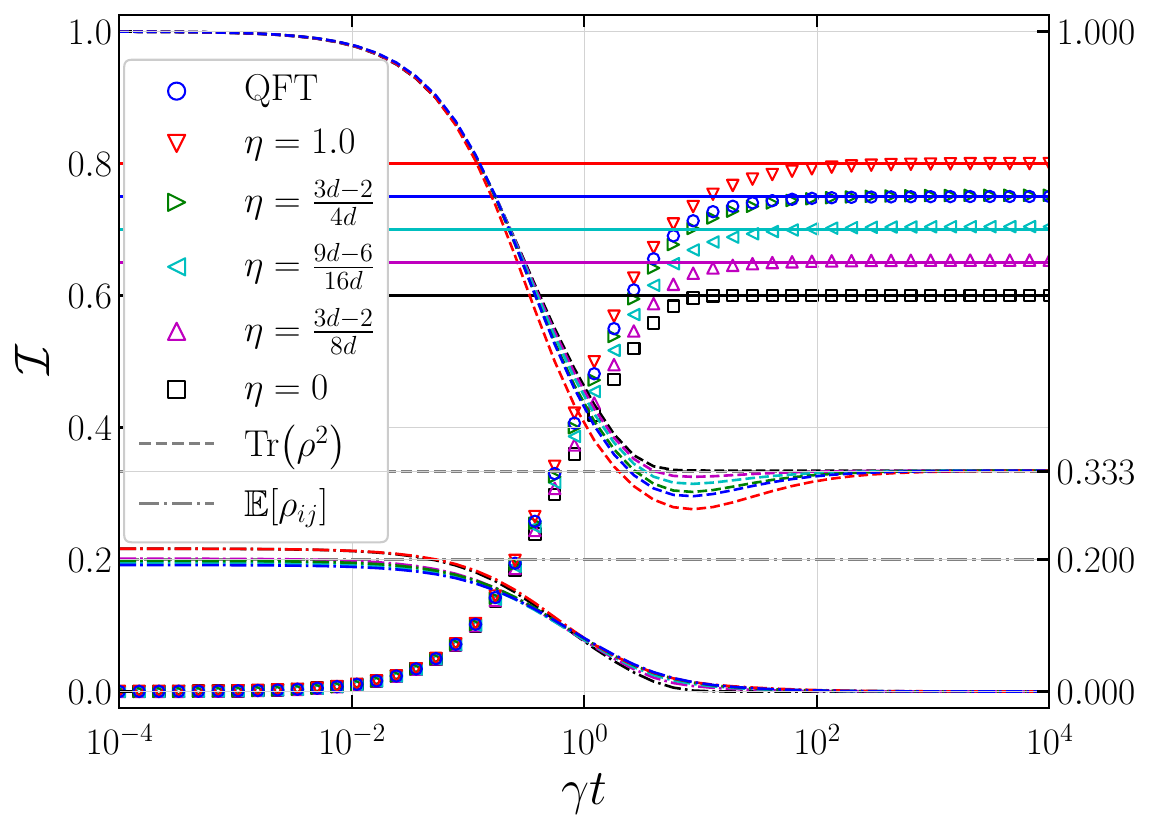}
    \caption{\justifying \textbf{Gate-dependence of the AGIs and qudit state evolution as functions of noise strength.} (Data points) AGI simulations of a $d=4$ qudit under pure dephasing were time-evolved for each of the listed quantum gates: the (blue circles) QFT gate, and (other points) interpolated $X^{\eta}$ gates for $\eta \in \left[ 0, 1 \right]$ as shown. Additionally shown for each gate: the (solid horizontal lines) plateau value $\mathcal{I}^*$ of the AGI, (dashed curves) state purity $\left( \Tr (\rho^2) \right)$, and (dashed-dotted curves) mean coherences $\mathbb{E}[\rho_{ij}]= \frac{2}{d(d-1)}\sum_{i>j} \abs{\rho_{i,j}}$ for $i>j$. Note that: (i) the AGI data points of the QFT and $X^{\eta}$ gate for $\eta = \frac{3d - 2}{4d}$ overlap and converge to the same plateau value (for any $d$), (ii) the $\eta$ values were chosen as they produce equally spaced AGI plateau values (for any $d$), (iii) the final state purity and coherences for each gate were averaged over a set of 2400 random pure Hermitian initial states, and (iv) their curves converge to $\frac{1}{d-1}$ and $0$ respectively, following the behaviour of the AGIs as they transition from the linear to nonlinear to plateau-like regimes.}
    \label{fig:04}
\end{figure}

In \cref{fig:04}, we present simulations of the AGI, as a function of $\gamma t$, on a $d=4$ qudit for a set of gates that do indeed reflect this behaviour when undergoing pure dephasing. The AGI values (discrete data points) are accompanied by dashed and dotted curves representing the state purity and mean coherences, while the horizontal lines show the $\mathcal{I}^*$ values of the associated gates. The purities and coherences were calculated over a set of $n=2400$ randomly generated pure Hermitian density matrices. The purity $\Tr{\rho^2}$ decays from $1$ to a constant $\frac{1}{d-1}$, while the averages of the $\frac{d(d-1)}{2}$ coherences decay to zero, confirming the dephasing action of the $J_z$ collapse operator. 

Regarding the set of gates, in addition to the QFT gate, we introduce the interpolated-$X$ gate (see \cref{app:B}), $X^{\eta}$, where the matrix is raised to the $\eta$-th power for $\eta \in \left[ 0, 1 \right]$. Clearly, $\eta = 1$ or $0$ produces the $X$ or $\mathbb{1}_d$ gate, respectively, while for $0 < \eta < 1$ the action interpolates between these two extremes.

We find that 
\begin{align}
    \mathcal{I}^*(\mathcal{E}, \mathbb{1}_d) &= \mathcal{I}^*_{\mathrm{min}} , \\ 
    \mathcal{I}^*(\mathcal{E}, F_d) &= \bar{\mathcal{I}}^* , \\ 
    \mathcal{I}^*(\mathcal{E}, X_d) &= \mathcal{I}^*_{\mathrm{max}} .
\end{align}
Specifically, the $\mathcal{I}^*$ are extremised above and below by the $X$ and identity gates, respectively, while the mean value $\bar{\mathcal{I}}^*$ indeed corresponds to the action of the QFT gate that creates a superposition state. Furthermore, by careful choice of the values of $\eta$, we can obtain intermediate values of $\mathcal{I}^*$. For example, the $\eta$ values chosen and shown in \cref{fig:04} lead to curves with plateau values that are equally spaced within the bounding region. We observed that for the value $\eta = \frac{3d - 2}{4d}$, the $\mathcal{I}^*$ matches that of the QFT gate, also converging to the mean.

The effect of these gates on the AGI can be understood by following the arguments in \cref{app:A:subsec:4} regarding the operator $M$ and its form based on the choice of quantum gate. It is clear that for the identity gate, or more generally any diagonal unitary matrix, $\Tr{M}=d$ which minimises $\mathcal{I}^*=1-\frac{2}{d+1}$. Indeed, this result applies directly to the set of generalised-PHASE gates, including Clifford CLOCK gate $Z_d$ and the non-Clifford $\pi/8$ gate $T_d$, whose AGI curves $\mathcal{I}(\gamma t)$ follow that of the identity gate and plateau at $\mathcal{I}^*_{\mathrm{min}}$. 

Similarly, the QFT gate produces a superposition state with each of the $d$ basis elements of $\rho$ being $\dfrac{1}{d}$, thus $\Tr{M}=1$ and $\mathcal{I}^*=1 - \frac{1}{d}$. Finally the $X$ gate increments all states circularly, and, having no diagonal elements, results in traceless $\Tr{M}=0$ and thus $\mathcal{I}^*=1 - \frac{1}{d+1}$.

\begin{figure}[t]
    \centering
    \includegraphics[width=\columnwidth]{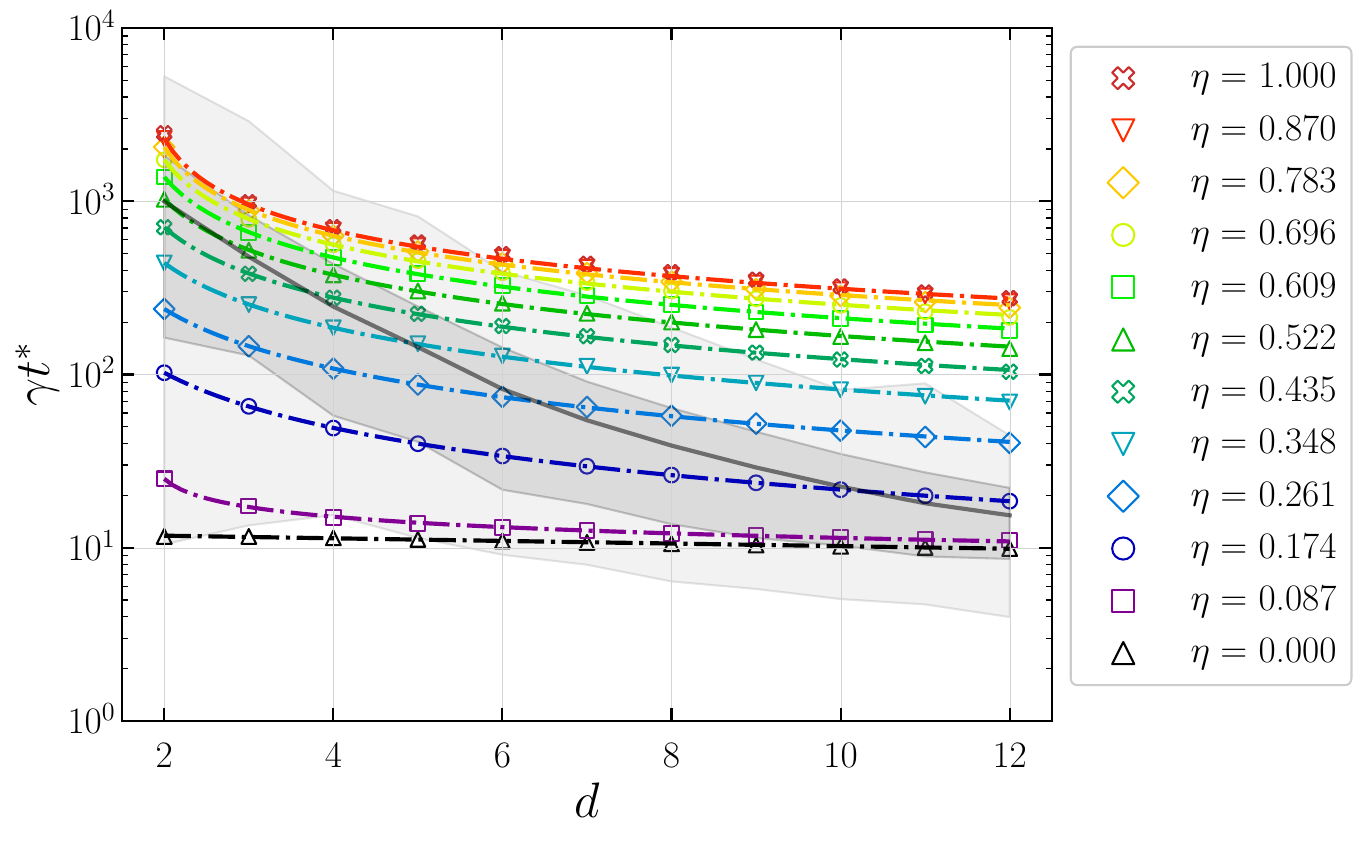}
    \caption{\justifying \textbf{Saturation points ($(\gamma t)^*$) of the AGIs for interpolated $X^{\eta}$ gates as functions of qudit dimension.} AGIs were calculated for a set of $12$ interpolated $X^{\eta}$ gates for $0\leq \eta \leq 1$, at dimensions $d\in\left[ 2, 12\right]$. For each dimension and gate, the AGI curve over $\gamma t$ was interpolated by a cubic spline. The saturation points $(\gamma t)^*$ were identified by root-finding algorithm of the points at which they converged (within $\varepsilon < 10^{-8}$) to their AGI plateau values. The plotted data points, on log-linear axes, indicate these root values as functions of $d$ for each gate, and the dashed lines represent their associated power-law fits. This analysis was repeated for a set of $n=4800$ Haar-random gates at each dimension. The light shaded areas are bounded by the maximum and minimum saturation points found at each dimension. The darker shaded areas represent a $1\sigma$ deviation about the mean (solid gray line).}
    \label{fig:05}
\end{figure}

Let us now consider how the choice of quantum gate affects the saturation point $(\gamma t)^*$ at which the AGI attains the plateau value $\mathcal{I}^*$. \Cref{fig:05} presents simulations of the plateau saturation points as a function of the qudit dimension for set of $X^{\eta}$ gates over a range of $\eta$ values from $0$ to $1$. In addition, the calculations were repeated over a set of $n=4800$ Haar random gates at each dimension from $d=2$ to $12$. The grey curve represents the mean value of the saturation point at each dimension, while the dark grey and light gray shaded regions indicate the standard deviations and max-min values, respectively.

For each of the simulated $X^{\eta}$ gates, the dashed-dotted lines represent power-law fits to the discrete data points at each dimension. The power-law model is given by
\begin{align} \label{eq:powerlaw}
    (\gamma t)^* = \alpha (d - \beta)^{\delta},
\end{align}
and, fitted by nonlinear least-squares regression, resulted in $R^2>0.999$ for each fit. It is interesting to note that for the case of the identity gate, the saturation point remains constant independent of dimension. 

Continuing the investigation, we performed regression analysis of the fitted model parameters $\alpha$, $\beta$ and $\delta$ across the set of $\eta$ values. The fits are presented in \cref{app:C:subsec:3}. Parameter $\alpha$ was best described by a sigmoid model while $\beta$ and $\delta$ by exponential models,
\begin{align}
    \alpha(\eta) &= \frac{\alpha_0}{(1 + e^{\alpha_1 (\eta - \alpha_2)})} + \alpha_3 , \label{eq:powerlawalpha} \\
    \beta(\eta) &= \beta_0 e^{\beta_1 (\eta - \beta_2)} + \beta_3 , \label{eq:powerlawbeta} \\
    \delta(\eta) &= \delta_0 e^{\delta_1 (\eta - \delta_2)} + \delta_3 . \label{eq:powerlawdelta}
\end{align}

\begin{figure}[t]
    \centering
    \includegraphics[width=\columnwidth]{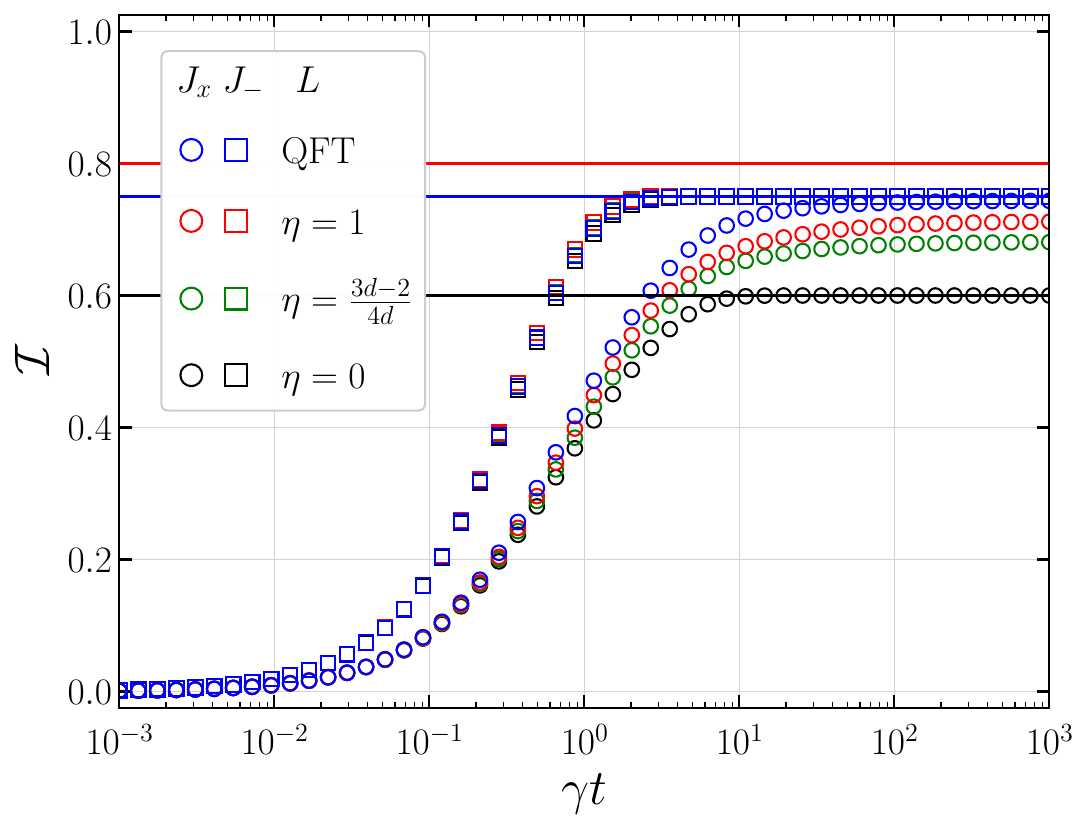}
    \caption{\justifying \textbf{Effect of bit-flip errors and relaxation processes on the AGI.} $X^{\eta}$ and QFT gate simulations were repeated on qudits for $d =4$ undergoing either (circles) bit-flip ($L = J_x$) errors or (squares) spin relaxation ($L = J_-$), rather than pure dephasing noise. The following gates were simulated:  $\mathbb{1}=X^0$ (black, $\eta=0$), $X^{\eta}$ (green, $\eta = \frac{3d - 2}{4d}$), $X$ (red, $\eta = 1$) and QFT (blue). The solid horizontal lines show the upper ($1 - \frac{1}{d + 1}$, red) and lower ($1 - \frac{2}{d + 1}$, black) bounds of the AGIs, as well as the mean ($1 - \frac{1}{d}$, blue). For $L=J_-$, the AGIs for all matrix powers of the $X^{\eta}$ gate behaved almost exactly as the QFT gate, while for $L=J_x$ the plateau AGI values for the non-zero matrix powers were observed to deviate upwards from the identity gate, but not to the same extent as for the case of pure dephasing. On the other hand, the QFT gate response was the same for all types of noise.}
    \label{fig:06}
\end{figure}


Thus far in our investigation, we have restricted analysis to the effect of pure dephasing by the operator $J_z$ on a qudit. Let us conclude this section by studying the effects of the $L=J_x$ and $L=J_-$ operators, corresponding to bit-flip errors and relaxation processes. In \cref{fig:06}, we present simulations of 4 gates: the QFT gate $F$ and 3 $X^{\eta}$ gates for $\eta={0, \frac{3d-2}{4d}}, 1$, for $d=4$ qudits interacting with the environment via each of the collapse operators separately. We may make the following remarks based on these results: (1) the plateau values $\mathcal{I}^*$ for the gates remain bounded between $\mathcal{I}^*_{\rm min}$ and $\mathcal{I}^*_{\rm max}$, (2) the QFT gate plateaus at the same mean value $\bar{\mathcal{I}}^*$ for all $L\in\lbrace J_z,J_x,J_- \rbrace$, (3) for bit-flip errors the behaviour of the identity gate remains the same, while the $X^{\eta}$ gates plateau below the QFT gate, and (4) for relaxation processes all gates have nearly the same behaviour as the QFT gate, and all plateau at $\bar{\mathcal{I}}^*$.

\subsection{\label{sec:2:subsec:5} First- and Second-Order Correction Terms to the AGI}

\Cref{sec:2:subsec:2} presented our general result for the perturbative expansion of the AGI to arbitrary order $m$ in \cref{eq:AGFcorrection}. Now, let us investigate in further detail the first and second-order correction terms. 

\begin{figure}[t]
    \centering
    \includegraphics[width=\columnwidth]{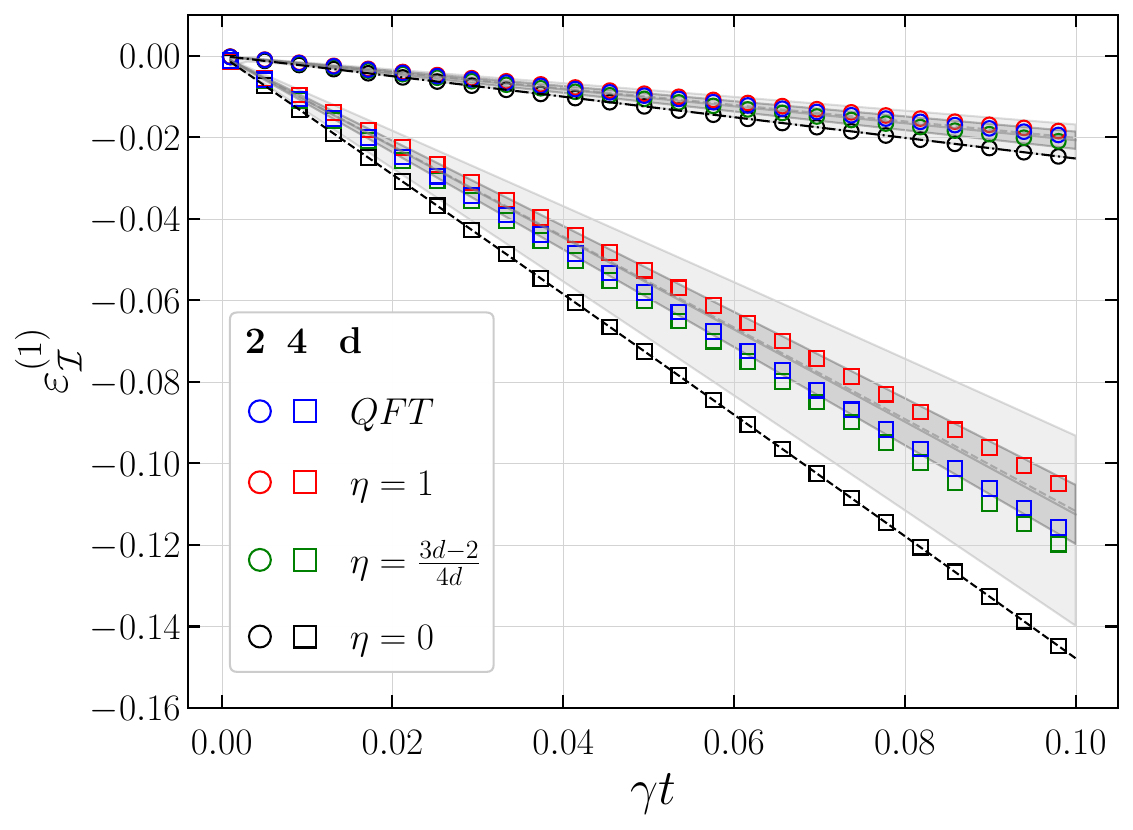}
    \caption{\justifying \textbf{Relative error of the first-order correction to the AGI, and gate-dependent deviation from linearity.} AGIs were simulated for $4$ gates acting on qubits (circles) and $d=4$ qudits (squares) undergoing $L=J_z$ pure dephasing. The relative error $\varepsilon^{(1)}_{\mathcal{I}} = \frac{\mathcal{I} - \mathcal{I}^{(1)}}{\mathcal{I}}$ was calculated for the first-order analytical correction. The dashed-dotted and dashed lines represent the first-order term of the perturbative expansion of the AGI, for $d=2$ and $d=4$, respectively. These gate-independent terms correspond to the behaviour of the identity gate, $\mathbb{1}=X^0$ (black, $\eta=0$). The $X^{\eta}$ (green, $\eta=\frac{3d-2}{4d}$), $X^1$ (red, $\eta=1$) and $F$ (blue, QFT) gates show significant deviation from the gate-independent term, and for $d=4$ are roughly $10$ times larger than for $d=2$. The solid and dashed grey lines represent the mean and median relative errors over a sample of $n=4800$ Haar-random gates, with the light and dark shaded areas indicating the min-max bounds and standard deviations.}
    \label{fig:07}
\end{figure}

Beginning with the first-order correction, at $m=1$, it can be seen that (see \cref{app:A:subsec:5}) the trace over $M^{(1)}$ simplifies to
\begin{align}\label{eq:firstorderAGIcorrection}
    \mathcal{I}^{(1)} = -\frac{(\gamma t) }{d(d+1)} \Tr{\mathcal{L}},
\end{align}
which confirms the gate-independent nature of the first-order term as the AGI depends only on the noise superoperator $\mathcal{L}$. Additionally, this trace can be expressed in terms of the regular operator $L$ as
\begin{align}\label{eq:TrL}
    \Tr{\mathcal{L}} =  \left|\Tr{L}\right|^2 - d\Tr{L^\dag L},
\end{align}
reproducing the results of \cite{jankovic_noisy_2024}. For the case of pure dephasing, we obtain the familiar result in \cref{eq:Jzfirstorderexplicit}
\begin{align} \label{eq:firstorderJzcorrection}
    \mathcal{I}^{(1)} = \frac{\gamma t}{12} d (d-1),
\end{align}
using the element-wise definition of the dephasing operator in \cref{eq:Jz}
\begin{align} \label{eq:Jzelements}
    \left( J_z \right)_{ii} = \frac{d - (2i - 1)}{2}.
\end{align}
Now, if we compare this analytical result with numerical simulations, we observe in \cref{fig:07} the relative error of the first-order correction to the AGI for dimensions $d={2, 4}$ over $\gamma t \in \left[ 0, 0.1 \right]$. This relative error is expressed by
\begin{align}
    \varepsilon^{(1)}_{\mathcal{I}} = \frac{\mathcal{I} - \mathcal{I}^{(1)}}{\mathcal{I}} .
\end{align}
The simulations were performed for the QFT, identity, $X$ and $X^{\eta}$ gates for $\eta = \frac{3d-2}{4d}$. The dashed lines show the expected analytic results for each dimension. The dark and light shaded regions represent the standard deviations and max-min values taken over a sample of Haar random gates

Clearly, the gate independence of the first-order correction term matches precisely the identity gate. The remaining gates show the expected deviation from linearity, with the analytic value underestimating the AGI values. We observe that the relative error for $d=2$ is approximately one order of magnitude smaller than for $d=4$.

Moving on to the second-order correction term, we have from \cref{eq:AGFcorrection}
\begin{align} \label{eq:secondorderfull}
    &\mathcal{I}^{(2)} = \frac{-(\gamma t)^2}{d(d+1)} \times \notag \\
    &\sum_{n_1,n_2=0}^{\infty} \frac{(-t)^{n_1 + n_2}\Tr{\comm{(S)^{n_1}}{\mathcal{L}}\comm{(S)^{n_2}}{\mathcal{L}}}}{n_1 ! (n_2 + 1)! (n_1 + n_2 + 2)},
\end{align}
which we reduce (see \cref{app:A:subsec:6}) to a single iterated commutator using the relation $n_1+n_2+2=s$,
\begin{align}
    \mathcal{I}^{(2)} = -\frac{(\gamma t)^2}{d(d+1)} \sum_{s=0}^\infty \Tr{\mathcal{L}\left[(\mathcal{S})^{s}, \mathcal{L}\right]}\frac{(-t)^s}{(s+2)!}.
\end{align}
For $s=0$, the trace term reduces further to $\Tr{\mathcal{L}^2}$ by definition of the iterated commutator. Additionally, we show (see \cref{app:A:subsec:7}) that, using the property \cite{volkin1968iterated},
\begin{align}
    \comm{(A)^s}{B} = \sum_{k=0}^s (-1)^k \binom{s}{k} A^{s-k} B A^k,
\end{align}
the trace over the iterated commutator is zero for all odd values of $s$. Therefore,
\begin{align} \label{eq:secondorderAGIcorrection}
     &\mathcal{I}^{(2)} = -\frac{(\gamma t)^2}{d(d+1)} \times \notag \\
     &\left( \frac{\Tr{\mathcal{L}^2}}{2} + \sum_{s=2}^\infty \frac{\Tr{\mathcal{L}\comm{(\mathcal{S})^{s}}{\mathcal{L}}}(-t)^s}{(s+2)!}\right) .
\end{align}
This shows us that the AGI remains gate-independent at order $\gamma^2 t^2$. Indeed the first appearance of gate-dependence through the $\mathcal{S}$ operator occurs at order $\gamma^2 t^4$, when $s=2$ and the iterated commutator term contains $\mathcal{S}$ and is non-zero.

\begin{figure}[t]
    \centering
    \includegraphics[width=\columnwidth]{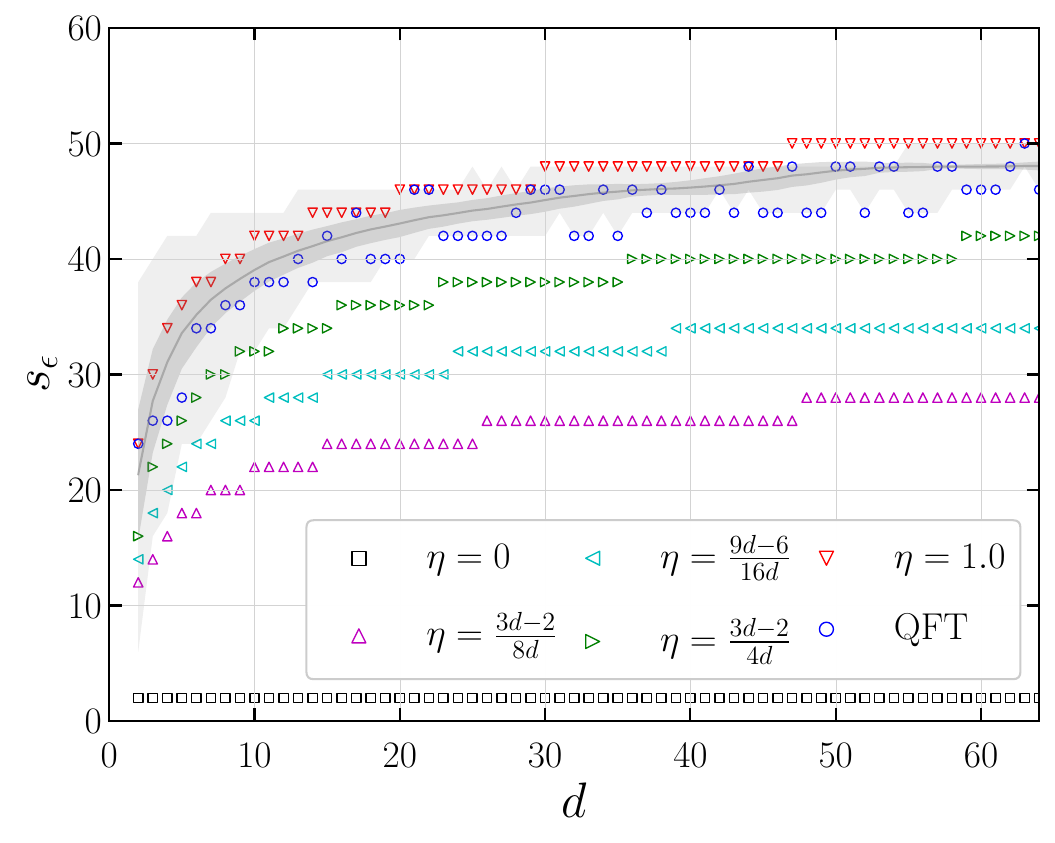}
    \caption{\justifying \textbf{Iteration-dependent convergence of the second-order AGI correction against qudit dimension.} The critical order $s_{\varepsilon}$ was calculated as the number of iterations over the summation term in the second-order AGI correction such that the absolute difference $\abs{f_2(s_{\varepsilon}) - f_2(s_{\varepsilon} - 1)} < \varepsilon$ is less than the error threshold $\varepsilon = 1\times 10^{-8}$. Calculations were performed for a set of interpolated $X^{\eta}$ gates and the QFT gate, as well as $n=2500$ Haar-random gates, over dimensions $d\in \left[ 2, 64 \right]$. The identity gate ($\eta=0$) only required the minimal $s=2$ iterations, independent of dimension, since the identity commutes with the collapse operator. Of the interpolated gates, the number of iterations required scales with the interpolation parameter $\eta$, while the QFT gate shows marked variation over $d$, without exceeding the critical order for $\eta=1$. The mean of the Haar-random gates (dark grey) also lies below the $\eta=1$ curve, while the min-max interval, (light grey), shows gates at lower dimension requiring a larger number of iterations to converge.}
    \label{fig:08}
\end{figure}

Since the summation of $s$ is over all even integers, we can introduce a maximum cutoff value to render this calculation computationally tractable. Indeed, for the correction term to not be unbound, we must have that the sum is convergent, allowing us to identify such a value. We can understand this computationally by identifying the $s$-th order of the summation,
\begin{align} \label{eq:sconvergenceterm}
    f(s) = \Tr{\mathcal{L}\left[(\mathcal{S})^{s}, \mathcal{L}\right]}\frac{(-t)^s}{(s+2)!} .
\end{align}
Then, for a given error threshold $\varepsilon$, the cutoff order $s_{\varepsilon}$ is simply the $s$ at which the absolute difference from the preceding order $s_{\varepsilon}-2$ is below this threshold,
\begin{align}\label{eq:sconvergenceerror}
    \abs{f(s_{\varepsilon}) - f(s_{\varepsilon} - 2)} < \varepsilon .
\end{align}
In \cref{fig:08} we present this iteration-dependent convergence of the second-order AGI correction term. Our set of interpolated $X^{\eta}$ gates and the QFT gate were simulated over dimensions $d\in\left[ 2, 64 \right]$ for a uniform gate time of $t=1$. We calculated for each of them the cutoff order $s_{\varepsilon}$ within an error threshold $\varepsilon = 1\times 10^{-8}$. This analysis was repeated for a set of $n=2500$ Haar random gates, and shown by the grey curve (mean value) and shaded region (bounds). It is useful to note that the number of iterations required grows approximately logarithmically with $d$. Furthermore we have a fairly clear range of validity on $s_{\varepsilon}$ for the gates, although the $X$ gate does not give a consistent upper bound here for smaller dimensions. On the other hand, the identity gate $\mathbb{1}=X^{0}$ always gives the lower bound of $s_{\varepsilon}=2$, since then $\comm{\mathbb{1}}{\mathcal{L}}=\mathbb{0}$.

\begin{figure}[t]
    \centering
    \includegraphics[width=\columnwidth]{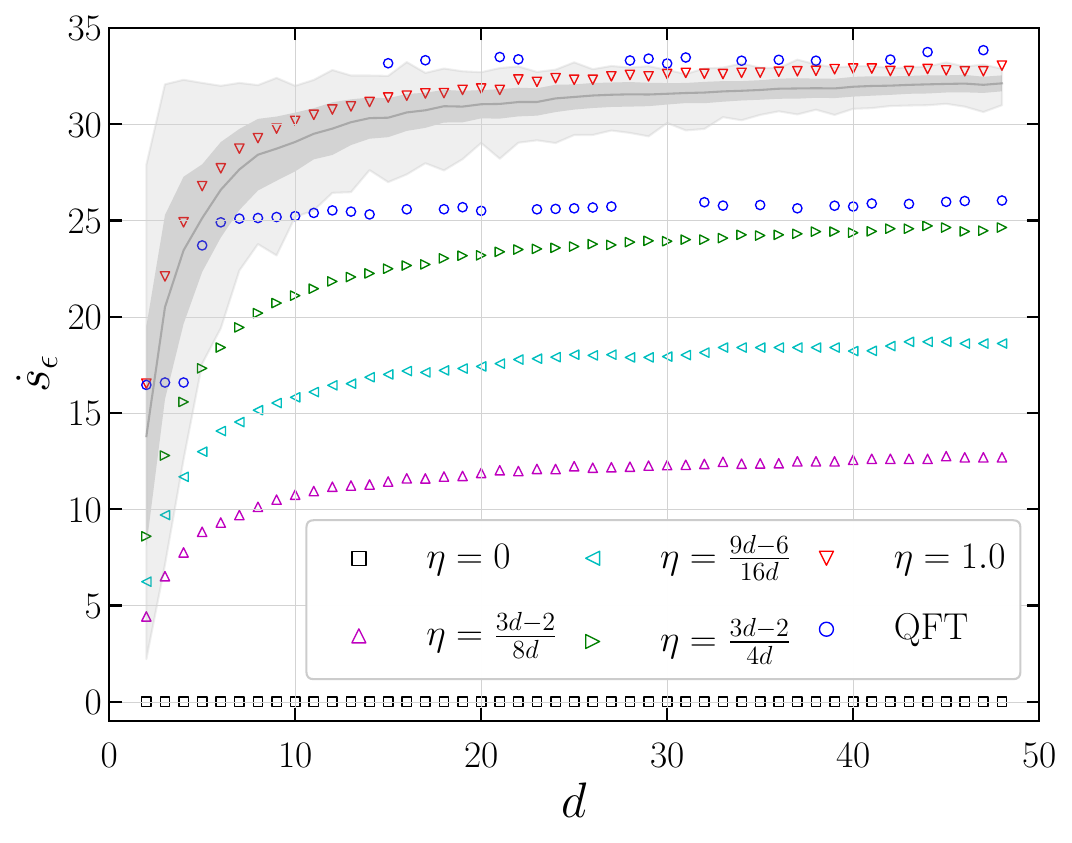}
    \caption{\justifying \textbf{Time derivative of the convergence of the second-order AGI correction over qudit dimension.} Linear gradients of the critical order $s_{\varepsilon}$ as a function of $t$ were fitted by linear regression for each of a set of gates at each dimension $d\in \left[ 2, 48 \right]$. The gate set included the QFT, identity ($X^0$), $X$, and uniformly interpolated $X^{\eta}$ gates and $n=200$ Haar-random gates. For the identity gate this is constant at zero-gradient, while the gradients for the interpolated gates show rapid initial growth that levels off as $d$ increases. The gradients of the QFT gates for $d> 10$ appear to populate two distinct regimes around $25$ and $33$. The sampled Haar-random gates exhibit behaviour similar to that of the $X$ gate within one standard deviation of the mean, and the large variation at lower dimensions stabilises as $d$ increases.}
    \label{fig:09}
\end{figure}
Next, we need to investigate the effect of the gate time parameter $t$ on the second-order correction term, since in the summation it is raised to the power of $s$. We choose our set of quantum gates, $U_g\in\lbrace X^{\eta}, F \rbrace$ with $\eta\in\lbrace 0, \frac{3d-2}{8d}, \frac{9d-6}{16d}, \frac{3d-2}{4d}, 1 \rbrace$, and simulate them on qudits of dimension $d\in\left[ 2, 48 \right]$. Instead of varying the noise parameter $\gamma$ with constant gate time $t=1$, we fix $\gamma =1$ and then vary the gate time $t\in \left[ 0, 5 \right]$. Note that, in order to implement the full quantum gate during this modified gate time, the amplitude of the control-pulse Hamiltonian matrix must be modulated by the inverse of the gate time, $H_c \rightarrow \frac{1}{t} H_c$. We observed (see \cref{app:C:subsec:4} for an example of $s_{\varepsilon}$ against $t$ for the $X$ gate) that for each gate at each dimension, the number of iterations required to reach convergence was roughly linear in time: $s_{\varepsilon} \propto t$. Hence to each such set of data we fitted using linear regression the model
\begin{align} \label{eq:linearlaw}
    s_{\varepsilon} = m t + c,
\end{align}
where $m = \frac{\text{d} s_{\varepsilon}}{\text{d}t}$. \Cref{fig:09} shows this curve of fitted gradients for each gate as a function of the qudit dimension. This analysis is also extended to a set of $n=200$ Haar random gates at each dimension, indicated by the grey curve and shaded regions. The interpolated gates all appear to evolve smoothly to a nearly constant value at large $d$. On the other hand, the QFT gate for $d> 6$ appears to oscillate randomly between two constant regimes near $\frac{\text{d} s_{\varepsilon}}{\text{d}t} \simeq 25$ and $\frac{\text{d} s_{\varepsilon}}{\text{d}t} \simeq 33$.

Based on these two analyses, we can identify a safe bound on $s_{\varepsilon}$ for precise calculation of the second-order correction. Recall our standard approach of normalising the Hamiltonian such that the gate time $t=1$, varying $\gamma$ only, and generally considering $d<64$. Then, we observe that approximately 50 iterations are sufficient to calculate the second-order correction to a precision of $\varepsilon < 1\times 10^{-8}$.

\begin{figure}[t]
    \centering
    \includegraphics[width=\columnwidth]{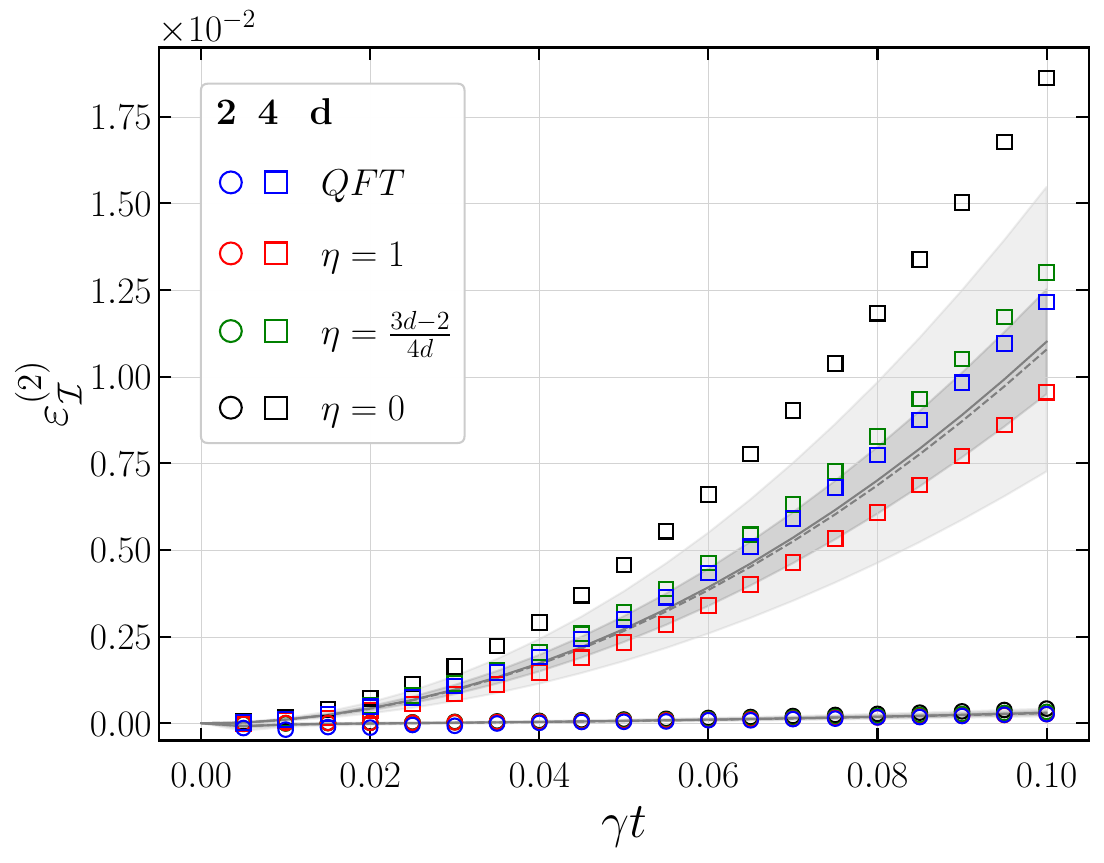}
    \caption{\justifying \textbf{Relative error of the AGI up to second-order.} AGIs were simulated for $4$ fixed gates and $n=1000$ Haar-random gates acting on qubits (circles) and $d=4$ qudits (squares) undergoing $L=J_z$ pure dephasing. The relative error $\varepsilon^{(2)}_{\mathcal{I}} = \frac{\mathcal{I} - \mathcal{I}^{(1)} - \mathcal{I}^{(2)}}{\mathcal{I}}$ was calculated for the second-order perturbative correction. The error for $d=4$ is roughly $1\%$ while for qubits this is reduced to $0.01\%$, two orders of magnitude lower than the relative error at first-order.}
    \label{fig:10}
\end{figure}

We now present in \cref{fig:10} our results for the relative error of the AGI up to second-order. The results presented follow the same approach as for those up to first-order shown in \cref{fig:07}. Choosing the same set of gates, $\lbrace X^0, X^{\eta}, X^1, F \rbrace$ with $\eta=\frac{3d-2}{4d}$ and evolving over $\gamma t \in \left[ 0, 0.1 \right]$, we show the scaling of the relative error for $d=2,4$,
\begin{align}
    \varepsilon^{(2)}_{\mathcal{I}} = \frac{\mathcal{I} - \mathcal{I}^{(1)} - \mathcal{I}^{(2)}}{\mathcal{I}}.
\end{align}
Comparing these results with those of just the first-order in \cref{fig:07}, we see that the error for $d=4$ is reduced by one order of magnitude to $\sim1\%$, while for $d=2$ the error of $\sim0.01\%$ is lower by two orders of magnitude. Additionally, the sign of the relative error has changed from negative to positive, indicating that despite the increased precision, the correction up to second order will tend to understate the actual AGI value, with this error growing quadratically in $\gamma t$.

Finally, as for the first-order case, it is interesting to express the trace of the repeated commutator of superoperators $\mathcal{S}$ and $\mathcal{L}$ in terms of regular operators $H$ and $L$. This will give a clear representation of the effect of the control Hamiltonian on the AGI, as well as its interaction with the collapse operator. Let us consider separately the two trace terms of $\mathcal{I}^{(2)}$ in \cref{eq:secondorderAGIcorrection}: $\Tr{\mathcal{L}^2}$ and $\Tr{\mathcal{L}\comm{(\mathcal{S})^{s}}{\mathcal{L}}}$. We find (see \cref{app:A:subsec:5}) that the gate-independent expression can be written as
\begin{align} \label{eq:secondordergateindependent}
    \Tr{\mathcal{L}^2} &= \left|\Tr{L^2}\right|^2+ \frac{1}{2} \Tr{L^\dag L}^2 +\frac{d}{2} \Tr{(L^\dag L)^2},
\end{align}
which, in the case of pure dephasing, $L = J_z$, reduces to
\begin{align}
    \Tr{J_z^2} &= \frac{d^2(3 - 5d^2 + 2d^4)}{120},
\end{align}
using the element-wise definition in \cref{eq:Jzelements}. This allows us express the $\gamma^2 t^2$ correction in terms of the first-order correction term from \cref{eq:firstorderJzcorrection},
\begin{align}\label{eq:Jzsecondrecursive}
    -\frac{(\gamma t)^2}{d(d+1)} \frac{\Tr{J_z^2}}{2} = -(\gamma t)\frac{2d^2 - 3}{20} \mathcal{I}^{(1)}.
\end{align}
However, it is not necessarily the case that any $m$-th order expresssion can be written recursively in terms of lower orders. \Cref{eq:Jzsecondrecursive} for the second-order appears to be a special case.

For the iterated commutator, we find that (see \cref{app:A:subsec:6})
\begin{align}\label{eq:iteratedcommutatoroperator}
    &\Tr{\mathcal{L}\comm{(\mathcal{S})^s}{\mathcal{L}}} \notag \\
    &= \sum_{k=0}^s \sum_{j=0}^{k} \sum_{l=0}^{s-k} \binom{s}{k} \binom{k}{j} \binom{s-k}{l} (-1)^{s+k-j-l} \times \notag\\
    &\times \Re \Biggl[\Tr{L H^{k-j}LH^{s-k-l}}^*\Tr{L H^jLH^l} \Biggr. \notag \\
    &+ \frac{1}{2} \Tr{L^TL^* H^{s-j-l}}^*\Tr{L^TL^* H^{j+l}} \notag \\
    &+ \frac{1}{2} \Tr{L^TL^* H^{k-j}L^TL^* H^{s-k-l}}^*\Tr{H^{j+l}} \notag \\
    &- \Tr{L H^{k-j}L^TL^* H^{s-k-l}}^*\Tr{LH^{j+l}} \notag \\
    &- \Biggl. \Tr{L^TL^* H^{k-j}L H^{s-k-l}}^*\Tr{LH^{j+l}} \Biggr] \, .
\end{align}
This gives a complete description of the second-order gate-dependence in terms of the explicit Hamiltonian terms, without resorting to the superoperator representation. The superoperator and operator expressions have complexity $\mathcal{O}(sd^6)$ and $\mathcal{O}((sd)^3)$, respectively. Since we observed in \cref{fig:08} that the maximal $s_{\varepsilon}$ needed to reach convergence is roughly logarithmic in $d$, the complexities can then be reduced to $\mathcal{O}(d^6 \log{d})$ and $\mathcal{O}((d\log{d})^3)$. Therefore, for $d\gg 2$, it is computationally preferred to use the operator-based expression, while for smaller qudits with $d\sim2$, the superoperator representation is generally preferable for numerical simulations.


%% file: sections/section3.tex

\section{\label{sec:3} Conclusions and Outlook}

In this work, we have performed a comprehensive analysis of the AGF for single qudit open quantum systems coupled to Markovian noise environments in the Lindblad superoperator formalism. Our primary contributions and findings are as follows:

We have expanded the AGF perturbatively in powers of the dimensionless coupling constant $\gamma t$. From this we derived our main result in \cref{eq:AGFcorrection} for the general expressions of the correction terms to arbitrary order. The result was expressed in terms of the iterated commutators of the gate and noise superoperators, $\mathcal{S}$ and $\mathcal{L}$, respectively, based on the expansion obtained in \cref{eq:Mseparated}. It is particularly significant that these corrections were expressed in a form that may be directly implemented in numerical calculations. We found that, at each order above the first, the correction term could be separated into a gate-independent term $\mathcal{O}((\gamma t)^m)$ depending only on $\mathcal{L}$, and gate-dependent terms containing $\mathcal{S}$ and higher powers of $t$. The first-order correction term, $\mathcal{I}^{(1)}$, is gate-independent and depends solely on the noise superoperator $\mathcal{L}$. The gate-dependence appears for the first time at $\gamma^2 t^4$, in the second-order correction term $\mathcal{I}^{(2)}$, highlighting the significant role of the interaction between the control Hamiltonian $\mathcal{S}$ and the noise operator $\mathcal{L}$. Explicit expressions for the first- and second-order correction terms in the operator representation were also derived. 

Our numerical simulations under pure dephasing noise revealed a clear transition from linear to nonlinear behaviour in the AGI as the noise coupling strength $\gamma t$ increases. For $\gamma t \sim 1$, the AGI deviates significantly from the linear approximation and eventually reaches a plateau at a stable value. The plateau values, $\mathcal{I}^*$, as well as their corresponding saturation points, $(\gamma t)^*$, were found to be dependent not only on the qudit dimension $d$ but also on the specific quantum gate implemented.

Indeed, utilising our theoretical framework we were able to derive in \cref{eq:upperlowerbounds} a significant result for universal upper and lower bounds of the AGI plateau values in the strong coupling regime, depending only on the dimensionality of the system. Furthermore we were able to identify specific gates that saturate these bounds: The identity gate ($\mathbb{1}_d$) and the generalised-NOT ($X$) achieve the lower and upper limits, respectively, while the QFT gate ($F_d$) was found to plateau at the mean AGI value of the Haar measure. Additionally, we introduced the interpolated $X^{\eta}$ gates to demonstrate the range of AGI values between these bounds. The saturation points of these interpolated gates were studied numerically, and were well-described by a power-law model of $(\gamma t)^*$ vs $d$.

We analyzed the convergence behavior of the second-order correction term, identifying that approximately 50 iterations are sufficient for precise calculation up to a qudit dimension of $d < 64$ within an absolute tolerance of $\varepsilon < 1\times10^{-8}$. The number of iterations to convergence was found to scale logarithmically with the system dimension $d$, and linearly with the gate time $t$.

Numerical simulations confirmed the analytical predictions, demonstrating that the relative error of the AGI is significantly reduced when including the second-order correction term. For $d=4$, the error is reduced to the order of $1\%$, and for $d=2$, it is reduced to $0.01\%$, indicating the necessity of higher-order corrections for accurate fidelity modeling, particularly as the dimension of the system increases.

The detailed insights gained from this study have several important applications and implications for the field of quantum computing, particularly given the growing interest in qudits. Understanding the behavior of the AGI and its dependence on the type of quantum gate and environmental noise facilitates the optimization of quantum gate design. Our approach provides a methodology for identifying quantum gates that have favourable fidelity characteristics. Thus, through appropriate choice or design of basis gates it may be possible to enhance the performance of quantum circuits. This could lead to significant improvements for optimising the robustness of quantum algorithms on near-term noisy platforms.

The results underscore the importance of incorporating higher-order corrections into error correction protocols. This is particularly relevant for qudits of higher dimensions, where noise effects are more pronounced. We envisage that the detailed perturbative expansions provided in this work could be leveraged to develop advanced error correction techniques, for example by incorporating the higher-order correction terms into the cost function of optimal control methods. In particular, our correction terms can be used to quantify and mitigate the error in logical qubit embedding protocols up to arbitrary order. Future research is needed to show how this could result in protection against such errors above first-order, and how this might incorporate the gate-dependencies identified.

The derived bounds and behaviours of AGI for different gates and dimensions also serve as benchmarks for assessing the performance of quantum systems. These benchmarks can guide experimentalists in evaluating and improving their quantum hardware and gate implementations. Indeed further study is needed to extend these results from the idealised single-pulse (time-independent) Hamiltonian to realistic pulse-based optimal control techniques for gate generation.

While this study focused primarily on pure dephasing and bit-flip errors, future research should explore further the impact of other noise models, such as amplitude damping and depolarizing noise. Understanding how different types of noise affect the AGI will provide a more comprehensive framework for designing noise-resilient quantum systems. 

Extending the framework to multi-qudit systems is a necessity. These introduce additional complexity due to inter-qudit interactions and correlated noise effects. Investigating these factors will be crucial for the scalability of qudit-based architectures, but could prove fruitful in reducing the circuit complexity of long range entangling operations.

Finally, it would be beneficial to have experimental verification of these predictions. Testing the derived correction terms and transition points in real quantum systems will help to validate and refine the models presented.

In conclusion, this study provides a detailed theoretical foundation for understanding the fidelity of quantum gates in noisy environments. The findings contribute to the ongoing efforts to develop robust, high-fidelity quantum operations, paving the way for practical and scalable quantum computing.


%% file: sections/acknowledgements.tex

\begin{acknowledgments}
This work was funded by the French National Research Agency (ANR) through the Programme d’Investissement d’Avenir under contract ANR-11-LABX-0058\_NIE and ANR-17-EURE-0024 within the Investissement d’Avenir program ANR-10-IDEX-0002-02. D.J. and M.R. gratefully acknowledge financial support from the Deutsche Forschungsgemeinschaft (DFG, German Research Foundation) through the Collaborative Research Centre “4f for Future” (CRC 1573, project number 471424360) project B3. J-G.H. also acknowledges QUSTEC funding from the European Union’s Horizon 2020 research and innovation program under the Marie Skłodowska-Curie Grant Agreement No. 847471. The authors would like to acknowledge the High Performance Computing Center of the University of Strasbourg for supporting this work by providing scientific support and access to computing resources. Part of the computing resources were funded by the Equipex Equip@Meso project (Programme Investissements d’Avenir) and the CPER Alsacalcul/Big Data.

\end{acknowledgments}

\section*{Code and Data Availability}
All code and data utilised in the simulations, analyses, and generation of the results presented in this study are publicly available via Zenodo \cite{hartmann_code_2025} under a Creative Commons Attribution (CC BY) license.


%% file: sections/appendix.tex

\appendix

\section{\label{app:A} Complementary Derivations}
Here, we present the detailed derivations of the analytical results in the main text.

\subsection{\label{app:A:subsec:1} Integration of the $m$-th Order Solution to the Master Equation}

\begin{theorem}
Consider an open quantum system of a single qudit of dimension $d$, initiated in a pure state $\Tr{\rho_0^2}=1$, under the influence of a time-independent noise superoperator $\mathcal{L}$ with coupling constant $\gamma$, and evolving via a time-independent unitary superoperator $\mathcal{S}$ over time $t$. The solution to the master equation in \cref{eq:GKSL} is given by the quantum channel in \cref{eq:supersol}, and can be expressed in a perturbative expansion in $\gamma$ by \cref{eq:rhoexpansion}, where the $m$-th order correction term is given by \cref{eq:rhocorrection}. Then, the nested integral can be evaluated, by induction, and results in \cref{eq:Mexpansion}:
\begin{align}
&\Tilde{M}^{(m)}(t) \notag \\
&= \int_0^t \cdots \int_0^{t_{m - 1}} \left( \prod_{i=1}^m  e^{-\mathcal{S}t_i} \mathcal{L} e^{\mathcal{S}t_i} \right) dt_{m} \cdots dt_1 \\
     &= t^m \sum_{n_1, \cdots, n_m = 0}^{\infty}\left( \prod_{i = 1}^{m} \frac{(-t)^{n_i}\comm{(\mathcal{S})^{n_i}}{\mathcal{L}}}{n_i! \sum_{j=i}^{m}\left( n_j + 1 \right)} \right) . \label{eq:Msol}
\end{align}
\end{theorem}

\begin{proof}
In the calculations that follow, we make use of Campbell's lemma, based on the Baker-Campbell-Hausdorff Formula \cite{campbell_law_1896, hall_lie_2015}, as stated in \cref{eq:BCH},
\begin{align}
    e^X Y e^{-X} = \sum_{n=0}^{\infty} \frac{\comm{(X)^n}{Y}}{n!} ,
\end{align}
with $\comm{(X)^n}{Y} = \comm{X}{\comm{(X)^{n-1}}{Y}}$ and $\comm{(X)^0}{Y} = Y$, and proceed with proof by induction.
\newline
\textit{Base Case:} To first order, $m=1$,
\begin{align}
    &\int_0^t e^{-\mathcal{S}t_1} \mathcal{L} e^{\mathcal{S}t_1} dt_1 \notag \\
    &= \int_0^t \sum_{n_1 = 0}^{\infty} \frac{\comm{(-\mathcal{S}t_1)^{n_1}}{\mathcal{L}}}{n_1 !}  dt_1 \\
    &= \sum_{n_1 = 0}^{\infty} \frac{(-1)^{n_1}}{n_1 !} \comm{(\mathcal{S})^{n_1}}{\mathcal{L}} \int_0^t t_1^{n_1} dt_1 \\
    &= t \sum_{n_1 = 0}^{\infty} \frac{(-t)^{n_1}\comm{(\mathcal{S})^{n_1}}{\mathcal{L}}}{n_1 ! (n_1 + 1)}  \\
    &= \Tilde{M}^{(1)}(t) .
\end{align}
\textit{Induction Step:} Assume that the result in \cref{eq:Msol} holds at $m$-th order, and relabel indices $t, t_1, \cdots, t_{m-1} \rightarrow t_1, t_2, \cdots, t_{m}$ and $t \rightarrow t_1$:
\begin{align}
    &\Tilde{M}^{(m)}(t_1) \notag \\
    &=  \int_0^{t_1} \cdots \int_0^{t_m} \left( \prod_{2=1}^{m+1}  e^{-\mathcal{S}t_i} \mathcal{L} e^{\mathcal{S}t_i} \right) dt_{m+1} \cdots dt_2 \\
     &= t_1^m \sum_{n_2=0}^{\infty} \cdots \sum_{n_{m+1}=0}^{\infty}\left( \prod_{i = 2}^{m+1} \frac{(-t_1)^{n_i}\comm{(\mathcal{S})^{n_i}}{\mathcal{L}}}{n_i! \sum_{j=i}^{m+1}\left( n_j + 1 \right)} \right) .
\end{align}
Then, to order $m+1$:
\begin{widetext}
\begin{align}
    & \int_0^t \Tilde{M}^{(m)}(t_1) \left( e^{-\mathcal{S}t_1} \mathcal{L} e^{\mathcal{S}t_1} \right) dt_1 \notag \\
    &= \int_0^t \left( t_1^m \sum_{n_2=0}^{\infty} \cdots \sum_{n_{m+1}=0}^{\infty}\prod_{i=2}^{m+1} \frac{(-t_1)^{n_i}\comm{(\mathcal{S})^{n_i}}{\mathcal{L}}}{n_i ! \sum_{j=i}^{m+1}\left( n_j + 1 \right)} \right) \left( \sum_{n_1=0}^{\infty} \frac{(-t_1)^{n_1}\comm{(\mathcal{S})^{n_1}}{\mathcal{L}}}{n_1 !} \right) dt_1 \\
    &= \sum_{n_2=0}^{\infty} \cdots \sum_{n_{m+1}=0}^{\infty} \frac{(-1)^{n_1}\comm{(\mathcal{S})^{n_1}}{\mathcal{L}}}{n_1 !} \prod_{i=2}^{m+1} \left(\frac{(-1)^{n_i}\comm{(\mathcal{S})^{n_i}}{\mathcal{L}}}{n_i ! \sum_{j=i}^{m+1}\left( n_j + 1 \right)} \right) \int_0^t t_1^m t_1^{n_1} \prod_{i=2}^{m+1} t_1^{n_i} dt_1 \\
    &= \sum_{n_2=0}^{\infty} \cdots \sum_{n_{m+1}=0}^{\infty} \frac{(-1)^{n_1}\comm{(\mathcal{S})^{n_1}}{\mathcal{L}}}{n_1 !} \prod_{i=2}^{m+1} \left(\frac{(-1)^{n_i}\comm{(\mathcal{S})^{n_i}}{\mathcal{L}}}{n_i ! \sum_{j=i}^{m+1}\left( n_j + 1 \right)} \right) \frac{t^{m+1+\sum_{i=1}^{m+1} n_i}}{m+1+\sum_{j=1}^{m+1} n_j} \\
    &= t^{m+1} \sum_{n_2=0}^{\infty} \cdots \sum_{n_{m+1}=0}^{\infty} \frac{(-t)^{n_1}\comm{(\mathcal{S})^{n_1}}{\mathcal{L}}}{n_1 ! \sum_{j=1}^{m+1}\left( n_j + 1 \right)} \prod_{i=2}^{m+1}\frac{(-t)^{n_i}\comm{(\mathcal{S})^{n_i}}{\mathcal{L}}}{n_i ! \sum_{j=i}^{m+1}\left( n_j + 1 \right) } \\
    &= t^{m+1} \sum_{n_2=0}^{\infty} \cdots \sum_{n_{m+1}=0}^{\infty} \prod_{i=1}^{m+1}\frac{(-t)^{n_i}\comm{(\mathcal{S})^{n_i}}{\mathcal{L}}}{n_i ! \sum_{j=i}^{m+1}\left( n_j + 1 \right) } \\
    &= \Tilde{M}^{(m+1)}(t) .
\end{align}
\end{widetext}
\end{proof}

\subsection{\label{app:A:subsec:2} Perturbative Expansion of the AGI}
We define the Average Gate Infidelity as
\begin{align}
    \mathcal{I}(\mathcal{E}, \mathcal{U}) &= 1 - \mathcal{F}(\mathcal{E}, \mathcal{U}) \\
    &= 1 - \left( 1 - \sum_{m=1}^{\infty} \gamma^m F^{(m)} \right),
\end{align}
where the quantum channel is given by
\begin{align}
    \mathcal{E} &= \mathcal{U} + \sum_{m=1}^{\infty} \gamma^m \mathcal{E}^{(m)} \\
    &= \mathcal{U} + \sum_{m=1}^{\infty} \gamma^m \mathcal{U} \Tilde{M}^{(m)} .
\end{align}
Then, from \cref{eq:AGF},
\begin{align}
    &\mathcal{F}(\mathcal{E}, \mathcal{U}) \notag \\ 
    &= \int_{\mathscr{H}} \left\langle\left(\mathcal{U^\dag \circ E}\right)[\rho_0]\right\rangle_0 d \rho_0 \\
    &= \int_{\mathscr{H}} \ev{\left(\mathcal{U}^\dag \circ \left( \mathcal{U} + \sum_{m=1}^{\infty} \gamma^m \mathcal{U} \Tilde{M}^{(m)}\right)\right)[\rho_0]}_0 d \rho_0 \\
    &=  \int_{\mathscr{H}} \ev{\rho_0 + \sum_{m=1}^{\infty} \gamma^m \Tilde{M}^{(m)}[\rho_0]}_0 d \rho_0 ,
\end{align}
and evaluating the expectation value over $\rho_0$ of initial states with $\ev{\mathcal{A}}_0 = \Tr{\mathcal{A} \rho_0}$,
\begin{align}
     &1 - \sum_{m=1}^{\infty} \gamma^m F^{(m)} \notag \\
     &= \int_{\mathscr{H}} \Tr{\rho_0^2 +  \sum_{m=1}^{\infty} \gamma^m \Tilde{M}^{(m)}[\rho_0] \rho_0} d \rho_0 \\
    &= 1 + \sum_{m=1}^{\infty} \gamma^m \int_{\mathscr{H}} \Tr{\Tilde{M}^{(m)}[\rho_0] \rho_0} d\rho_0,
\end{align}
since $\Tr{\rho_0^2}=1$ and $\int_{\mathscr{H}} d\rho_0 = 1$, and using \cref{eq:agfexpansion}. The $m$-th order term of the AGF is therefore
\begin{align}
    F^{(m)} = - \int_{\mathscr{H}} \Tr{\Tilde{M}^{(m)}[\rho_0] \rho_0} d\rho_0 ,
\end{align}
and we may write the AGI in terms of $t^m M^{(m)} = \Tilde{M}^{(m)}$,
\begin{align} \label{eq:AGIappendix}
    \mathcal{I}(\mathcal{E}, \mathcal{U}) = - \sum_{m=1}^{\infty} (\gamma t)^m \int_{\mathscr{H}} \Tr{M^{(m)}[\rho_0] \rho_0}d\rho_0 .
\end{align}

\subsection{\label{app:A:subsec:3} Integral over the Fubini-Study Measure}
We can evaluate \cref{eq:AGIappendix} to arrive at the result in \cref{eq:AGFcorrection}. The integral over the Fubini-Study measure is calculated using methods from \cite{collins_weingarten_2022} relating to Weingarten calculus. Specifically, we make use of the following result,
\begin{align}
    &\int_{\mathscr{H}} U_{ka} U_{ic} \bar{U}_{jb} \bar{U}_{rd} dU \notag \\
    &= \frac{1}{(d^2 - 1)}\left[ \left( \delta_{kj} \delta_{ab}\delta_{ir}\delta_{cd} + \delta_{kr} \delta_{ad}\delta_{ij}\delta_{cb} \right)\right] - \notag\\
    &- \frac{1}{d(d^2 - 1)}\left[ \left( \delta_{kj} \delta_{ad}\delta_{ir}\delta_{cb} + \delta_{kr} \delta_{ab}\delta_{ij}\delta_{cd} \right)\right], \label{eq:weingarten}
\end{align}
where, and for the remainder of this section, the indices represent the matrix elements using Einstein sum notation. 

Now, beginning with the element-wise definition of the superoperator term,
\begin{align}
    \left( M[\rho] \right)_{pi} &= M_{kj, pi} \rho_{kj}, 
\end{align}
where we identify $\rho$ and $\rho_0$ for convenient usage of the element-wise sum notation, this extends to
\begin{align}
        \left( M[\rho] \rho \right)_{pq} &= \left( M[\rho] \right)_{pi} \rho_{iq} \\
    &= M_{kj, pi} \rho_{kj} \rho_{iq} .
\end{align}
Taking the trace of this term corresponds to
\begin{align}
    \Tr{M[\rho] \rho} &= \delta_{pq} \left( M[\rho] \rho \right)_{pq} \\
    &= M_{kj, ri} \rho_{kj} \rho_{ir},
\end{align}
where we define $r := p=q$. Rewriting the integral of this expression in terms of the associated unitary operator $U$ leads to
\begin{align}
    & \int_{\mathscr{H}} \Tr{M\left[U \rho U^{\dag} \right] U \rho U^{\dag}}dU \notag\\ 
    &= \int_{\mathscr{H}}  M_{kj, ri} \left( U_{ka} \rho_{ab} \bar{U}_{jb} \right) \left( U_{ic} \rho_{cd} \bar{U}_{rd} \right) dU \\
    &=  \rho_{ab} M_{kj, ri} \rho_{cd} \int_{\mathscr{H}} \left( U_{ka}\bar{U}_{jb} U_{ic}\bar{U}_{rd} \right) dU .
\end{align}
If we now substitute the result of this integral from \cref{eq:weingarten}, and evaluate all of the delta function terms, we arrive at
\begin{align}
    &\int_{\mathscr{H}} \Tr{M[\rho] \rho}d\rho \notag \\
    &= \frac{1}{(d^2 - 1)} \left[ \rho_{aa}\rho_{cc} M_{jj,ii} + \rho_{ac}\rho_{ac} M_{ki,ki}  \right] - \notag \\
    &- \frac{1}{d(d^2 - 1)} \left[ \rho_{ac}\rho_{ac} M_{jj,ii} + \rho_{aa}\rho_{cc} M_{ki,ki} \right]  \\
    &= \frac{1}{(d^2 - 1)} \left[ \left( \Tr{\rho} \right)^2 \Tr{M\left[ \mathbb{1} \right]} + \Tr{\rho^2}\Tr{M} \right] - \notag \\
    &- \frac{1}{d(d^2 - 1)} \left[ \Tr{\rho^2} \Tr{M\left[ \mathbb{1} \right]} + \left( \Tr{\rho} \right)^2\Tr{M} \right] .
\end{align}
Returning to the matrix forms from the Einstein sum notation, it can be seen that:
\begin{align}
    \rho_{aa}\rho_{cc} &= \left( \Tr{\rho} \right)^2 ,\\
    \rho_{ac}\rho_{ac} &= \Tr{\rho^2} ,\\
    M_{jj,ii} &=  \Tr{M\left[ \mathbb{1} \right]} ,\\
    M_{ki,ki} &= \Tr{M} ,
\end{align}
which results in
\begin{align}
    &\int_{\mathscr{H}} \Tr{M[\rho] \rho} d\rho \notag \\   
    &= \frac{1}{(d^2 - 1)} \left[ \left( \Tr{\rho} \right)^2 \Tr{M\left[ \mathbb{1} \right]} + \Tr{\rho^2}\Tr{M} \right] - \notag \\
    &- \frac{1}{d(d^2 - 1)} \left[ \Tr{\rho^2} \Tr{M\left[ \mathbb{1} \right]} + \left( \Tr{\rho} \right)^2\Tr{M} \right] .
\end{align}
Finally, noting that $\Tr{\rho} = 1$, $\Tr{\rho^2} = 1$ and $\rho^2 = \rho$, we arrive at
\begin{align}
    \int_{\mathscr{H}} \Tr{M[\rho] \rho} d\rho &= \frac{\Tr{M} + \Tr{M\left[ \mathbb{1} \right]}}{d(d+1)} \label{eq:Mintegral} \\
    &=  \frac{\Tr{M}}{d(d+1)} ,
\end{align}
since for the cases where $M$ is composed of a traceless lindbladian operator $\mathcal{L}$, $\Tr{M[\mathbb{1}]} = 0$, and we arrive at \cref{eq:AGFcorrection}.

\subsection{\label{app:A:subsec:4} Boundedness of the AGI}
We make use of the relation in \cref{eq:Mintegral} to prove that, in the limit of strong dephasing, the AGI of a single qudit is bounded above and below, with the bounds depending only on the dimension of the system.

\begin{theorem}
    Consider an open quantum system of a single qudit of dimension $d$, initiated in a pure state $\Tr{\rho_0^2}=1$, under the influence of a time-independent noise superoperator $\mathcal{L}$ with coupling constant $\gamma$ and pure dephasing collapse operator $L=J_z$, and evolving via an arbitrary time-independent unitary superoperator $\mathcal{S}$ over time $t$. Then, in the limit of strong coupling where $(\gamma t) \gg 1$, the AGI $\mathcal{I}$ is bounded in the region
    \begin{align}
        1 - \frac{2}{d + 1} \leq \mathcal{I} \leq 1 - \frac{1}{d + 1} .
    \end{align}
\end{theorem}

\begin{proof}    
    In superoperator form, the quantum channel can be written as $\mathcal{E} = \mathcal{U} \circ M$. Now, since the channel is completely positive and trace-preserving (CPTP) and $\mathcal{U}$ is a unitary operator, $M$ must also be CPTP. Substituting this into the integral equation for the AGF in \cref{eq:AGF} and using \cref{eq:Mintegral}, we have
    \begin{align}
        \mathcal{F} = \frac{\Tr{M} + \Tr{M\left[ \mathbb{1} \right]}}{d(d+1)}.
    \end{align}
    Therefore, proving the bound on $\mathcal{I}$ is equivalent to proving that
    \begin{align}
        1 &\leq (d + 1)\mathcal{F} \leq 2 , \\
        d &\leq \Tr{M} + \Tr{M\left[ \mathbb{1} \right]} \leq 2d , \\
        0 &\leq \Tr{M} \leq d,
    \end{align}
    since $\Tr{M\left[ \mathbb{1} \right]} = \Tr{\mathbb{1}}=d$ due to the trace-preservation of $M$. 
    
    Now, to show this, consider the action of $M$ on the state $\rho$. The collapse operator $L=J_z$ of pure dephasing is a diagonal matrix which preserves the populations of $\rho$ while causing the coherences to decay to zero. Thus,
    \begin{align}
        M \; : \; \rho &\rightarrow \sum_{i=0}^{d-1} \rho^{\prime}_{ii} \ketbra{i}{i} \\
        \forall \rho \; , \; M^{kl}_{ij} \rho_{kl} &\propto \delta_{ij} \\
        M^{kl}_{ij} &= \lambda_i^{kl} \delta_{ij}.
    \end{align}
    In the Kraus representation, using $M=\sum_{k=1}^K E_k^* \otimes E_k$, the matrix elements are given, trivially, by
    \begin{align}
        M_{ij}^{nm} = \sum_{k=1}^K E_{k, jm}^* E_{k, in} ,
    \end{align}
    and, since $M$ is trace-preserving,
    \begin{align}
        \forall n,m, \; \sum_{i=0}^{d-1} M_{ii}^{nm} &= \sum_{k=1}^K \sum_{i=0}^{d-1} E_{k, im}^* E_{k, in}\\
        &= \delta_{nm}.
    \end{align}
    Furthermore,
    \begin{align}
    \Tr{M\rho} &= \Tr{\rho} \\
    \text{vec}(\mathbb{1})^{\text{T}} \cdot M \cdot \text{vec}(\rho) &= \text{vec}(\mathbb{1})^{\text{T}} \cdot \text{vec}(\rho) \\
    \therefore \sum_{n=0}^{d-1}M^{nn}_{ij} &= \delta_{ij} .
    \end{align}
    Thus, in particular, if $n=m$:
    \begin{align}
        \sum_{k=1}^K \sum_{i=0}^{d-1} \abs{E_{k,in}}^2 &= 1 \\
        \iff \sum_{k,\,i=n}\abs{E_{k,in}}^2 + \sum_{k,\,i\neq n}\abs{E_{k,in}}^2 &= 1 \\
        \therefore \forall n \; , \; 0 \leq \sum_{k=1}^K \abs{E_{k,nn}}^2 &\leq 1 \label{eq:KrausBound}
    \end{align}
    Moreover, since
    \begin{align}
         \Tr{M} &= \sum_{k, \, ij}E_{k,\,jj}^* E_{k,\,ii} \\
         &= \sum_k \abs{\Tr{E_k}}^2 ,
    \end{align}    
    the trace of $M$ can be equivalently expressed by
    \begin{align}
        \Tr{M} &= \sum_{i=0}^{d-1} \sum_{j=0}^{d-1} M^{ij}_{ij} \\        
        &= \sum_{i=0}^{d-1} M^{ii}_{ii} \\
        &= \sum_{k=1}^K\sum_{i=0}^{d-1} \abs{E_{k, \, ii}}^2 .
    \end{align}
    Therefore, combining this last expression with \cref{eq:KrausBound} where each of the $d$ elements $\abs{E_{k, ii}}^2$ are bounded above by $1$, it is clear that the sum of all $d$ of them, and thus $\Tr{M}$, must be bounded by $d$,
    \begin{align}
        0 \leq \sum_k^K \sum_{i=0}^{d-1} \abs{E_{k,\, ii}}^2 \leq d ,
    \end{align} 
    proving the boundedness of $\mathcal{F}$ and $\mathcal{I}$.
\end{proof}

\subsection{\label{app:A:subsec:5} Gate-Independent Correction Terms}
In \cref{sec:2:subsec:2}, we found in \cref{eq:mthorderAGIcorrection} that the $m$-th order AGI correction always contains a gate-independent term proportional to the trace of the $m$-th power of the collapse operator $\mathcal{L}$ in superoperator form. Specifically, in \cref{sec:2:subsec:5} \cref{eq:firstorderAGIcorrection,eq:secondorderAGIcorrection}, we presented the first- and second-order gate-independent AGI correction terms.

Since we have in \cref{eq:Lprepost} an expression for $\mathcal{L}$ in terms of the (regular) collapse operators $L$, we can use this to express the correction terms in terms of $L$ instead.
To first-order,
\begin{align}
    &\Tr{\mathcal{L}} \notag \\
    &= \Tr{L^* \otimes L - \frac{1}{2}\left(L^{\dag} L \otimes \mathbb{1}_d + \left( \mathbb{1}_d \otimes L^{\dag} L \right)^T \right)} \\
    &= \Tr{L^* \otimes L} - \frac{1}{2}\Tr{L^{\dag} L \otimes \mathbb{1}_d +\mathbb{1}_d \otimes \left(L^{\dag} L \right)^T} \\
    &= \Tr{L^*}\Tr{L} - \frac{d}{2}\left( \Tr{L^{\dag}L} + \Tr{\left(L^{\dag} L \right)^T} \right) \\
    &= \abs{\Tr{L}}^2 - d \Tr{L^{\dag}L},
\end{align}
giving the result of \cref{eq:TrL}.

By the same procedure, it is trivial to see that the second-order term in \cref{eq:secondordergateindependent} is given by,
\begin{align}
     \Tr{\mathcal{L}^2} &= \abs{\Tr{L^2}}^2 + \frac{1}{2}\abs{\Tr{L^{\dag}L}}^2 + \frac{d}{2}\Tr{(L^{\dag}L)^2} .
\end{align}

Furthermore, it is possible to make use of the multinomial expansion to find the $m$-th power trace to arbitrary order. In general, this produces a non-trivial sum over all $L$-words of length $m$. In the specific case of real and symmetric collapse operators $L=L^*=L^T=L^{\dag}$, these words can be factorised, resulting in the triangular sum
\begin{align}
   &\Tr{\mathcal{L}^m} \notag \\
   &= \Tr{\left( L^* \otimes L - \frac{1}{2}\left(L^{\dag} L \otimes \mathbb{1}_d + \left( \mathbb{1}_d \otimes L^{\dag} L \right)^T \right) \right)^m} \\
    &= m!\sum_{k_1+k_2+k_3=m}\frac{\Tr{(L^*)^{k_1}(L^{\dag}L)^{k_2}\otimes L^{k_1}((L^{\dag}L)^T)^{k_3}}}{(-2)^{k_2 + k_3}k_1!k_2!k_3!} \\
    &= m!\sum_{k_1+k_2+k_3=m}\frac{\Tr{L^{k_1+2k_2}}\Tr{L^{k_1+2k_3}}}{(-2)^{k_2+k_3} k_1!k_2!k_3!}.  \\
    &= m!\sum_{k_1=0}^m\sum_{k_2=0}^{m-k_1} \frac{\Tr{L^{k_1+2k_2}}\Tr{L^{2m-k_1-2k_2}}}{(-2)^{m -  k_1 - 1}k_1!k_2!(m-k_1-k_2)!} .    
\end{align}

\subsection{\label{app:A:subsec:6} Second-order Gate-Dependent Correction Term}
Beginning with the general expression for the second-order term in the AGI expansion in \cref{eq:AGFcorrection}, and combining with the $m=2$ term $M^{(2)}(t)$ in \cref{eq:Mseparated}, we obtain \cref{eq:secondorderfull} since 
\begin{align}
    &\Tr{M^{(2)}(t)} \notag \\
    &=\Tr{\frac{\mathcal{L}^2}{2!} + \sum_{n_1 = 1}^{\infty}\sum_{n_2 = 1}^{\infty} \left( \prod_{i = 1}^{2} \frac{(-t)^{n_i}\comm{(\mathcal{S})^{n_i}}{\mathcal{L}}}{n_i! \sum_{j=i}^{2}\left( n_j + 1 \right)} \right)} \\
    &=\Tr{\frac{\mathcal{L}^2}{2!}} + \notag\\
    &+\sum_{n_1 = 1}^{\infty}\sum_{n_2 = 1}^{\infty} \frac{(-t)^{n_1+n_2}\Tr{\comm{(\mathcal{S})^{n_1}}{\mathcal{L}}\comm{(\mathcal{S})^{n_2}}{\mathcal{L}}}}{n_1!n_2! (n_1 + n_2 + 2)(n_2 + 1)}
\end{align}
Focusing now on the gate-dependent terms only, we can express the double summation of the trace term using the relation $n_1+n_2+2=s$

\begin{align}
    &\sum_{n_1 = 1}^{\infty}\sum_{n_2 = 1}^{\infty} \frac{(-t)^{n_1+n_2}\Tr{\comm{(\mathcal{S})^{n_1}}{\mathcal{L}}\comm{(\mathcal{S})^{n_2}}{\mathcal{L}}}}{n_1!n_2! (n_1 + n_2 + 2)(n_2 + 1)} \\
    &=\sum_{s = 0}^{\infty} \sum_{n = 0}^{s-2}  \frac{(-t)^{s-2}\Tr{\comm{(\mathcal{S})^{n}}{\mathcal{L}}\comm{(\mathcal{S})^{s-2-n}}{\mathcal{L}}}}{n! s(s-1-n)!}  \\
    &=\sum_{s = 0}^{\infty} \frac{(-t)^{s-2}}{s(s-1)!}\times \notag \\&\times\sum_{n = 0}^{s-2}  \binom{s-1}{n}
     \Tr{\comm{(\mathcal{S})^{n}}{\mathcal{L}}\comm{(\mathcal{S})^{s-2-n}}{\mathcal{L}}} .\label{eq:lemmaresult}
\end{align}

\begin{lemma}
Given operators $A$ and $B$ the following binomial sum of the product of iterated commutators at order $n$ and $j-n$ is traceless,
\begin{align}
    \sum_{n=1}^{j} \binom{j+1}{n} \Tr{\comm{(A)^n}{B}\comm{(A)^{j-n}}{B}} = 0 .
\end{align}
\end{lemma}
\begin{proof}
    Using $\binom{j+1}{n} = \binom{j+1 - 1}{n - 1} + \binom{j+1 - 1}{n}$, we have
    \begin{align}
        &\sum_{n=1}^{j} \binom{j+1}{n} \Tr{\comm{(A)^n}{B}\comm{(A)^{j-n}}{B}} \notag\\&=\sum_{n=1}^{j} \binom{j}{n - 1} \Tr{\comm{(A)^n}{B}\comm{(A)^{j-n}}{B}}  + \notag\\&+\sum_{n=1}^{j} \binom{j}{n} \Tr{\comm{(A)^n}{B}\comm{(A)^{j-n}}{B}} . \label{eq:binomialtraces}
    \end{align}
    Now, for the first term with binomial $\binom{j+1-1}{n-1}$ and using the change of index $n \rightarrow n+1$,
    \begin{align}
        &\sum_{n=1}^{j} \binom{j}{n - 1} \Tr{\comm{(A)^n}{B}\comm{(A)^{j-n}}{B}} \notag \\
        &= \sum_{n=0}^{j-1} \binom{j}{n} \Tr{\comm{(A)^{n+1}}{B}\comm{(A)^{j-n-1}}{B}} \\
        &= \sum_{n=0}^{j-1} \binom{j}{n} \Tr{\comm{A}{\comm{(A)^{n}}{B}}\comm{(A)^{j-n-1}}{B}} \\
        &= -\sum_{n=0}^{j-1} \binom{j}{n} \Tr{\comm{(A)^{n}}{B}\comm{(A)^{j-n}}{B}} ,
    \end{align}
    by cyclic permutation of the trace. This summation is the negation of the second binomial term $\binom{j+1-1}{n}$ for all terms $1\leq n\leq j-1$. Therefore only the $n=0$ and $n=j$ terms remain. Moreover since $\binom{j}{j} = \binom{j}{0} = 1$, we have that
    \begin{align}
        &\sum_{n=1}^{j} \binom{j+1}{n} \Tr{\comm{(A)^n}{B}\comm{(A)^{j-n}}{B}} \notag\\
        &= \Tr{\comm{(A)^{j}}{B}B} -\Tr{B\comm{(A)^{j}}{B}}  \\
        &= 0 .
    \end{align}
\end{proof}
Now, applying this result to the summation in \cref{eq:lemmaresult}, where we identify $j=s-2$, we have that all terms for $n\geq 1$ are zero, and therefore only the $n=0$ term survives giving the following relation
\begin{align}
    &\sum_{n = 0}^{s-2}  \binom{s-1}{n}\Tr{\comm{(\mathcal{S})^{n}}{\mathcal{L}}\comm{(\mathcal{S})^{s-2-n}}{\mathcal{L}}} 
    \notag \\
    &= \Tr{\mathcal{L}\comm{(\mathcal{S})^{s-2}}{\mathcal{L}}} .
\end{align}
From this, with a change of index $s-2=s$, we can rewrite the full second-order correction as
\begin{align}
    &\mathcal{I}^{(2)} = \frac{-(\gamma t)^2}{d(d+1)} \times \notag \\
    &\times \left( \frac{\Tr{\mathcal{L}^2}}{2} +\sum_{s = 2}^{\infty} \frac{(-t)^{s} \Tr{\mathcal{L}\comm{(\mathcal{S})^{s}}{\mathcal{L}}}}{(s+2)!} \right) ,
\end{align}
giving the result in \cref{eq:secondorderAGIcorrection}.

Now, as for the gate-independent terms, we wish to express this gate-dependent trace of superoperators explicitly in terms of the operators $H$ and $L$.

Beginning with the binomial expansion of the trace of the iterated commutator \cite{volkin1968iterated},
\begin{align} \label{eq:tracesuperoperatorbinomial}
    &\Tr{\mathcal{L}\comm{(\mathcal{S})^s}{\mathcal{L}}} \notag \\ 
    &= \sum_{k=0}^s (-1)^k \binom{s}{k} \Tr{\mathcal{L} \mathcal{S}^{k} \mathcal{L} \mathcal{S}^{s-k}} \\
    &= i^s  \sum_{k=0}^s \sum_{j=0}^{k} \sum_{l=0}^{s-k} \binom{s}{k} \binom{k}{j} \binom{s-k}{l} (-1)^{s+k-j-l} \times \notag\\
    &\times \Tr{\mathcal{L} \left( (H^*)^{k-j} \otimes H^j \right) \mathcal{L} \left( (H^*)^{s-k-l} \otimes H^l \right)},
\end{align}
where we used the binomial expansion of the expression for $\mathcal{S}$ in terms of $H$ in \cref{eq:Sprepost},
\begin{align}
    \mathcal{S}^k = i^k \sum_{j=0}^k (-1)^{k-j}\binom{k}{j} (H^*)^{k-j} \otimes H^j .
\end{align}
Substituting now the expression for $\mathcal{L}$ in terms of $L$ in \cref{eq:Lprepost}, and expanding algebraically the products of the operator terms in the trace, we obtain the following nine expressions, the sum of which is equivalent to the superoperator trace (omitting the prefactors and summations for convenience):
\begin{align}
    &\Tr{L H^{k-j} L H^{s-k-l}}^*\Tr{L H^j L H^l} , \label{eq:Tr1} \\
    -\frac{1}{2} &\Tr{L H^{k-j} L^TL^* H^{s-k-l}}^*\Tr{L H^{j+l}} , \label{eq:Tr2} \\
    -\frac{1}{2} &\Tr{L H^{s-j-l}}^*\Tr{L H^j L^TL^* H^l} , \label{eq:Tr3} \\
    -\frac{1}{2} &\Tr{L^TL^* H^{k-j} L H^{s-k-l}}^*\Tr{L H^{j+l}} , \label{eq:Tr4} \\
    \frac{1}{4} &\Tr{L^TL^* H^{k-j} L^TL^* H^{s-k-l}}^*\Tr{H^{j+l}} , \label{eq:Tr5} \\
    \frac{1}{4} &\Tr{L^TL^* H^{s-j-l}}^*\Tr{L^TL^* H^{j+l}} , \label{eq:Tr6} \\
    -\frac{1}{2} &\Tr{L H^{s-j-l}}^*\Tr{L^{\dag}L H^j L H^l} , \label{eq:Tr7} \\
    \frac{1}{4} &\Tr{L^TL^* H^{s-j-l}}^*\Tr{L^{\dag}L H^{j+l}} , \label{eq:Tr8} \\
    \frac{1}{4} &\Tr{H^{s-j-l}}^*\Tr{L^{\dag}L H^j L^TL^* H^l} . \label{eq:Tr9}
\end{align}

Now, we can simplify these expressions by making the following observations: Firstly, $s$ must be even since (i) heuristically, the trace must be real since we are calculating the fidelity (a real quantity) and therefore the prefactor $i^s$ in \cref{eq:tracesuperoperatorbinomial} cannot be imaginary, and (ii) formally, we prove this in \cref{app:A:subsec:7}. Secondly, by symmetry of the binomial expansions, we are free to make the following change of variables:
\begin{align*}
    j \rightarrow k - j, \\
    l \rightarrow s - k - l . \\
\end{align*}
For clarity, note that this change of variables does not affect the prefactor
\begin{align*}
    (-1)^{s + k - j - l} \rightarrow (-1)^{j + k + l}, \\
\end{align*}
since, with $s$ even, the ratio of these two terms is always a power of 2 and therefore equal to 1. 

As an example to illustrate the effect of this change of variables, for the first term \cref{eq:Tr1}, the forward and backwards sums are complex conjugates and therefore only the real part can remain (omitting prefactors of the full expression):
\begin{widetext}
    \begin{align}
    \sum_{j=0}^{k} \sum_{l=0}^{s-k} \Tr{L H^{k-j}LH^{s-k-l}}^*\Tr{L H^jLH^l} &\rightarrow \sum_{j=0}^{k} \sum_{l=0}^{s-k} \Tr{L H^{j}LH^{l}}^*\Tr{L H^{k-j}LH^{s-k-l}} \\
    \implies \sum_{j=0}^{k} \sum_{l=0}^{s-k} \Tr{L H^{k-j}LH^{s-k-l}}^*\Tr{L H^jLH^l} &= \sum_{j=0}^{k} \sum_{l=0}^{s-k} \Re{\Tr{L H^{k-j}LH^{s-k-l}}^*\Tr{L H^jLH^l}} .
\end{align}
\end{widetext}

Similarly, by comparing the remaining terms in \cref{eq:Tr2,eq:Tr3,eq:Tr4,eq:Tr5,eq:Tr6,eq:Tr7,eq:Tr8,eq:Tr9} with their counterparts following the change of variables, it is possible to identify the following sums as complex conjugates of one another:
\begin{align*}
    \sum_{j=0}^{k} \sum_{l=0}^{s-k}\left(\ref{eq:Tr2} + \ref{eq:Tr3}\right) &= \sum_{j=0}^{k} \sum_{l=0}^{s-k}2 \Re{\ref{eq:Tr2}} ,\\
    \sum_{j=0}^{k} \sum_{l=0}^{s-k}\left(\ref{eq:Tr4} + \ref{eq:Tr7}\right) &= \sum_{j=0}^{k} \sum_{l=0}^{s-k}2 \Re{\ref{eq:Tr4}} ,\\
    \sum_{j=0}^{k} \sum_{l=0}^{s-k}\left(\ref{eq:Tr6} + \ref{eq:Tr8}\right) &= \sum_{j=0}^{k} \sum_{l=0}^{s-k}2 \Re{\ref{eq:Tr6}} ,\\
   \sum_{j=0}^{k} \sum_{l=0}^{s-k} \left(\ref{eq:Tr5} + \ref{eq:Tr9}\right) &= \sum_{j=0}^{k} \sum_{l=0}^{s-k}2 \Re{\ref{eq:Tr5}} .
\end{align*}

Hence, by combining all of these terms into the full expression for $\Tr{\mathcal{L}\comm{(\mathcal{S})^s}{\mathcal{L}}}$ in \cref{eq:tracesuperoperatorbinomial}, we obtain the following result of \cref{eq:iteratedcommutatoroperator}:
\begin{equation}
   \begin{aligned}
    &\Tr{\mathcal{L}\comm{(\mathcal{S})^s}{\mathcal{L}}}\\
    &=\sum_{k=0}^s \sum_{j=0}^{k} \sum_{l=0}^{s-k} \binom{s}{k} \binom{k}{j} \binom{s-k}{l} (-1)^{s+k-j-l} \times \\
    &\times \Re \Biggl[\Tr{L H^{k-j}LH^{s-k-l}}^*\Tr{L H^jLH^l} \Biggr. \\
    &+ \frac{1}{2} \Tr{L^TL^* H^{s-j-l}}^*\Tr{L^TL^* H^{j+l}}  \\
    &+ \frac{1}{2} \Tr{L^TL^* H^{k-j}L^TL^* H^{s-k-l}}^*\Tr{H^{j+l}}  \\
    &- \Tr{L H^{k-j}L^TL^* H^{s-k-l}}^*\Tr{LH^{j+l}}\\
    &- \Biggl. \Tr{L^TL^* H^{k-j}L H^{s-k-l}}^*\Tr{LH^{j+l}} \Biggr] .
\end{aligned} 
\end{equation}

\subsection{\label{app:A:subsec:7} Tracelessness of the Odd-Order Iterated Commutator}
\begin{theorem}
    Given two operators $A$ and $B$, the product of $B$ with the iterated commutator of order $s$, $B\comm{(A)^s}{B}$, is traceless for all odd $s$,
    \begin{align}
        \forall \, s=2k+1 \, , \, k\in\mathbb{Z} \, ,\, \Tr{B\comm{(A)^s}{B}} = 0 .
    \end{align}
\end{theorem}

\begin{proof}
We begin with the trace of the binomial expansion of the iterated commutator \cite{volkin1968iterated},
\begin{align}
    B\comm{(A)^s}{B} &= \sum_{k=0}^s (-1)^k \binom{s}{k} B A^{s-k} B A^k \\
    \therefore \Tr{B\comm{(A)^s}{B}} &= \sum_{k=0}^s (-1)^k \binom{s}{k} \Tr{B A^{s-k} B A^k} .
\end{align}
By symmetry of the binomial expansion $\binom{s}{k} = \binom{s}{s-k}$, and considering $s$ odd $(-1)^{s-k} = -(-1)^k$, we can write the equivalent expansion
\begin{align}
    B\comm{(A)^s}{B} &= -\sum_{k=0}^s (-1)^k \binom{s}{k} B A^k B A^{s-k} \\
    \therefore \Tr{B\comm{(A)^s}{B}} &= -\sum_{k=0}^s (-1)^k \binom{s}{k} \Tr{B A^k B A^{s-k}} \\
    &= -\sum_{k=0}^s (-1)^k \binom{s}{k} \Tr{B A^{s-k} B A^k},
\end{align}
by cyclic permutation of the trace, $\Tr{CD} = \Tr{DC} \implies \Tr{\comm{C}{D}}=0$. Hence, we have the result for odd $s$
\begin{align}
    \Tr{B\comm{(A)^s}{B}} &= - \Tr{B\comm{(A)^s}{B}} \\
    \implies \Tr{B\comm{(A)^s}{B}} &= 0 .
\end{align}
\end{proof}

\section{\label{app:B} Gates and Operators}
Here we compile additional information and examples regarding the quantum gates and collapse operators used in the main text.

\subsection{\label{app:B:subsec:1} Quantum Gates}
Given the set of single-qudit gates $U_g \in \lbrace \mathbb{1}_d, X, Z, F \rbrace$ defined in \cref{eq:identity,eq:X,eq:Z,eq:F}, we present them in matrix form to illustrate their structure. The identity matrix in $d$ dimensions is trivial, and generated by a null control Hamiltonian $H_{\rm c} = \mathbb{0}_d$, corresponding to no action on the qudit state. 
The generalized Pauli-X or SHIFT gate $X = \Sigma_x$ is defined as
\begin{align}
    X = \begin{bmatrix}
    0 & 1 & 0 & 0 & \cdots & 0 \\
    0 & 0 & 1 & 0 & \cdots & 0 \\
    0 & 0 & 0 & 1 & \cdots & 0 \\
    \vdots & \vdots & \vdots & \vdots & \ddots & \vdots \\
    0 & 0 & 0 & 0 & \cdots & 1 \\    
    1 & 0 & 0 & 0 & \cdots & 0
    \end{bmatrix},
\end{align}
which cyclically permutes the basis states, incrementing them all by 1 level. An equivalent formulation is given by the transpose $X^T$, which cyclically permutes the states downwards,
 \begin{align}
    X^T = \begin{bmatrix}
    0 & 0 & 0 & \cdots & 0 & 1 \\
    1 & 0 & 0 & \cdots & 0 & 0 \\
    0 & 1 & 0 & \cdots & 0 & 0 \\
    0 & 0 & 1 & \cdots & 0 & 0 \\
    \vdots & \vdots & \vdots & \ddots & \vdots & \vdots \\        
    0 & 0 & 0 & \cdots & 1 & 0
    \end{bmatrix}.
\end{align}
The generalized Pauli-$Z$ or CLOCK gate $Z = \Sigma_z$ is defined as
\begin{align}
    Z = \begin{bmatrix}
    1 & 0 & 0 & \cdots & 0 \\
    0 & \omega & 0 & \cdots & 0 \\
    0 & 0 & \omega^2 & \cdots & 0 \\
    \vdots & \vdots & \vdots & \ddots & \vdots \\
    0 & 0 & \cdots & 0 & \omega^{d-1}
    \end{bmatrix},
\end{align}
where the $d$-th roots of unity, $\omega$, are also used to define the matrix form of the Quantum Fourier Transform in $d$ dimensions, being the generalisation of the Walsh-Hadamard (superposition) gate,
\begin{align}
    F &= \frac{1}{\sqrt{d}} \times \notag \\
    &\begin{bmatrix}
    1 & 1 & 1 & 1 & \cdots & 1 \\
    1 & \omega & \omega^2 & \omega^3 & \cdots & \omega^{d-1} \\
    1 & \omega^2 & \omega^4 & \omega^6 & \cdots & \omega^{2(d-1)} \\
    1 & \omega^3 & \omega^6 & \omega^9 & \cdots & \omega^{3(d-1)} \\
    \vdots & \vdots & \vdots & \vdots & \ddots & \vdots \\
    1 & \omega^{d-1} & \omega^{2(d-1)} & \omega^{3(d-1)} & \cdots & \omega^{(d-1)(d-1)}
    \end{bmatrix} .
\end{align}
Note that we may consider a further generalisation of the CLOCK gate from a phase shift of the roots of unity to an arbitrary phase $\phi$,
\begin{align}
    \mathrm{PHASE}(\phi) = \begin{bmatrix}
    1 & 0 & 0 & \cdots & 0 \\
    0 & e^{i \phi} & 0 & \cdots & 0 \\
    0 & 0 & e^{2 i \phi} & \cdots & 0 \\
    \vdots & \vdots & \vdots & \ddots & \vdots \\
    0 & 0 & \cdots & 0 & e^{(d-1) i \phi}
    \end{bmatrix},
\end{align}
where the choice of phase factor allows us to generate not only Clifford gates but also non-Clifford gates, such as for the $T$ gate with $\phi=\pi/8$.

Now, we introduce the interpolated $X$ (SHIFT) and $Z$ (CLOCK) gates, where we raise the $X$ and $Z$ gates to the power $\eta \in \left[ 0, 1 \right]$, which allows us to smoothly interpolate the action of the gate(s) between the identity ($\eta=0$) and the original gate ($\eta = 1$). As an example, consider $\eta=\frac{3d - 2}{4d}$ for $d=2$:
\begin{align}
X^{0.5} = \frac{1}{\sqrt{2}}
\begin{bmatrix}
e^{i \frac{\pi}{4}} & e^{-i \frac{\pi}{4}} \\
e^{-i \frac{\pi}{4}} & e^{i \frac{\pi}{4}} \\
\end{bmatrix} .
\end{align}
It can be seen that for $0 < \eta < 1$ the $X^{\eta}$ and $Z^{\eta}$ gates are non-Clifford.

\subsection{\label{app:B:subsec:2} Collapse Operators}
We present a brief example for the collapse operators of pure dephasing, bit-flip and relaxation, as defined in \cref{eq:Jz,eq:Jx,eq:Jm}. In $d=4$, these are precisely the generalised spin operators $J_z$, $J_x$ and $J_-$ for spin-$\frac{3}{2}$ systems, given by
\begin{align}
    J_z = \frac{1}{2}\begin{bmatrix}
3 & 0 & 0 & 0 \\
0 & 1 & 0 & 0 \\
0 & 0 & -1 & 0 \\
0 & 0 & 0 & -3
\end{bmatrix},
\end{align}
and
\begin{align}
J_x = \frac{1}{2} \begin{bmatrix}
0 & \sqrt{3} & 0 & 0 \\
\sqrt{3} & 0 & 2 & 0 \\
0 & 2 & 0 & \sqrt{3} \\
0 & 0 & \sqrt{3} & 0
\end{bmatrix},
\end{align}
and
\begin{align}
J_- = \begin{bmatrix}
0 & 0 & 0 & 0 \\
\sqrt{3} & 0 & 0 & 0 \\
0 & 2 & 0 & 0 \\
0 & 0 & \sqrt{3} & 0
\end{bmatrix}.
\end{align}

\section{\label{app:C} Complementary Results}
Here, we present further results and investigations that support the findings of the main text.

\subsection{\label{app:C:subsec:1} Uniformity of the CUE for Haar-random Unitaries}
As noted in \cref{sec:2:subsec:3}, the ability to generate ensembles of well-distributed random matrices for arbitrary qudit dimensions is of fundamental importance to the study of the statistical behaviour of the AGI, particularly regarding the aspect of gate-dependence. Given a qudit of dimension $d$, we identify the space of all possible quantum gates with the unitary group $\mathbf{U}(d)$ of $d \times d$ unitary matrices, associated to skew-Hamiltonian control terms generated through the matrix exponential map of the Lie algebra $\mathfrak{u}(d)$ . Therefore, the aim is to sample from a uniform distribution over this group. 

\begin{figure}[t!]
    \centering
    \begin{subfigure}[b]{0.23\textwidth}
        \centering
        \includegraphics[width=\textwidth]{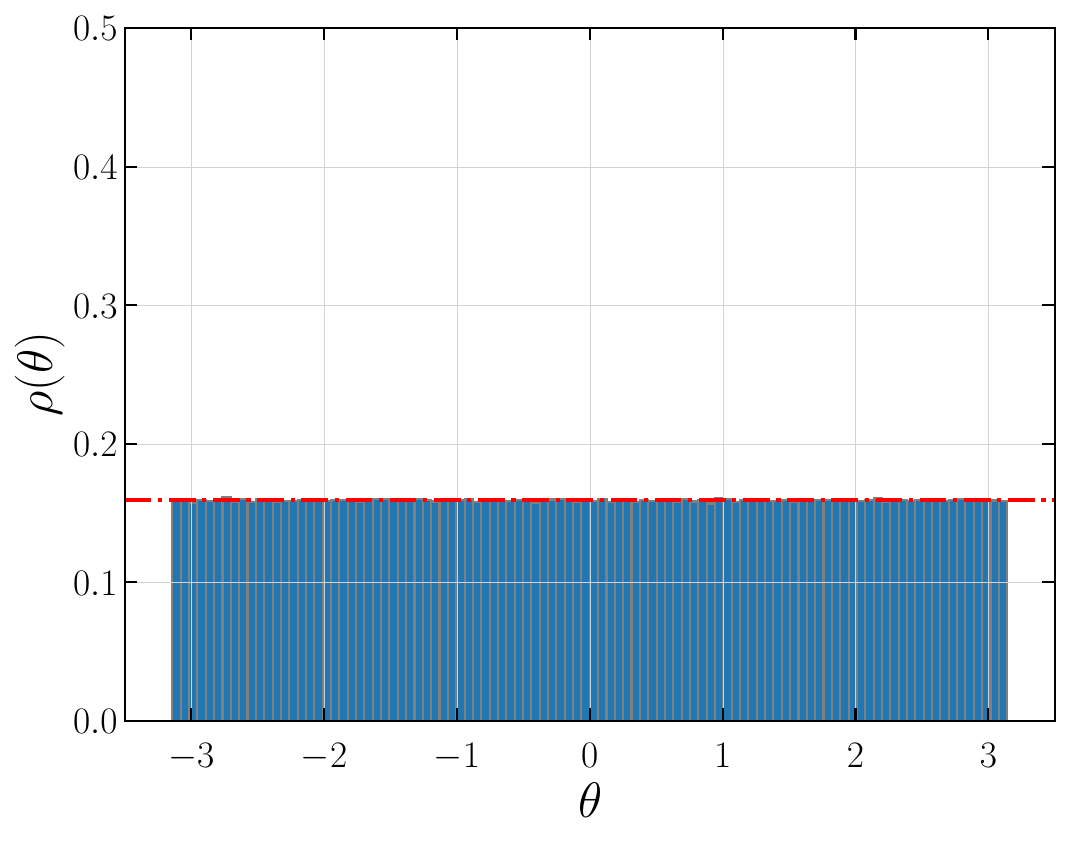}
        \caption{$d=100$ level density}
        \label{subfig:app:1}
    \end{subfigure}
    \hfill
    \begin{subfigure}[b]{0.23\textwidth}
        \centering
        \includegraphics[width=\textwidth]{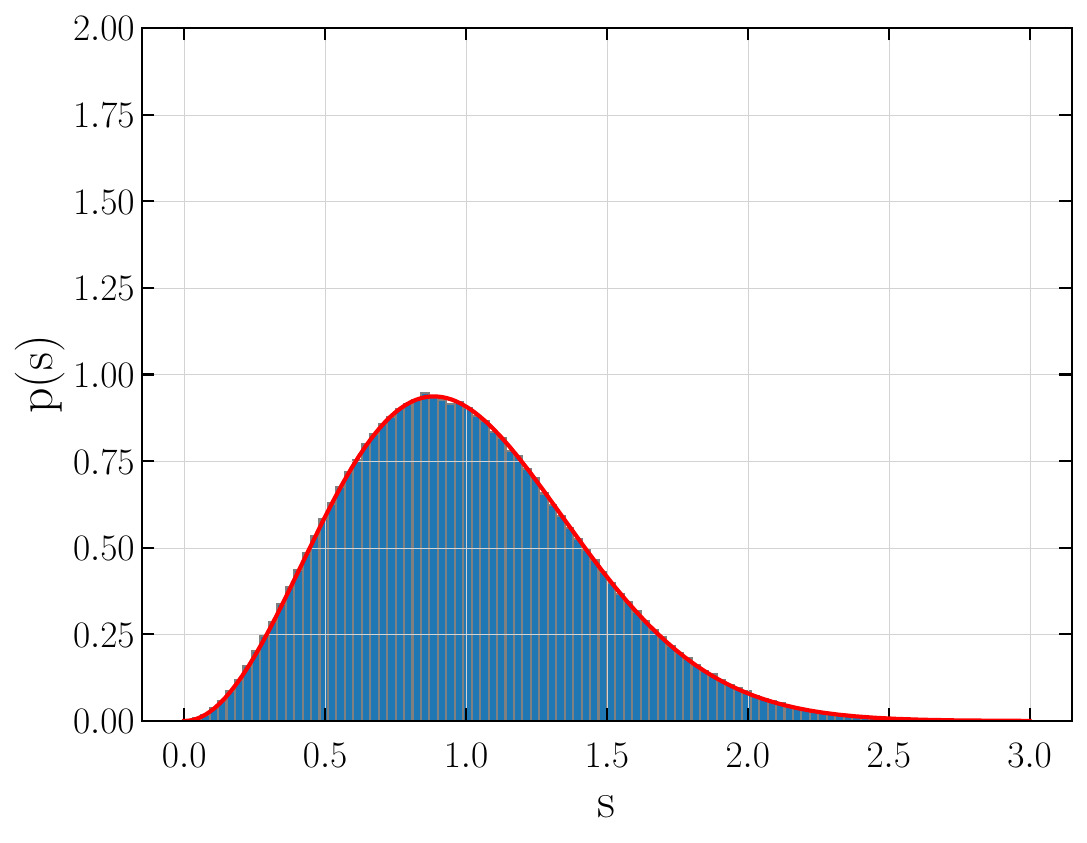}
        \caption{$d=100$ spacings}
        \label{subfig:app:2}
    \end{subfigure}    
    \vfill
    \begin{subfigure}[b]{0.23\textwidth}
        \centering
        \includegraphics[width=\textwidth]{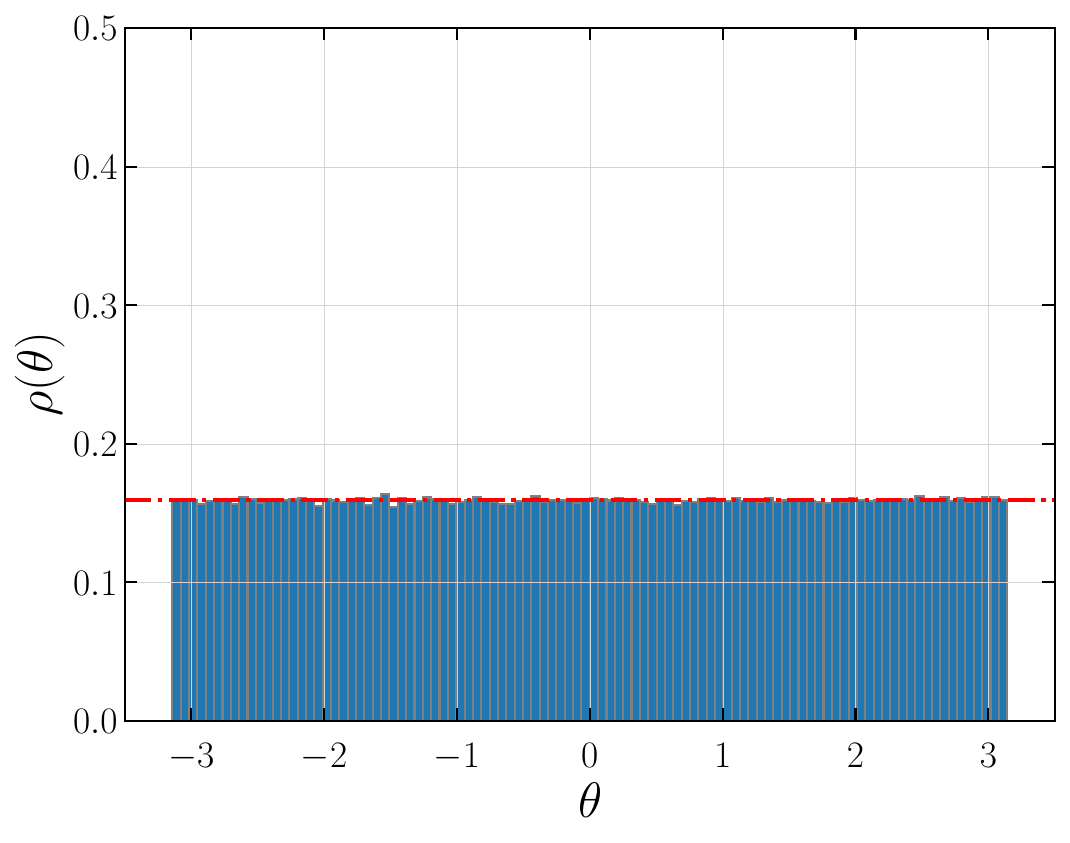}
        \caption{$d=2$ level density}
        \label{subfig:app:3}
    \end{subfigure}
     \hfill
    \begin{subfigure}[b]{0.23\textwidth}
        \centering
        \includegraphics[width=\textwidth]{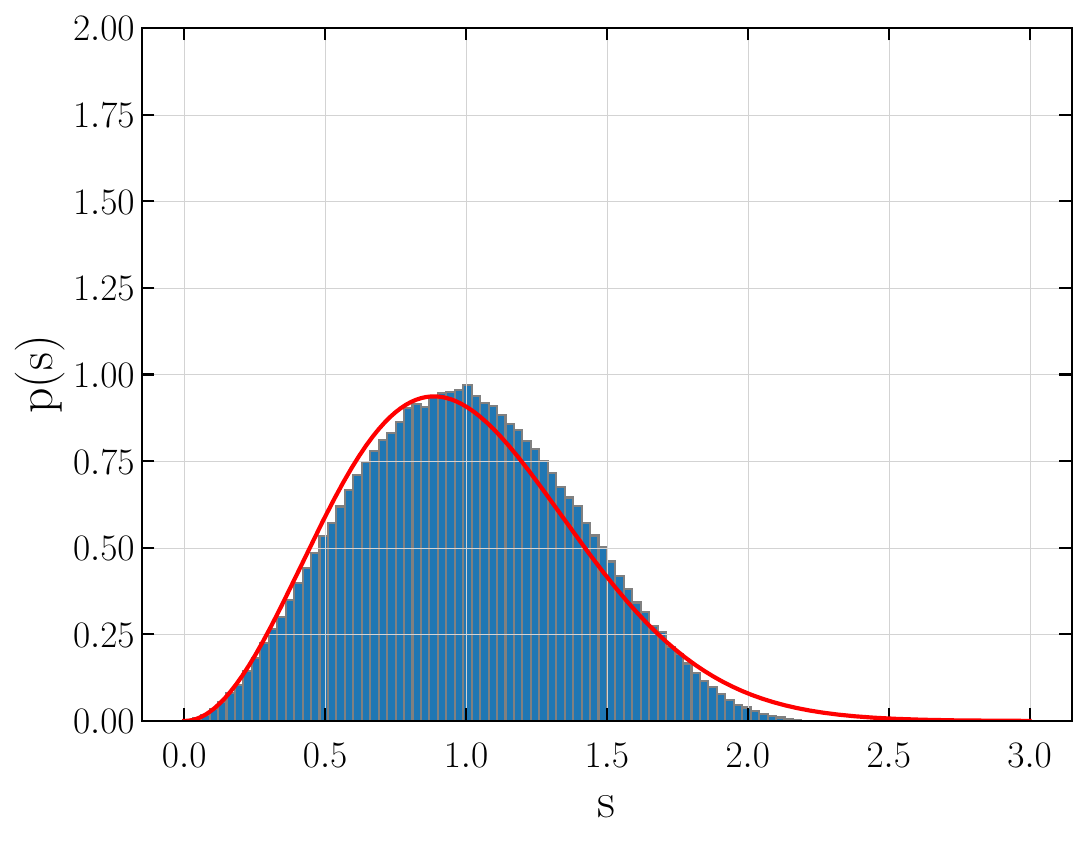}
        \caption{$d=2$ spacings}
        \label{subfig:app:4}
    \end{subfigure}    
    \caption{Method \textsc{QuTiP}.\texttt{rand\_unitary\_haar} (and equivalently \textsc{Cirq}.\texttt{rand\_unitary}), based on \cite{mezzadri_how_2007}.}
    \label{fig:app:1-4}
\end{figure}

Since $\mathbf{U}$ is a compact, connected Lie group, it has a unique (up to positive scalar multiplication) and operation-invariant measure, referred to as the Haar measure \cite{mezzadri_how_2007}. When normalised to one, it produces a uniform probability measure on $\mathbf{U}$, with the associated probability space referred to by the name Circular Unitary Ensemble (CUE), as originally introduced by Dyson \cite{dyson_threefold_1962}. Elements of the CUE are thus typically referred to as Haar-random unitary matrices. The properties of these random matrices have been well-studied, particularly in the standard reference text on random matrix theory by Mehta \cite{mehta_random_2004}, to which we shall refer. For the specific case of the generation of random unitary matrices, we shall refer to the work of Mezzadri \cite{mezzadri_how_2007}.

The spectral statistics form the foundation for the study of random matrices, and in particular we shall concern ourselves with the two primary quantities: (1) the eigenvalue distribution (level density), and (2) the nearest-neighbour eigenvalue ($n$-level, for $n=1$) spacing distribution. 

\begin{figure}[t!]
    \centering
    \begin{subfigure}[b]{0.23\textwidth}
        \centering
        \includegraphics[width=\textwidth]{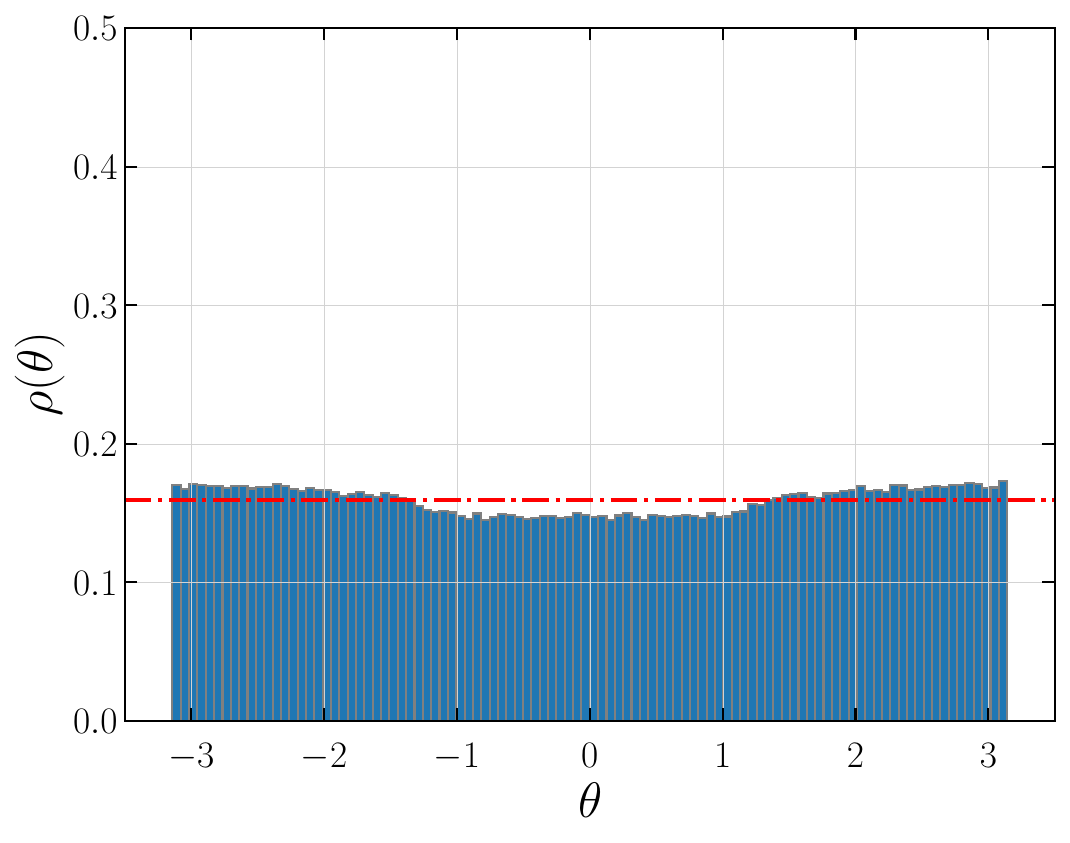}
        \caption{$d=100$ level density}
        \label{subfig:app:5}
    \end{subfigure}
    \hfill
    \begin{subfigure}[b]{0.23\textwidth}
        \centering
        \includegraphics[width=\textwidth]{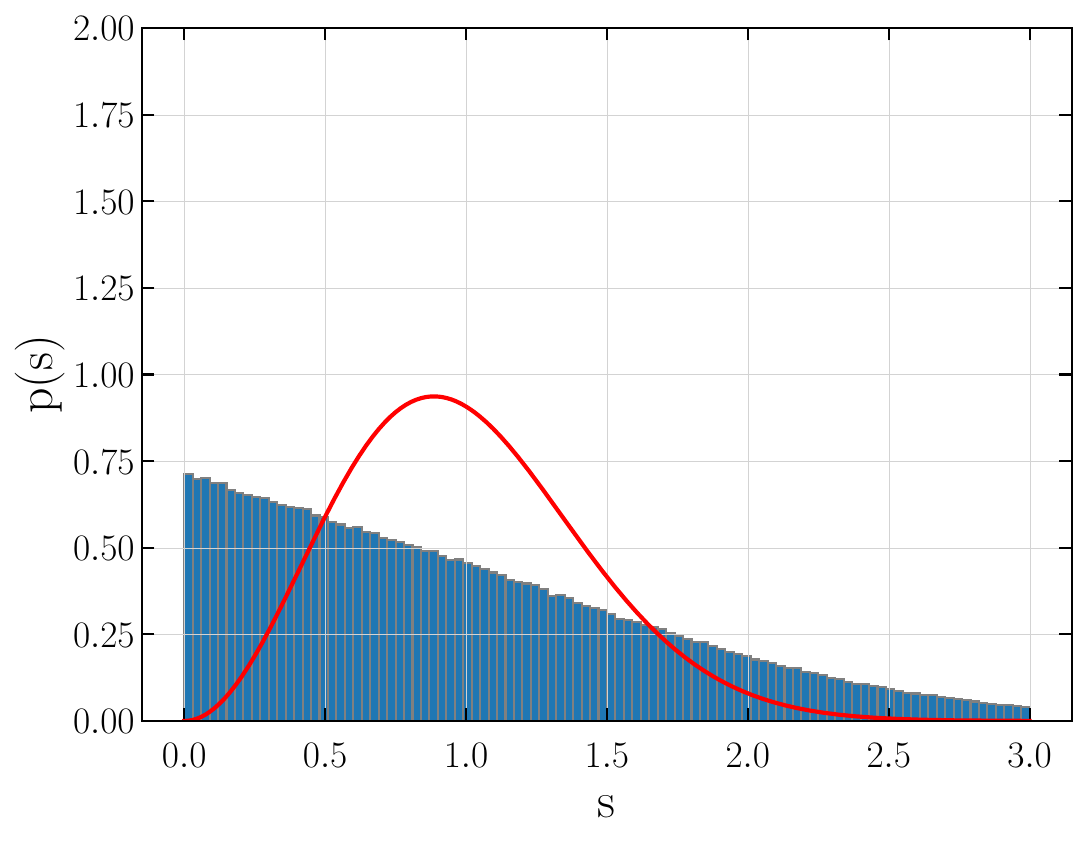}
        \caption{$d=100$ spacings}
        \label{subfig:app:6}
    \end{subfigure}    
    \vfill
    \begin{subfigure}[b]{0.23\textwidth}
        \centering
        \includegraphics[width=\textwidth]{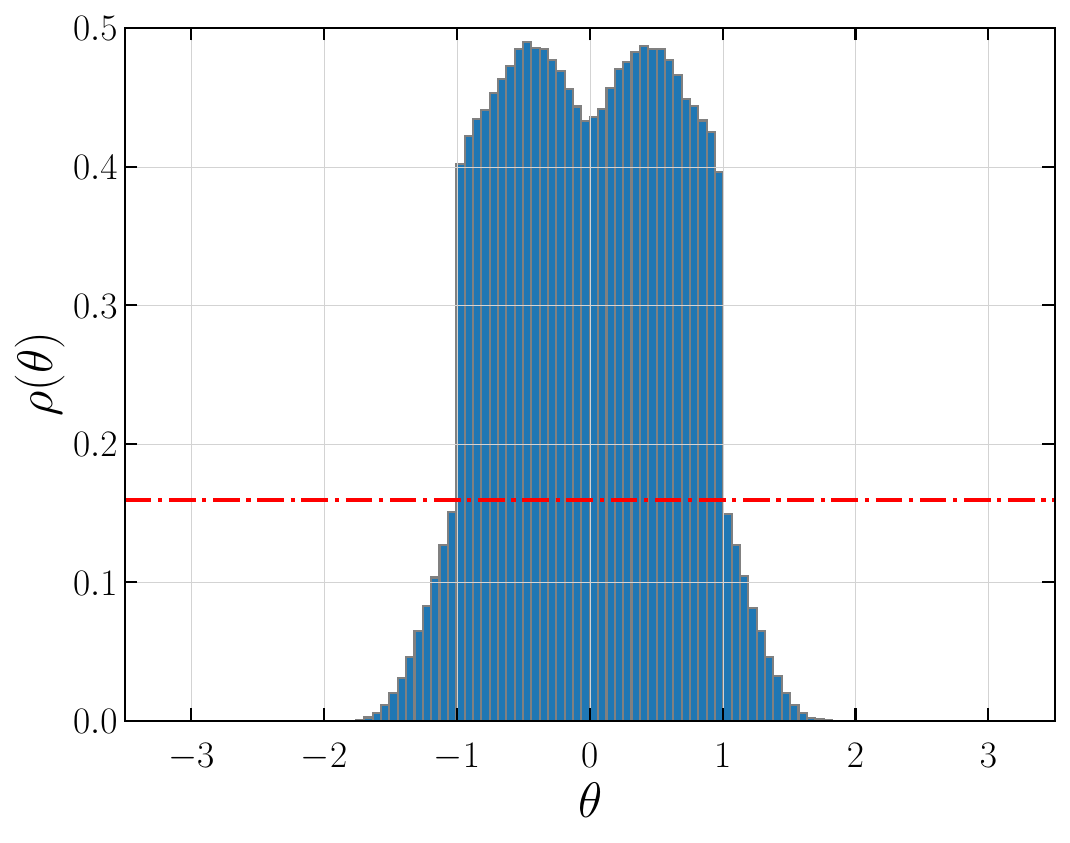}
        \caption{$d=2$ level density}
        \label{subfig:app:7}
    \end{subfigure}
     \hfill
    \begin{subfigure}[b]{0.23\textwidth}
        \centering
        \includegraphics[width=\textwidth]{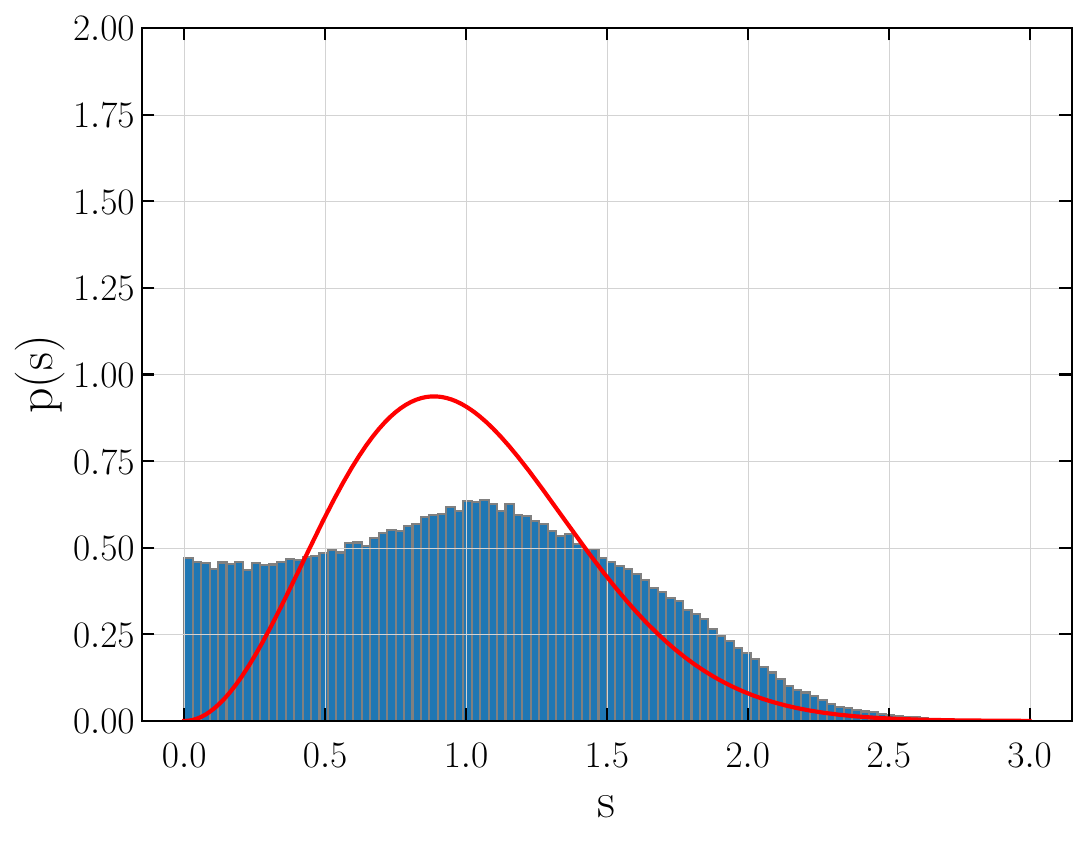}
        \caption{$d=2$ spacings}
        \label{subfig:app:8}
    \end{subfigure}    
    \caption{Method \textsc{QuTiP}.\texttt{rand\_unitary} (and equivalently \textsc{QuTiP}.\texttt{rand\_herm}).}
    \label{fig:app:5-8}
\end{figure}

Since we are studying unitary matrices whose eigenvalues all lie on the unit circle in the complex plane, $\lambda_j = e^{i \theta_j}$, the eigenvalue distribution is thus the distribution of the complex phases $\theta_j$ of the $d$ eigenvalues:
\begin{align}
    -\pi \leq \theta_j < \pi \quad \forall \quad 1 \leq j \leq d .
\end{align}
In the large matrix limit where $d\rightarrow\infty$ the eigenvalue phases will tend towards a uniform distribution on the domain $\left[-\pi, \pi \right)$ with probability density $\rho (\theta) = \frac{1}{2 \pi}$.

Next, the nearest-neighbour eigenvalue spacing distribution $p(s)$ is obtained by: (1) sorting the spectrum of phases $\theta_j$ from smallest to largest, (2) unfolding this spectrum by the factor $\frac{d}{2 \pi}$ to normalise the mean spacing to 1, and then (3) taking their successive differences $s_j = \frac{d}{2 \pi}(\theta_{j+1} - \theta_j)$, resulting in a list of $d-1$ elements. This distribution for the unitary ensembles is well approximated by the Wigner Surmise $p_w(s)$,
\begin{align}
    p_{\rm w}(s) &= \frac{32 s^2}{\pi^2}e^{\frac{-4 s^2}{\pi}},
\end{align}
where $s$ is taken in the continuous limit of $s_j$. In the original work of Wigner \cite{wigner_characteristic_1955} on the Gaussian Unitary Ensemble (GUE) of random Hamiltonian (Hermitian) matrices, this is the exact analytical form for the case of $2\times2$ matrices, which applies to very good approximation for large $d$. Additionally since the GUE and CUE are related functionally, these two distributions become equivalent for both ensembles for large $d$.

Given our requirements of sampling random matrices uniformly from the unitary group, we note that there exist numerous (\textsc{Python}-based) packages with methods for generating unitary matrices. Some common, but non-exhaustive, examples include: \textsc{QuTiP} \cite{johansson_qutip_2012, johansson_qutip_2013} (\texttt{rand\_unitary}, \texttt{rand\_unitary\_haar}, \texttt{rand\_herm}), \textsc{Cirq} \cite{cirq_developers_cirq_2023} (\texttt{rand\_unitary}, \texttt{rand\_special\_unitary}), and \textsc{Bristol} \cite{suzen_spectral_2017} (\texttt{gen\_cue}). 

Both \textsc{QuTiP}'s \texttt{rand\_unitary\_haar} and \textsc{Cirq}'s \texttt{rand\_unitary} methods implement precisely the algorithm provided by Mezzadri in \cite{mezzadri_how_2007}. The \textsc{Cirq}.\texttt{rand\_special\_unitary} method augments the original \texttt{rand\_unitary} method by normalising the determinant of the randomly generated matrix to 1. The \textsc{QuTiP}.\texttt{rand\_unitary} method uses the \texttt{rand\_herm} method to generate a random Hermitian matrix and then computes its unitary equivalent by matrix exponentiation.

\begin{figure}[t!]
    \centering
    \begin{subfigure}[b]{0.23\textwidth}
        \centering
        \includegraphics[width=\textwidth]{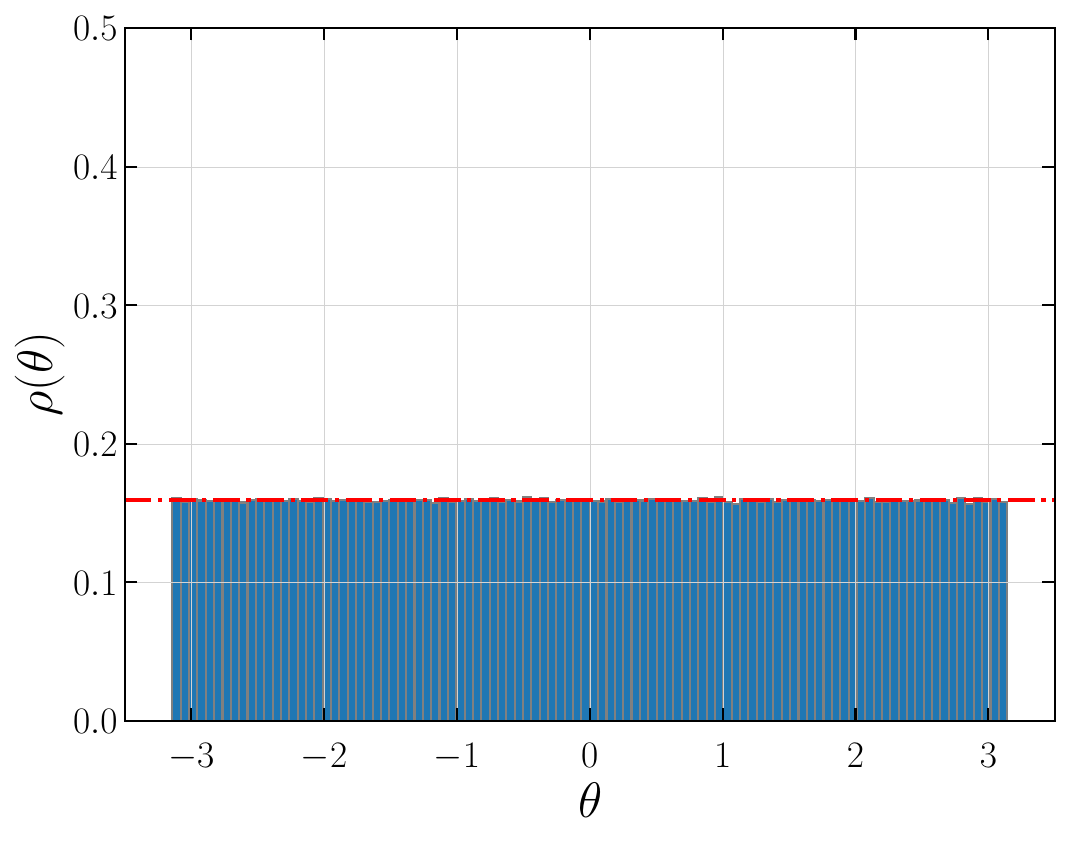}
        \caption{$d=100$ level density}
        \label{subfig:app:9}
    \end{subfigure}
    \hfill
    \begin{subfigure}[b]{0.23\textwidth}
        \centering
        \includegraphics[width=\textwidth]{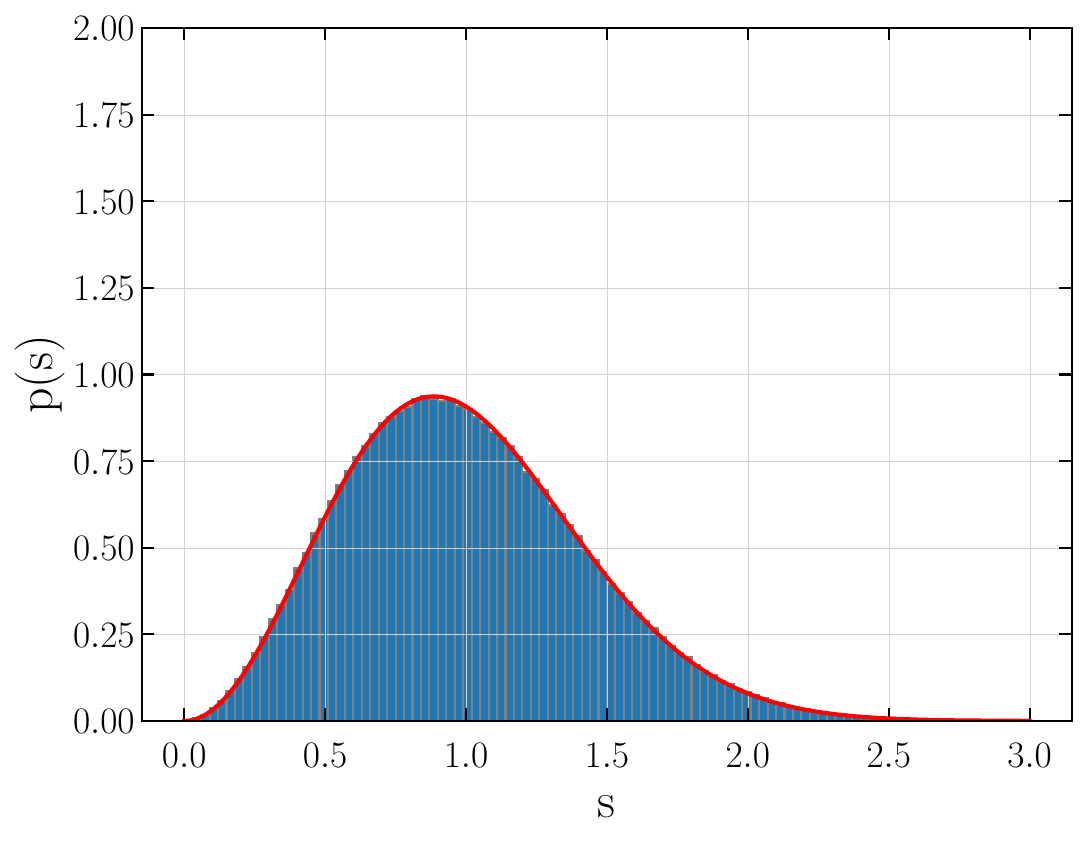}
        \caption{$d=100$ spacings}
        \label{subfig:app:10}
    \end{subfigure}    
    \vfill
    \begin{subfigure}[b]{0.23\textwidth}
        \centering
        \includegraphics[width=\textwidth]{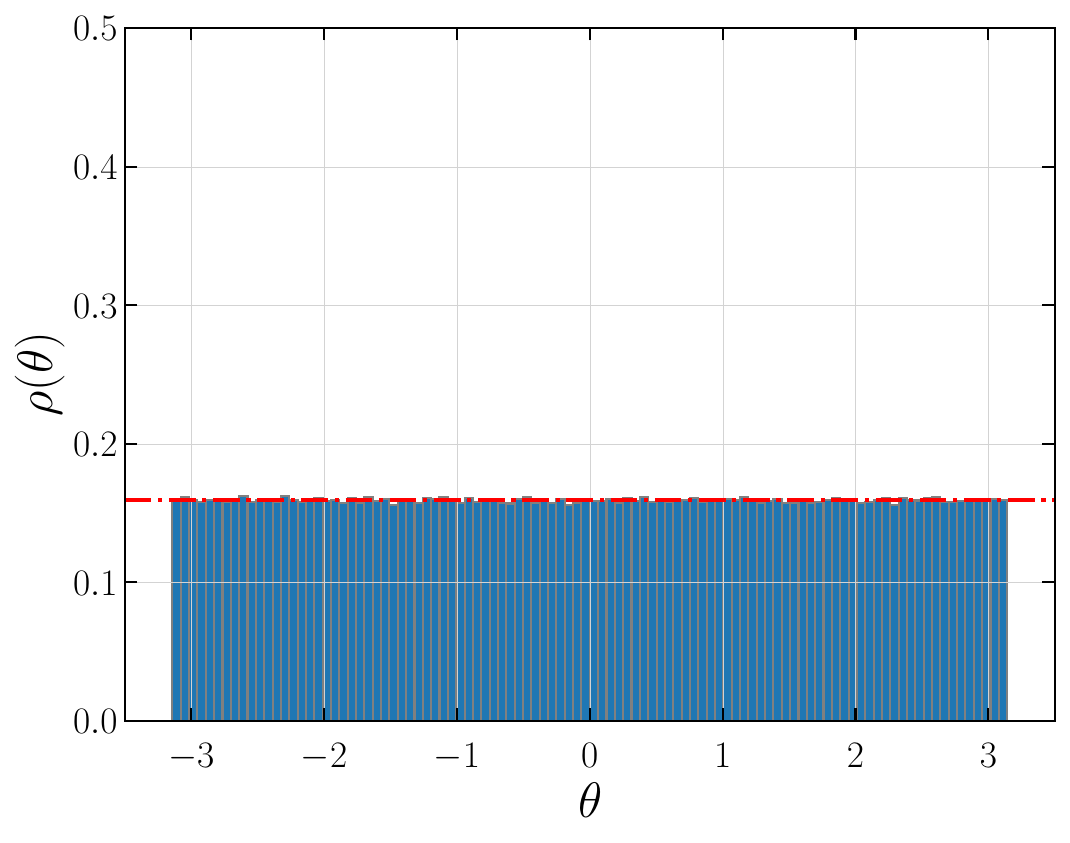}
        \caption{$d=2$ level density}
        \label{subfig:app:11}
    \end{subfigure}
     \hfill
    \begin{subfigure}[b]{0.23\textwidth}
        \centering
        \includegraphics[width=\textwidth]{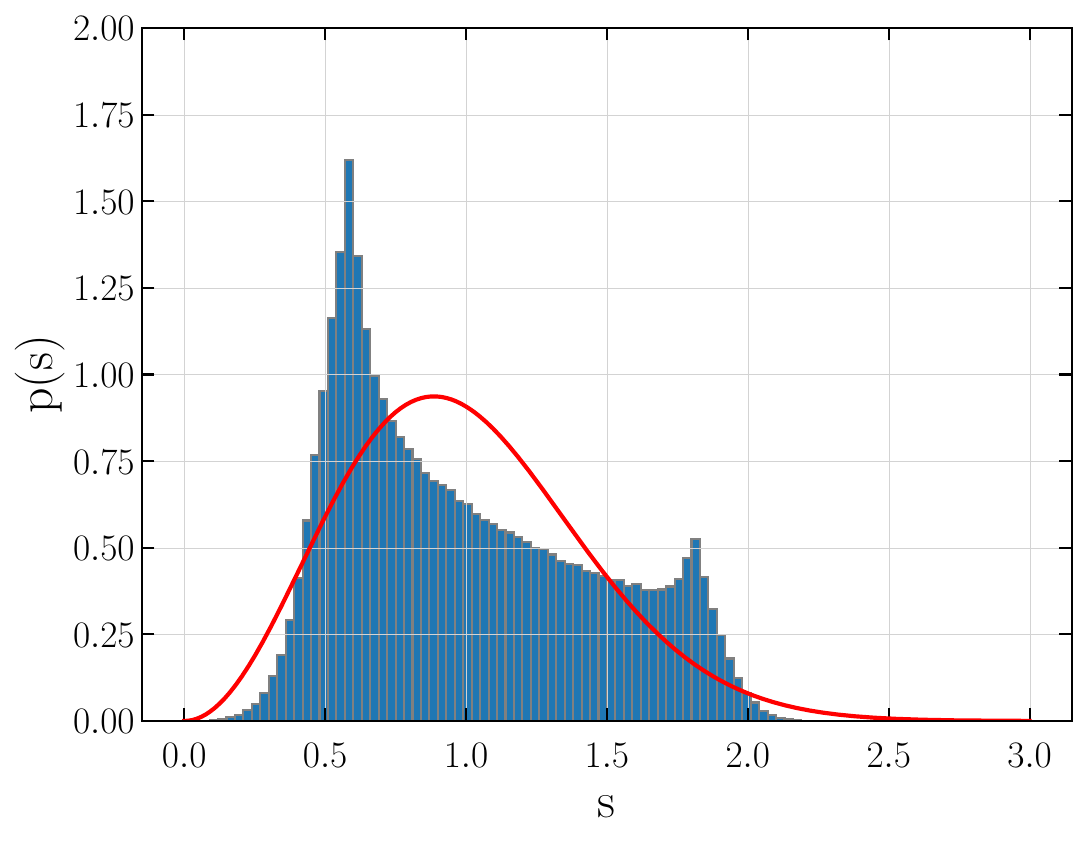}
        \caption{$d=2$ spacings}
        \label{subfig:app:12}
    \end{subfigure}    
    \caption{Method \textsc{Bristol}.\texttt{gen\_cue}}
    \label{fig:app:9-12}
\end{figure}

We tested each of these different methods by sampling one million eigenvalues in each of 2 matrix size configurations, $d\in\lbrace 2, 100 \rbrace$ for each method. These two values were chosen to represent the two regimes of large and small matrix dimensions. The number of eigenvalues was chosen such that, for the algorithm of Mezzadri the standard deviation of the mean for the level density of the eigenvalue phases of the $d=100$ matrices fell below the threshold of $0.001$. The number of eigenvalues was kept constant at each dimension by varying the number of quantum gates generated, specifically 500000 and 10000 for $d=2$ and $d=4$ respectively.

\begin{figure}[t!]
    \centering
    \begin{subfigure}[b]{0.23\textwidth}
        \centering
        \includegraphics[width=\textwidth]{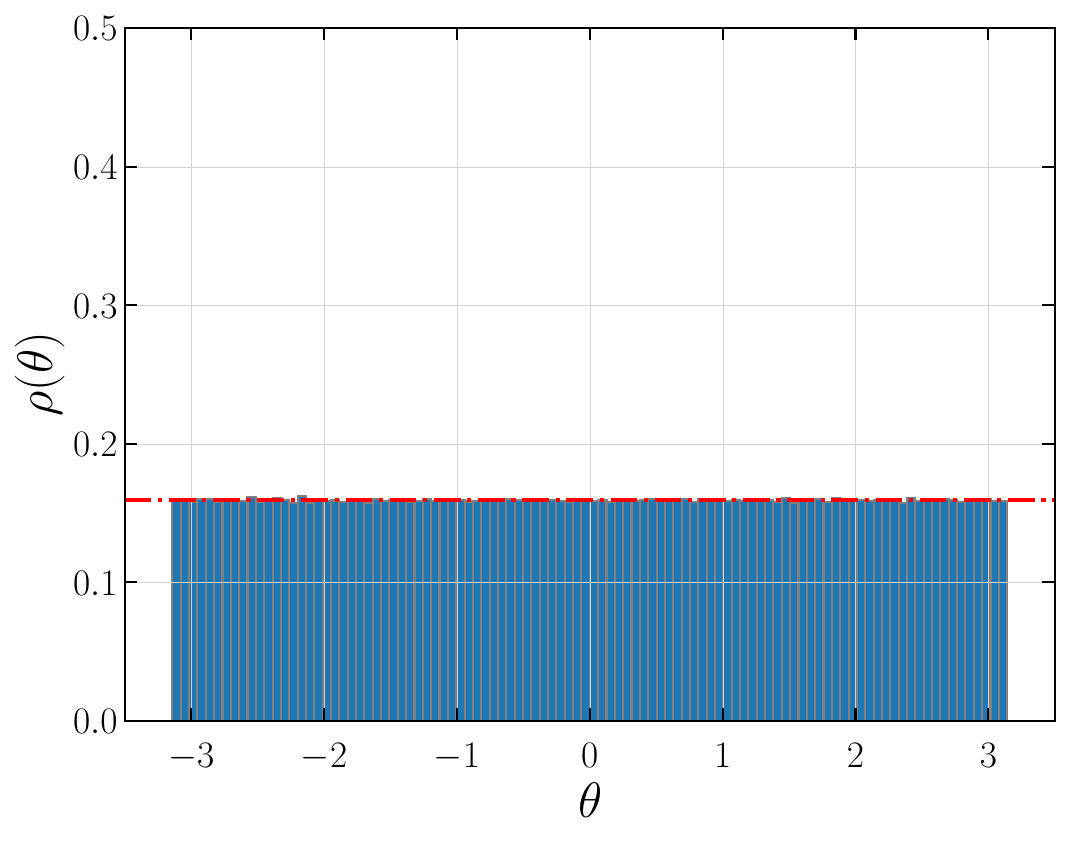}
        \caption{$d=100$ level density}
        \label{subfig:app:17}
    \end{subfigure}
    \hfill
    \begin{subfigure}[b]{0.23\textwidth}
        \centering
        \includegraphics[width=\textwidth]{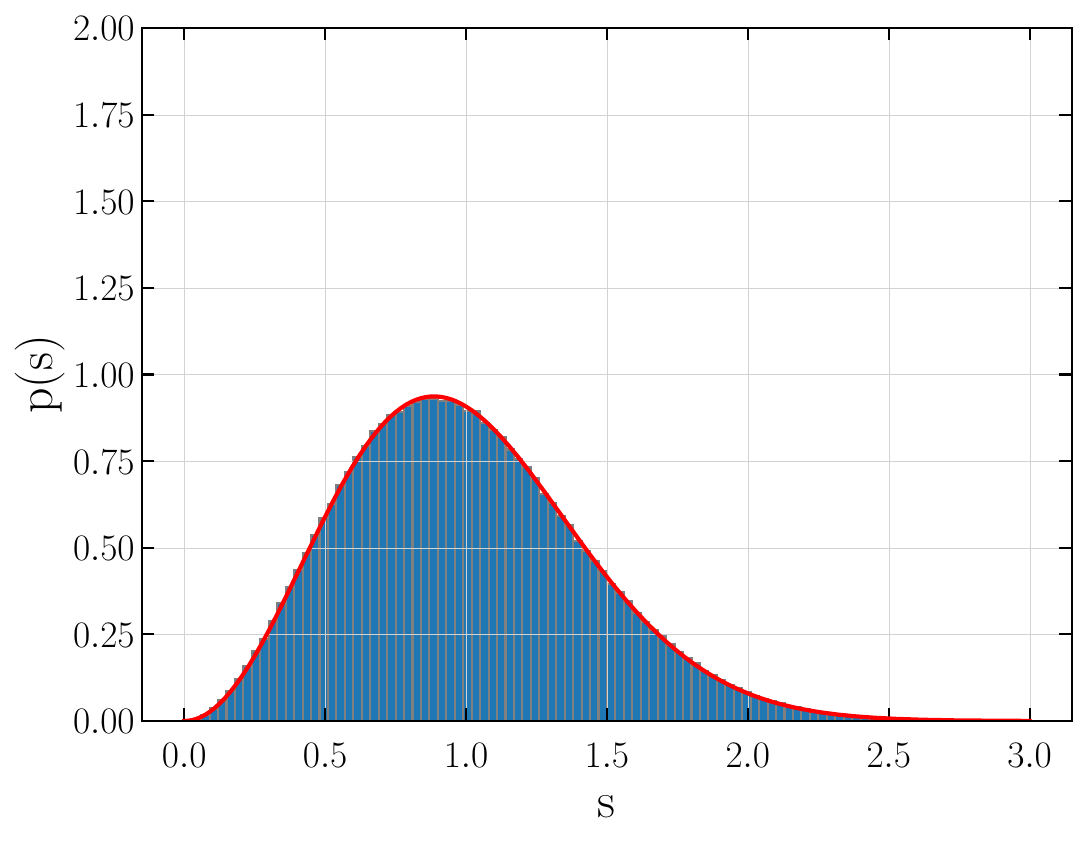}
        \caption{$d=100$ spacings}
        \label{subfig:app:18}
    \end{subfigure}    
    \vfill
    \begin{subfigure}[b]{0.23\textwidth}
        \centering
        \includegraphics[width=\textwidth]{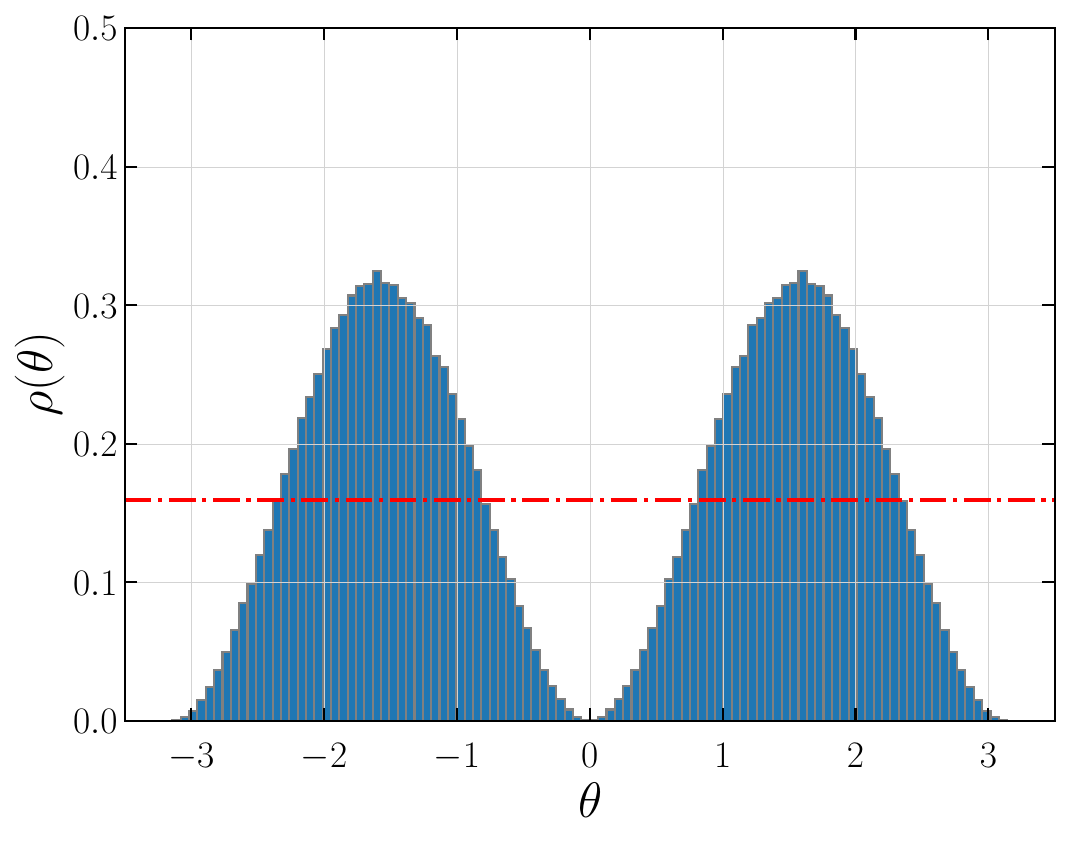}
        \caption{$d=2$ level density}
        \label{subfig:app:19}
    \end{subfigure}
     \hfill
    \begin{subfigure}[b]{0.23\textwidth}
        \centering
        \includegraphics[width=\textwidth]{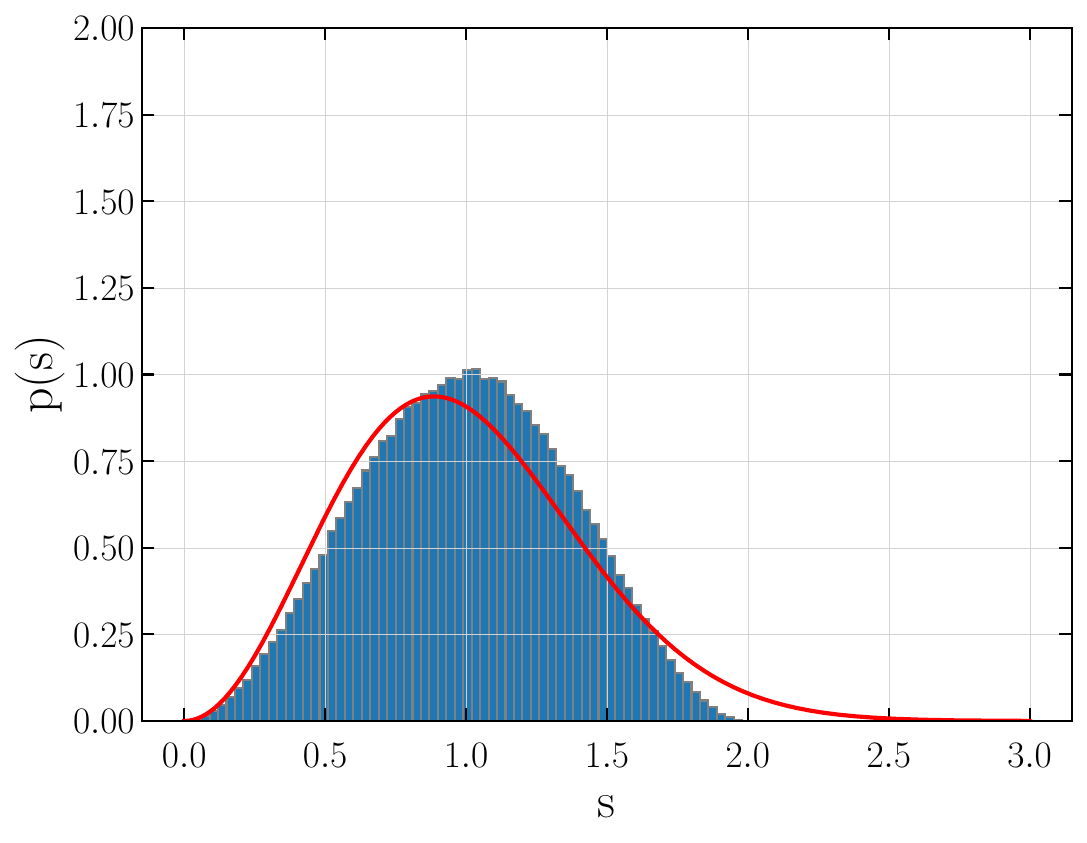}
        \caption{$d=2$ spacings}
        \label{subfig:app:20}
    \end{subfigure}    
    \caption{Method \textsc{Cirq}.\texttt{rand\_special\_unitary}}
    \label{fig:app:17-20}
\end{figure}

In \cref{fig:app:1-4,fig:app:5-8,fig:app:9-12,fig:app:17-20}, the simulated data of eigenvalue and spacing distributions (blue) are compared to their analytically expected results (red). We can see that the method of Mezzadri in \cref{fig:app:1-4} performs the best for generating uniformly distributed Haar-random gates. Hence this is our method of choice. However, even for this method the spacing distribution does show reduced accuracy for lower dimensions, despite the level densities remaining uniform. This could be attributed to the fact that the Wigner Surmise is exact for $2\times2$ Hermitian matrices (but still very accurate for large $d$), but it is only in the large $d$ limit that the distributions for the CUE are identical to that of the GUE \cite{mehta_random_2004}. Thus some deviation is to be expected for small unitary matrices. 

The \textsc{QuTiP}.\texttt{rand\_unitary} and \textsc{QuTiP}.\texttt{rand\_herm} methods in \cref{fig:app:5-8} do not appear suitable for sampling uniformly from the CUE. Indeed, inspection of the sampled matrices showed that approximately 50\% of the generated unitary matrices were diagonal. On the other hand, the \textsc{Bristol}.\texttt{gen\_cue} in \cref{fig:app:9-12} is well optimised for high-dimensional systems but is unsuitable for small dimensions. Finally, it is interesting to note that the special unitary matrices of \textsc{Cirq}.\texttt{rand\_special\_unitary} in \cref{fig:app:17-20} are similarly distributed for large dimension, but show singular behaviour for small $d$.

\subsection{\label{app:C:subsec:2} Distribution of the asymptotic AGIs}
\begin{figure*}[htbp!]
    \centering
    \begin{subfigure}[b]{0.45\textwidth}
        \centering
        \includegraphics[width=\textwidth]{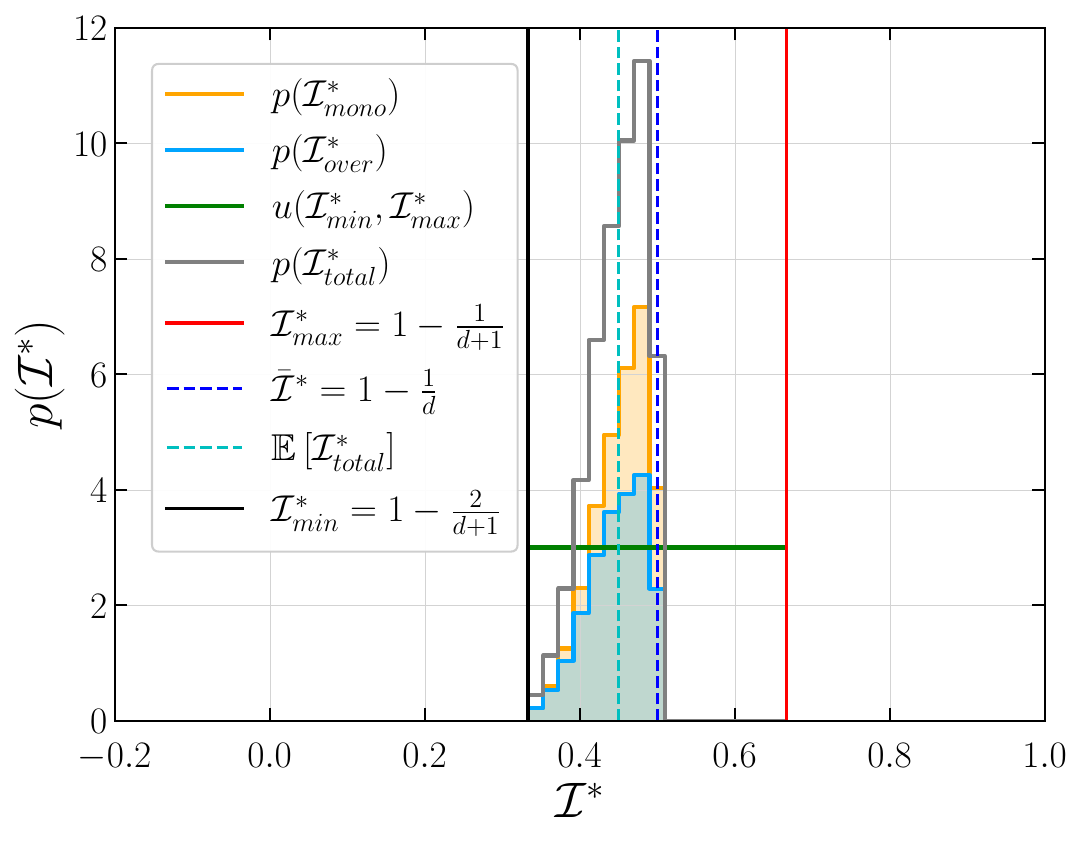}
        \caption{\textsc{Bristol}.\texttt{gen\_cue}}
        \label{subfig:app:22}
    \end{subfigure}
    \hfill
    \begin{subfigure}[b]{0.45\textwidth}
        \centering
        \includegraphics[width=\textwidth]{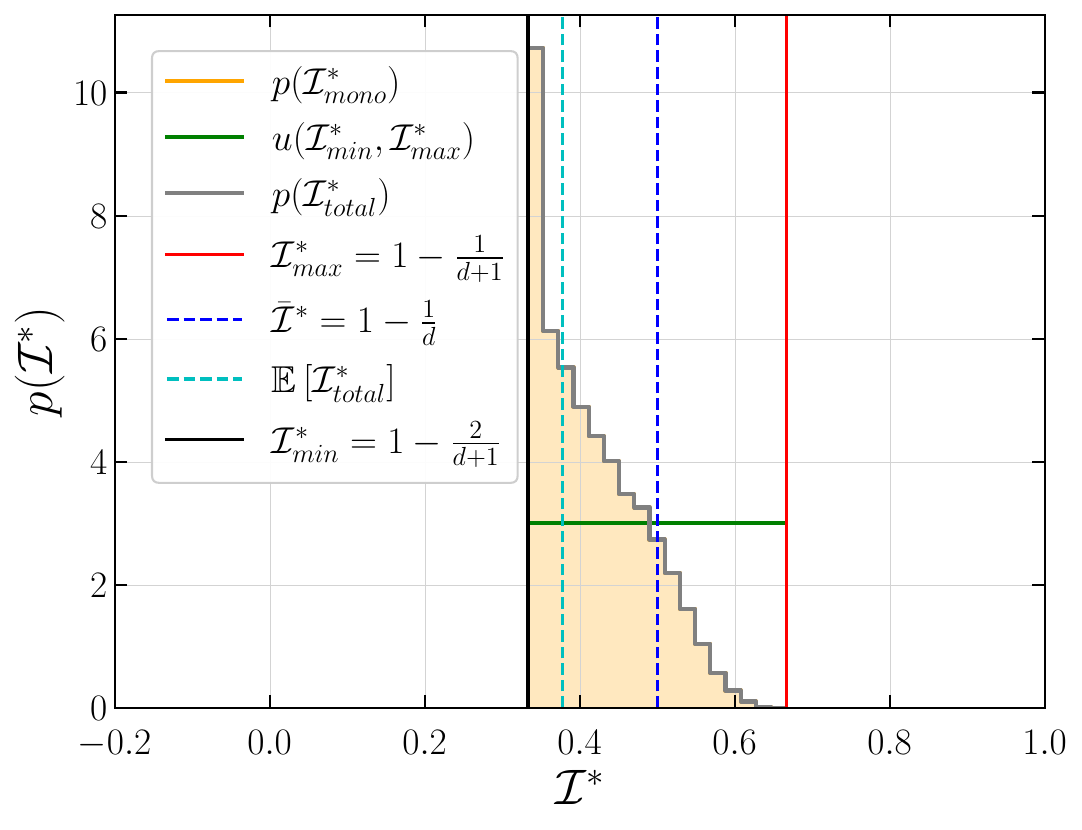}
        \caption{\textsc{QuTiP}.\texttt{rand\_unitary}}
        \label{subfig:app:23}
    \end{subfigure}    
    \vfill
    \begin{subfigure}[b]{0.45\textwidth}
        \centering
        \includegraphics[width=\textwidth]{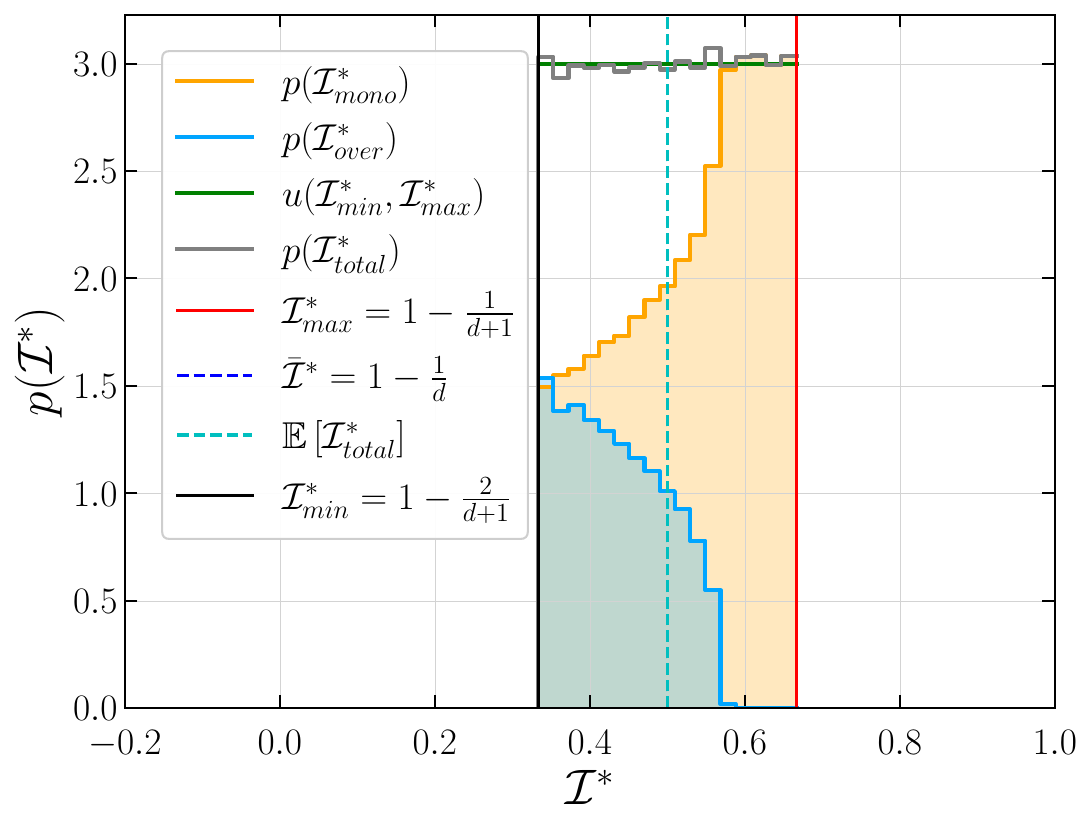}
        \caption{\textsc{Cirq}.\texttt{rand\_unitary}}
        \label{subfig:app:24}
    \end{subfigure}
     \hfill
    \begin{subfigure}[b]{0.45\textwidth}
        \centering
        \includegraphics[width=\textwidth]{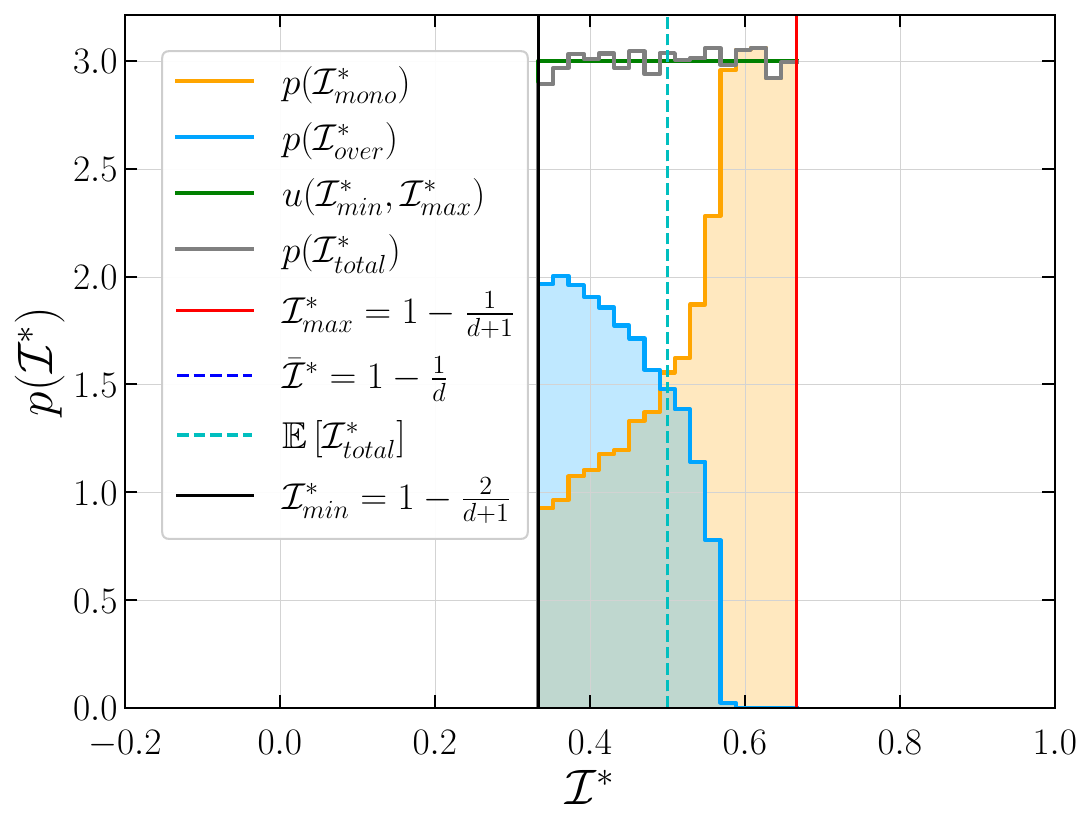}
        \caption{\textsc{Cirq}.\texttt{rand\_special}}
        \label{subfig:app:25}
    \end{subfigure}    
    \caption{\justifying \textbf{Probability distributions of the AGI plateaus for different gate generation methods.} 100000 gates were generated using each of the four methods for $d=2$: \textsc{Bristol}.\texttt{gen\_cue} in \cref{subfig:app:22}, \textsc{QuTiP}.\texttt{rand\_unitary} in \cref{subfig:app:23}, \textsc{Cirq}.\texttt{rand\_unitary} in \cref{subfig:app:24} and \textsc{Cirq}.\texttt{rand\_special} in \cref{subfig:app:25}. For each gate, the $\mathcal{I}^*$ was calculated and categorised according to whether the curves were monotonic (orange) or overshooting (blue). The distributions of the counts of $\mathcal{I}^*$ in the domain $\left[\mathcal{I}^*_{\rm min}, \mathcal{I}^*_{\rm max}\right]$ (black and red vertical lines, respectively) were binned according to Sturge's formula $n_{\rm bins} = \log_2 (n_{\rm gates} + 1)$ and normalised such that the total distribution $p(\mathcal{I}^*_{\rm total})$ (grey) had unit area. The mean plateau value $\mathbb{E}\left( \mathcal{I}^*_{\rm total} \right)$ (cyan, dotted vertical line), was plotted in comparison with the expected mean $\bar{\mathcal{I}}^*=1 - \frac{1}{d}$ (blue, dashed vertical line). The green line represents the expected uniform distribution $u\left(\mathcal{I}^*_{\rm min}, \mathcal{I}^*_{\rm max} \right)$ in the domains.}
    \label{fig:app:22-25}
\end{figure*}

\begin{figure*}[htbp!]
    \centering
    \begin{subfigure}[b]{0.45\textwidth}
        \centering
        \includegraphics[width=\textwidth]{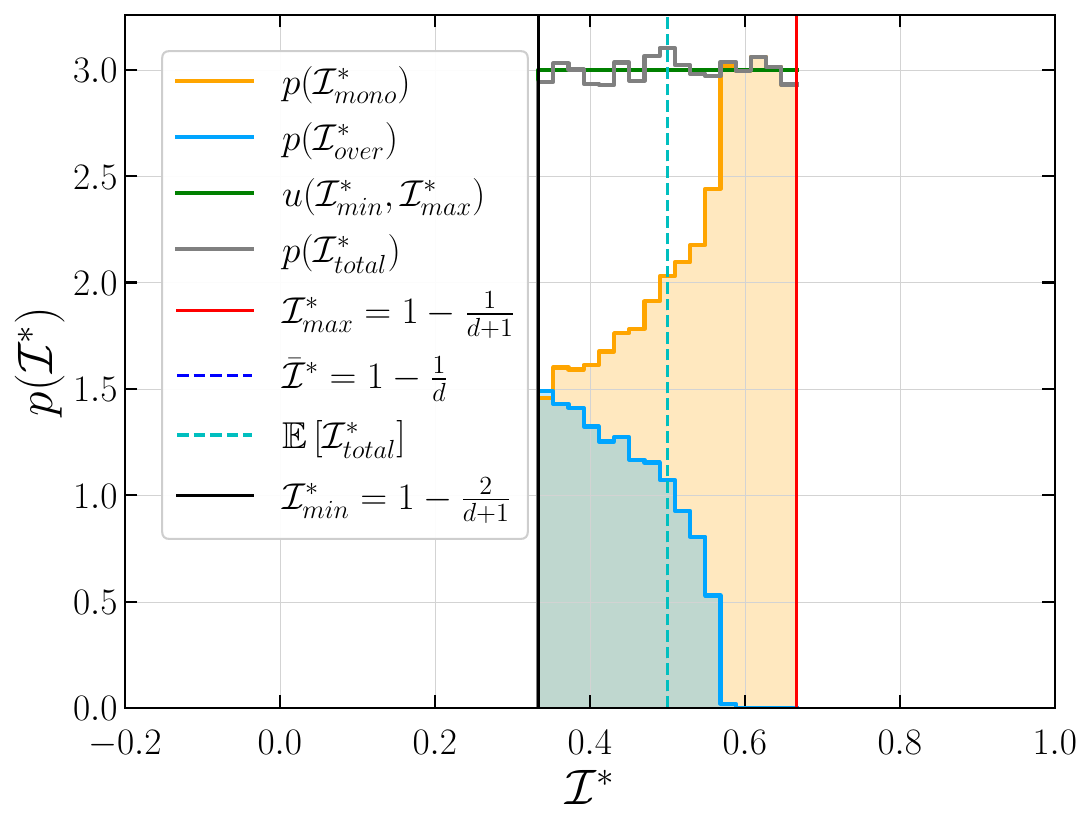}
        \caption{$d=2$}
        \label{subfig:app:21}
    \end{subfigure}
    \hfill
    \begin{subfigure}[b]{0.45\textwidth}
        \centering
        \includegraphics[width=\textwidth]{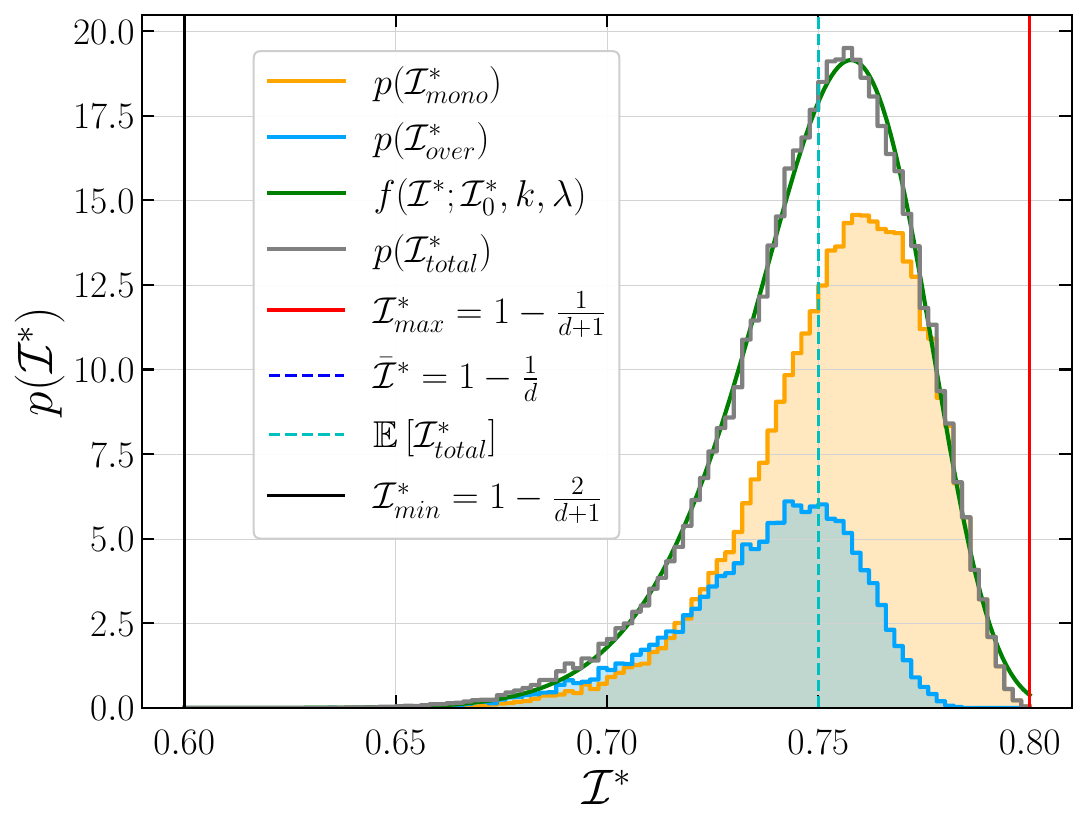}
        \caption{$d=4$}
        \label{subfig:app:26}
    \end{subfigure}    
    \vfill
    \begin{subfigure}[b]{0.45\textwidth}
        \centering
        \includegraphics[width=\textwidth]{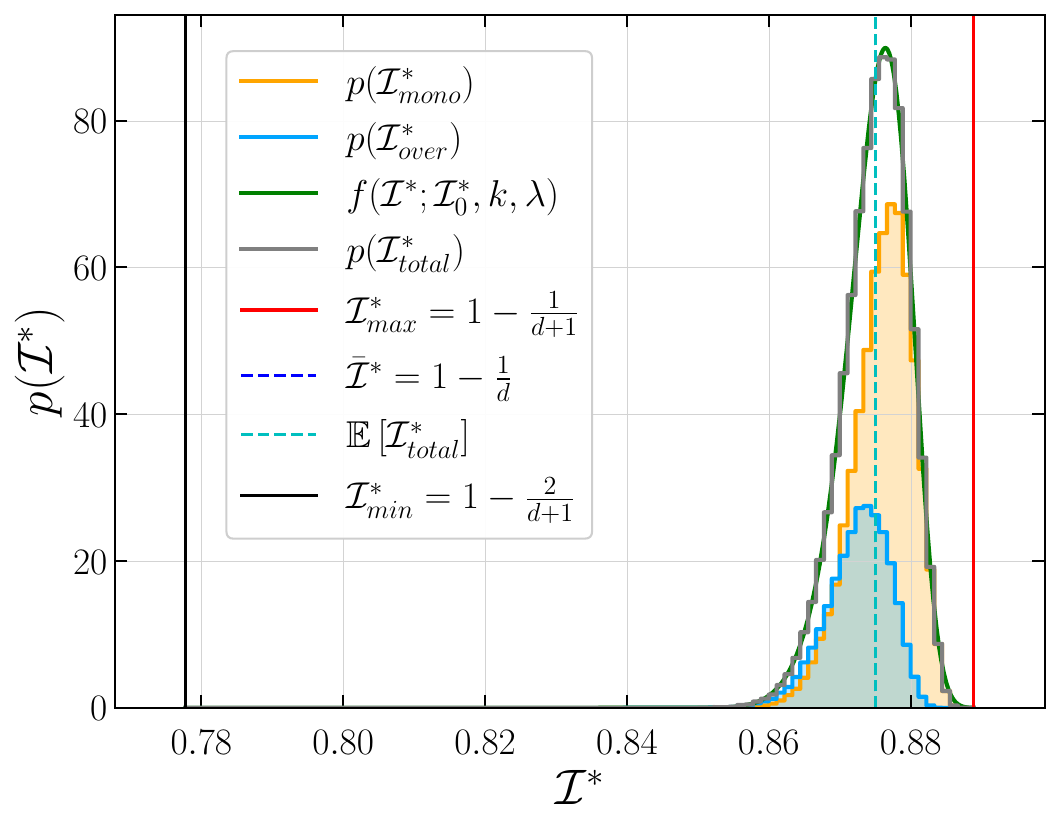}
        \caption{$d=8$}
        \label{subfig:app:27}
    \end{subfigure}
     \hfill
    \begin{subfigure}[b]{0.45\textwidth}
        \centering
        \includegraphics[width=\textwidth]{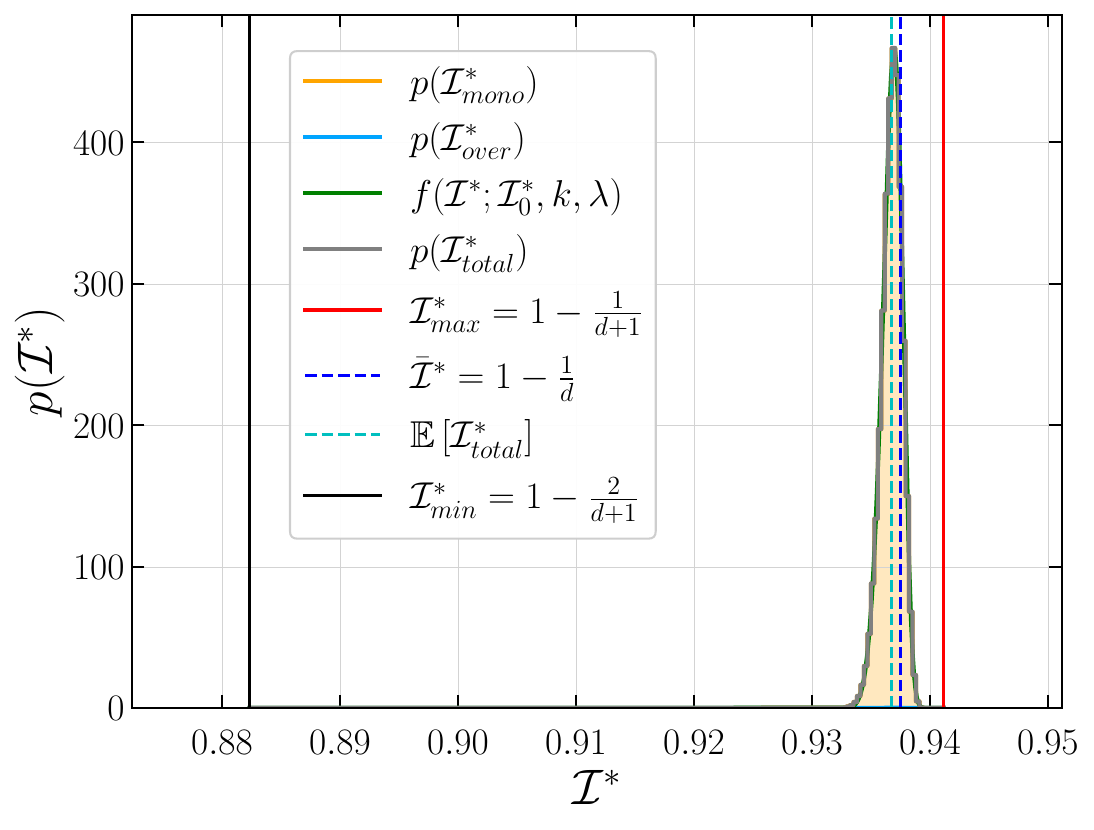}
        \caption{$d=16$}
        \label{subfig:app:28}
    \end{subfigure}    
    \caption{\justifying \textbf{Probability distributions of the AGI plateaus for dimensions $d\in\lbrace 2,4,8,16\rbrace$.} 100000 gates were generated using the method \textsc{QuTiP}.\texttt{rand\_unitary\_haar} of \cite{mezzadri_how_2007} at each dimension: $d=2$ in \cref{subfig:app:21}, $d=4$ in \cref{subfig:app:26}, $d=8$ in \cref{subfig:app:27}, $d=16$ in \cref{subfig:app:28}. For each gate, the $\mathcal{I}^*$ was calculated and categorised according to whether the curves were monotonic (orange) or overshooting (blue). The distributions of the counts of $\mathcal{I}^*$ in the domain $\left[\mathcal{I}^*_{\rm min}, \mathcal{I}^*_{\rm max}\right]$ (black and red vertical lines, respectively) were binned according to: Sturge's formula $n_{\rm bins} = \log_2 (n_{\rm gates} + 1)$ for $d=2$, $n_{\rm bins}=100$ for $d=4,8$ and $n_{\rm bins}=200$ for $d=16$, due to the increasing concentration of the counts for the larger dimensions. The counts were normalised such that the total distribution $p(\mathcal{I}^*_{\rm total})$ (grey) had unit area. The mean plateau value $\mathbb{E}\left( \mathcal{I}^*_{\rm total} \right)$ (cyan, dotted vertical line), was plotted in comparison with the expected mean $\bar{\mathcal{I}}^*=1 - \frac{1}{d}$ (blue, dashed vertical line). The green curve represents the fitted Weibull distributions $f\left(\mathcal{I}^*; \mathcal{I}^*_0, k, \lambda \right)=\frac{k}{\lambda}\left( \frac{\mathcal{I}^* - \mathcal{I}^*_0}{\lambda} \right)^{k-1} e^{-\left( \frac{\mathcal{I}^* - \mathcal{I}^*_0}{\lambda} \right)^k}$ in the domain.}
    \label{fig:app:21-28}
\end{figure*}
We continue our investigations on the different \textsc{Python} methods for generating Haar-random gates, namely: \textsc{Bristol}.\texttt{gen\_cue}, \textsc{QuTiP}.\texttt{rand\_unitary}, \textsc{QuTiP}.\texttt{rand\_unitary\_haar}, \textsc{Cirq}.\texttt{rand\_unitary} and \textsc{Cirq}.\texttt{rand\_special\_unitary}.

In \cref{sec:2:subsec:3}, and \cref{fig:03} specifically, we discussed the distribution of the plateau values $\mathcal{I}^*$ of the AGI in the large-$\gamma t$ regime. Recall that we refer to these plateau values $\mathcal{I}^*$ at the large-$\gamma t$ limit, $\lim_{\gamma t \gg 1}{\mathcal{I}} \rightarrow \mathcal{I}^*$. We showed that they must be bounded above and below by \cref{eq:upperlowerbounds} for all gates. We also observed the appearance of two distinct behaviours ($\mathcal{I}^*_{\rm mono}$ and $\mathcal{I}^*_{\rm over}$) in the sampled Haar-random gates: (1) Monotonic growth of the AGI towards $\mathcal{I}^*$, and (2) overshoot with a single turning point above $\mathcal{I}^*$. 

We note that, as for all the numerical results in the main text, the \textsc{QuTiP}.\texttt{rand\_unitary\_haar} method was used to generate the data shown in \cref{fig:03}. We complement this work by repeating these simulations using the remaining methods (for $d=2$) and also with \textsc{QuTiP}.\texttt{rand\_unitary\_haar} for $d\in\lbrace 2,4,8,16\rbrace$. For each test method and dimension, 100000 gates were generated, their $\mathcal{I}^*$ calculated numerically and their behaviour categorised as either monotonic or overshooting as shown in \cref{tab:app:1}.

\Cref{fig:app:22-25,fig:app:21-28} display the data of these experiments by plotting histograms of the probability densities of the $\mathcal{I}^*$ for each category of gate, $\mathcal{I}^*_{\rm mono}$ in orange and $\mathcal{I}^*_{\rm over}$ in blue, as well as their combined distribution $\mathcal{I}^*_{\rm total}$ in grey. The histograms were normalised such that the area under the curve of $p(\mathcal{I}^*_{\rm total})$ was equal to $1$. The red and black vertical lines show the upper and lower bounds on $\mathcal{I}^*$, respectively. The dashed blue vertical lines show the expected mean value for each dimension.

Looking closer at \cref{fig:app:22-25}, the two \textsc{Cirq} methods in \cref{subfig:app:24,subfig:app:25} appear to have distributions matching most closely with that of \textsc{QuTiP}.\texttt{rand\_unitary\_haar} in \cref{subfig:app:21}, although the normalisation procedure for the special unitary matrices appears to generate a larger proportion of gates with overshoot. It is interesting to note that the \textsc{Bristol}.\texttt{gen\_cue} method in \cref{subfig:app:22} generates an increasing number of gates near the expected mean, but only below, with no gates in the range $1-\frac{1}{d} \leq \mathcal{I}^* \leq 1 - \frac{1}{d + 1}$. As mentioned in \cref{app:C:subsec:1}, \textsc{QuTiP}.\texttt{rand\_unitary} in \cref{subfig:app:23} generates a large number of diagonal unitary matrices, whose $\mathcal{I}^*=\mathcal{I}^*_{\rm min} = 1 - \frac{2}{d+1}$ are equivalent to that of the identity matrix $\mathbb{1}_d$.

\Cref{fig:app:21-28} shows the preferred \textsc{QuTiP}.\texttt{rand\_unitary\_haar} method for different dimensions $d\in\lbrace 2,4,8,16\rbrace$. We note that at each dimension the computed mean over all sampled $\mathcal{I}^*$ is equal to (within statistical error) the expected mean shown by the dashed blue vertical line, $\bar{\mathcal{I}}^*=1 - \frac{1}{d}$. However, most striking is the fact that for $d>2$ the total probability density is no longer uniformly distributed between $\mathcal{I}^*_{\rm min}$ and $\mathcal{I}^*_{\rm max}$, but becomes increasingly concentrated about the mean. 

\begin{table*}[t!]
    \centering
    \begin{tabular}{|c|c|c|c|c|c|}
        \hline
        $d$ & Method & $\mathcal{I}^*_{\rm mono}$ (\%) & $\mathcal{I}^*_{\rm over}$ (\%) & $\epsilon_{\rm rel}$ (\%) & $D_{\rm KL}$ \\
        \hline
        2 & \textsc{Bristol}.\texttt{gen\_cue} & $59.56$ & $40.44$ & $0.100$ & $0.885$ \\
        2 & \textsc{QuTiP}.\texttt{rand\_unitary} & $100.00$ & $0.00$ & $0.243$ & $0.438$ \\
        2 & \textsc{Cirq}.\texttt{rand\_unitary} & $73.06$ & $26.94$ & $1.02\times 10^{-3}$ & $5.92\times 10^{-5}$ \\
        2 & \textsc{Cirq}.\texttt{rand\_special\_unitary} & $61.66$ & $38.34$ & $6.83\times 10^{-4}$ & $1.32\times 10^{-4}$ \\
        2 & \textsc{QuTiP}.\texttt{rand\_unitary\_haar} & $75.51$ & $24.49$ & $2.56\times 10^{-4}$ & $7.15\times 10^{-5}$ \\
        4 & \textsc{QuTiP}.\texttt{rand\_unitary\_haar} & $71.24$ & $28.76$ & $4.38\times 10^{-5}$ & $2.92\times 10^{-3}$ \\
        8 & \textsc{QuTiP}.\texttt{rand\_unitary\_haar} & $70.21$ & $29.79$ & $1.27\times 10^{-5}$ & $1.21\times 10^{-3}$ \\
        16 & \textsc{QuTiP}.\texttt{rand\_unitary\_haar} & $99.95$ & $0.05$ & $7.98\times 10^{-4}$ & $1.36\times 10^{-3}$ \\
        \hline
    \end{tabular}
    \caption{\justifying \textbf{Proportions of Random Gates with and without Turning Points.} 100000 gates were generated for each of the method and dimension pairs shown. The $\mathcal{I}^*_{\rm mono}$ (\%) and $\mathcal{I}^*_{\rm over}$ columns show the percentages of gates falling into either of the two categories: (1) monotonic growth, and (2) overshooting behaviour. The column $\epsilon_{\rm rel}$ gives the relative error between the measured mean of the overall data $\mathbb{E}(\mathcal{I}^*_{\rm total})$ and the expected mean $\bar{\mathcal{I}}^*=1 - \frac{1}{d}$ for the calculated AGI plateau values, given by $\epsilon_{\rm rel}=\abs{\frac{\mathbb{E}(\mathcal{I}^*_{\rm total}) - \bar{\mathcal{I}}^*}{\bar{\mathcal{I}}^*}}$. The quantity $D_{\rm KL}$ is the Kullback-Leibler divergence measuring the relative entropy between the distribution $p\left(\mathcal{I}^*_{\rm total}\right)$ and the reference distribution: $D_{\rm KL}\left(p(\mathcal{I}^*_{\rm total}) || u\left(\mathcal{I}^*_{\rm min}, \mathcal{I}^*_{\rm max} \right)\right)$ for $d=2$ uniform and $D_{\rm KL}\left(p(\mathcal{I}^*_{\rm total}) || f\left(\mathcal{I}^*; \mathcal{I}^*_0, k, \lambda \right)\right)$ for $d>2$ Weibull.}
    \label{tab:app:1}
\end{table*}

We remark that this observed behaviour is due to the concentration of measure phenomenon as the dimensionality of the system increases \cite{giannopoulos_concentration_2000,ledoux_concentration_2001}. To understand this, we first demonstrate the following property of the AGI:

\begin{lemma}
    The AGI function $\mathcal{I} \, : \, \mathbf{U}(d) \rightarrow \mathbb{R}$ is $L$-Lipschitz continuous with respect to the Frobenius norm $\norm{A}_F=\sqrt{\Tr{A^{\dag}A}}$ over the unitary group $\mathbf{U}(d)$ with Lipschitz constant $L = \frac{2}{d+1}$, such that
    \begin{align}
        \abs{\mathcal{I}(U) - \mathcal{I}(U^{\prime})} \leq L \norm{U - U^{\prime}}_F 
    \end{align}
    where $U, U^{\prime} \in \mathbf{U}(d)$.
\end{lemma}

\begin{proof}
    The AGF $\mathcal{F}$ can be written in terms of the process fidelity $F$ as \cite{horodecki_general_1999}
    \begin{align}
        \mathcal{F} = \frac{d F + 1}{d + 1},
    \end{align}
    where $F = \frac{1}{d^2} \abs{\Tr{U^{\dag}V}}^2$ for two unitaries $U, V \in \mathbf{U}(d)$. Then, with $\mathcal{I} = 1 -  \mathcal{F}$, we have that
    \begin{align}
        \mathcal{I} = \frac{d^2 - \abs{\Tr{U^{\dag}V}}^2}{d(d+1)}.
    \end{align}
    Considering the difference, we have that
    \begin{align}
        &\abs{\mathcal{I}(U) - \mathcal{I}(U^{\prime})} \notag \\
        &= \frac{1}{d(d+1)} \abs{\abs{\Tr{U^{\dag}V}}^2 - \abs{\Tr{U^{\prime \dag}V}}^2} \\
        &= \frac{1}{d(d+1)}\left|\abs{\Tr{U^{\dag}V}} - \abs{\Tr{U^{\prime \dag}V}}\right| \times \notag\\
        & \times \left|\abs{\Tr{U^{\dag}V}} + \abs{\Tr{U^{\prime \dag}V}}\right| ,
    \end{align}
    by difference of squares. Now,
    \begin{align}
        &\left|\abs{\Tr{U^{\dag}V}} + \abs{\Tr{U^{\prime \dag}V}}\right| \notag \\
        &\leq \abs{\Tr{U^{\dag}V}} + \abs{\Tr{U^{\prime \dag}V}}  \\
        &\leq 2d ,
    \end{align}
    by the Cauchy-Schwartz inequality and since the trace of a unitary is at most $d$. Similarly,
    \begin{align}
        &\left|\abs{\Tr{U^{\dag}V}} - \abs{\Tr{U^{\prime \dag}V}}\right| \notag \\
        &\leq \abs{\Tr{U^{\dag}V} - \Tr{U^{\prime \dag}V}} \\
        &\leq \norm{U^{\dag}V - U^{\prime \dag}V}_F \\
        &\leq \norm{\left(U - U^{\prime}\right)^{\dag}V}_F \\
        &\leq \norm{U - U^{\prime}}_F ,
    \end{align}
    and by unitary invariance of the Frobenius norm. Therefore, substituting these two expressions,
    \begin{align}
        \abs{\mathcal{I}(U) - \mathcal{I}(U^{\prime})} &\leq \frac{2}{d+1} \norm{U - U^{\prime}}_F ,
    \end{align}
    and thus $\mathcal{I}$ is Lipschitz with $L=\frac{2}{d+1}$.
\end{proof}
Now, we can apply this result for $\mathcal{I}$ in conjunction with Levy's Lemma extended to the classical compact group $\mathbf{U}(d)$ \cite{meckes_random_2019}, which we restate here without proof:
\begin{theorem}
    Let $f \, : \, X \rightarrow \mathbb{R}$ be $L$-Lipschitz continuous with Lipschitz constant $L$, and let $X$ be an element of one of the compact classical groups. Then, for each $t>0$,
    \begin{align} \label{eq:levylemma}
        \mathbb{P}\left[ \abs{f(X) - \mathbb{E}\left[f(X)\right]} \geq t \right] &\leq e^{-\frac{(d-2)t^2}{24L^2}} .
    \end{align}    
\end{theorem}
Using our result for $\mathcal{I}$, we have that
\begin{align} \label{eq:levylemma2}
    \mathbb{P}\left[ \abs{\mathcal{I}(U) - \mathbb{E}\left[\mathcal{I}(U)\right]} \geq t \right] &\leq e^{-\frac{(d-2)(d+1)^2 t^2}{96}} .
\end{align}
This therefore confirms our observation of the concentration of measure phenomenon in \cref{fig:app:21-28}; it tells us that the probability of observing a gate whose infidelity is further than $t$ from the expectation value of the AGI decays exponentially in the distance $t^2$ and the dimensionality $(d-2)(d+1)^2$ of the system. 

For the case $d=2$, the exponent is zero and thus the probability is constant for all $t$ and so we see in \cref{subfig:app:21} that the AGI plateau values are uniformly distributed. However, as $d>2$ for \cref{subfig:app:26,subfig:app:27,subfig:app:28}, the $\mathcal{I}^*$ become exponentially concentrated about the mean $\bar{\mathcal{I}}^*$.

This could also account for the fluctuation in the ratios of $\mathcal{I}^*_{\rm mono} / \mathcal{I}^*_{\rm over}$ at different system sizes. However, further work is needed to elucidate the precise reasons for these observations.

\subsection{\label{app:C:subsec:3} Power-law behaviour of the saturation points of the AGI}
\begin{figure}[t!]
    \centering
    \includegraphics[width=\columnwidth]{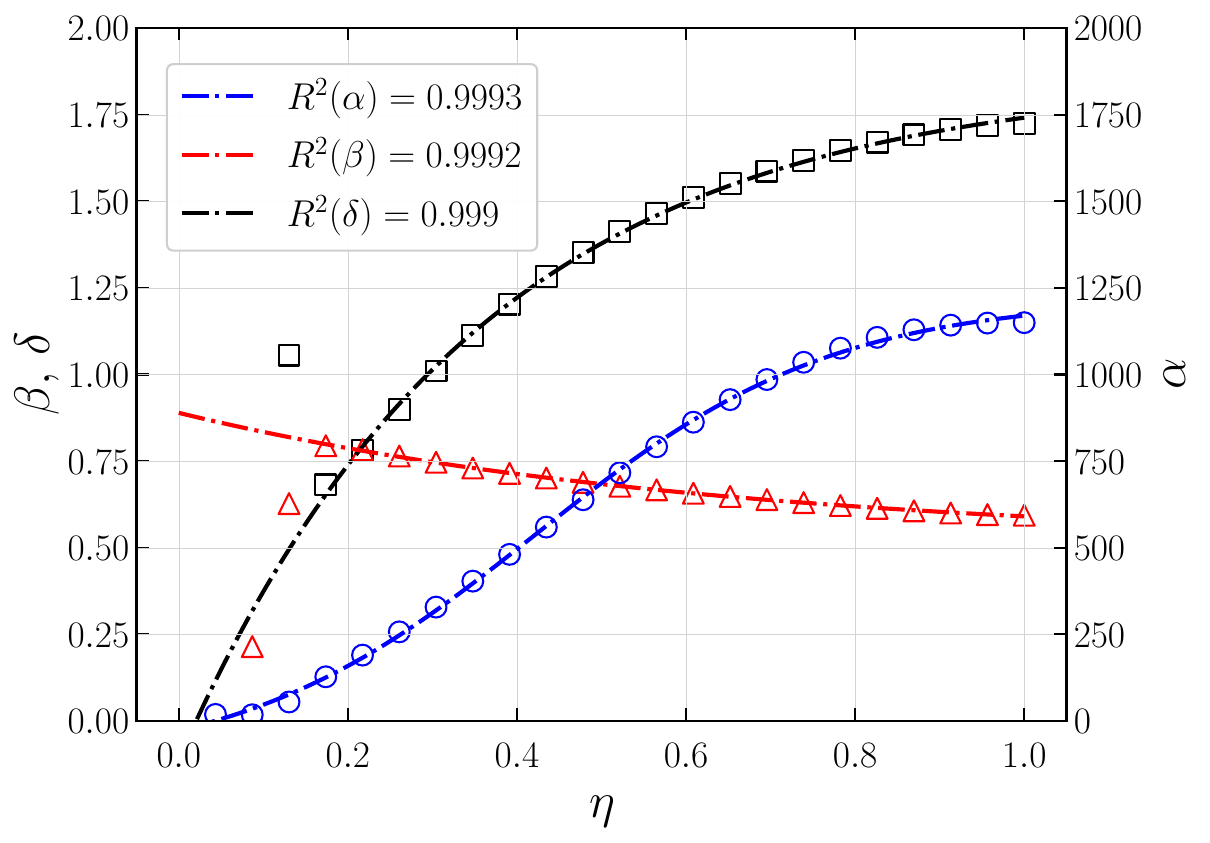}
    \caption{\justifying \textbf{Fitted power-law model parameters as functions of $\eta$ for the saturation points $(\gamma t)^*$ of the $X^{\eta}$ gates.} AGIs for $X^{\eta}$ with $\eta \in \left( 0, 1 \right]$ for $d\in\left[2, 12\right]$. Saturation points $(\gamma t)^*$ were calculated by interpolation and root-finding. A power law model was fitted with parameters $\alpha(\eta)$ (blue, circles), $\beta(\eta)$ (red, triangles), and $\delta(\eta)$ (black, squares) were then fitted to sigmoidal and exponential models (dashed lines) as functions of $\eta$.}
    \label{fig:app:29}
\end{figure}
As shown in \cref{fig:05} in \cref{sec:2:subsec:4}, the saturation points $(\gamma t)^*$ of the AGIs for the interpolated $X^{\eta}$ gates were observed to follow a power-law model described by \cref{eq:powerlaw}. The initial analysis was performed for each gate in a range of values $\eta=\left( 0, 1\right]$ for $d\in\left[2,12\right]$ by: (1) computing the AGIs numerically for a range of $\gamma t$ values in \textsc{Python} using \textsc{QuTiP}'s \texttt{propagator} and \texttt{process\_fidelity} methods, (2) interpolating the AGI curves using \textsc{SciPy}.\texttt{interpolate}.\texttt{CubicSpline}, and (3) finding the plateau saturation points $(\gamma t)^*$ by root-finding with \textsc{SciPy}.\texttt{optimize}.\texttt{root\_scalar}.

Then, this three-parameter power law model in \cref{eq:powerlaw} was fitted on $(\gamma t)^*$ as a function of $d$ for each $\eta$ curve by nonlinear least-squares regression using the \textsc{SciPy}.\texttt{optimize}.\texttt{curve\_fit} method. This produced three sets of data for the fitted parameters $\alpha$, $\beta$, $\gamma$ as functions of $\eta$. These three curves were then themselves fitted to their respective sigmoidal and exponential models in \cref{eq:powerlawalpha,eq:powerlawbeta,eq:powerlawdelta}, using the same \textsc{SciPy}.\texttt{optimize}.\texttt{curve\_fit} method. The models showed good agreement with the data, with $R^2(\alpha) = 0.9992$, $R^2(\beta) = 0.9992$ and $R^2(\delta) = 0.9989$. 

The fitted data points and resulting models are shown in \cref{fig:app:29}. We note that the outliers at small values of $\eta$ (including $\eta=0$ for the identity gate with constant $(\gamma t)^*$) are due to the near-constant saturation point values of these gates over the range of $d$, which slowly transition towards more power-law-like behaviour as $\eta$ increases. For example, for $\eta = 0$ the saturation point is constant in $d$, giving a horizontal line. Fitting a power-law curve to this dataset was possible but required unrealistic fit parameters like $\alpha \sim 10^{50}$. Nevertheless, this analysis hints at a potential approach for estimating the saturation points based only on the parameter $\eta$, without requiring the explicit numerical simulation.ut

\subsection{\label{app:C:subsec:4} Time dependence of the second-order correction to the AGI}
As alluded to in \cref{fig:09} in \cref{sec:2:subsec:5}, the convergence of the second-order (gate-dependent) AGI correction term as a function of $t$ was observed to be linear in \cref{eq:linearlaw}. More specifically, the number of iterations $s_{\varepsilon}$ required to reach convergence of two successive terms in the summation \cref{eq:secondorderAGIcorrection} was calculated according to \cref{eq:sconvergenceterm,eq:sconvergenceerror}.
\begin{figure}[t!]
    \centering
    \includegraphics[width=\columnwidth]{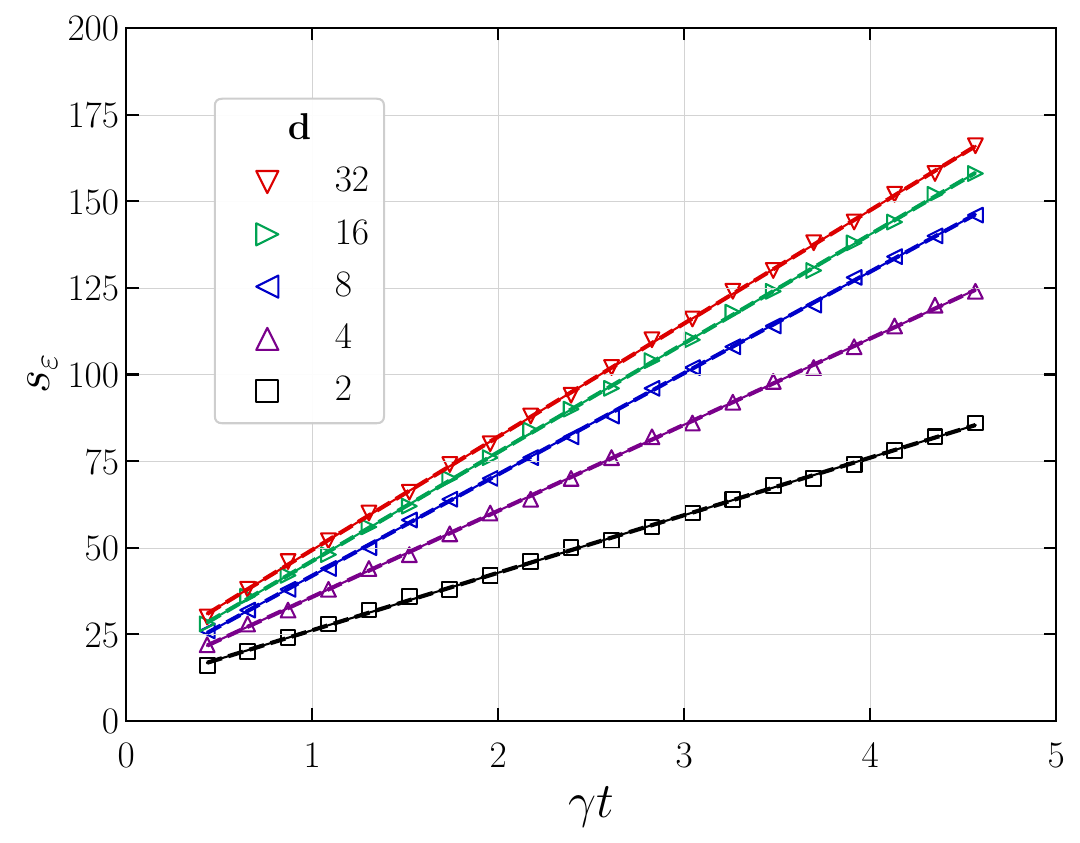}
    \caption{\justifying \textbf{Linear regression fits of the critical order $s_{\varepsilon}$ over gate time $t$ for the $X$ gate.} The $X$ gate ($\eta=1$) was simulated for dimensions $d\in\lbrace 2,4,8,16,32 \rbrace$ over gate times $\gamma t \in \left[ 0.5, 4.5 \right]$ with $\gamma = 1$ fixed. The second-order gate-dependent correction term was calculated at each $t$ for increasing order $s$ in the summation until convergence between successive terms was obtained at the critical value $s_{\varepsilon}$. For each $d$, a linear model was fitted to the data as shown by the dotted lines. }
    \label{fig:app:30}
\end{figure}

To understand the time-dependence of this, for each of the $\eta$ and $d$ values shown in \cref{fig:09}, the critical $s_{\varepsilon}$ was computed for a range of times $t\in\left[0, 5\right]$ with $\gamma$ fixed at $1$. Interestingly, we can see that in \cref{fig:app:30} the $s_{\varepsilon}$ does indeed appear linear in $t$ for each dimension shown for the $X$ gate. This behaviour extends to other $\eta$ values too. Therefore, a linear least-squares regression was performed on these data points using the \textsc{SciPy}.\texttt{optimize}.\texttt{curve\_fit}; the resulting fitted models are shown in \cref{fig:app:30} alongisde the data points, indicating good agreement. The resulting gradients $m = \frac{\text{d} s_{\varepsilon}}{\text{d}t}$ for each fitted curve are thus the data shown in \cref{fig:09}.


%% file: main.bbl
\begin{thebibliography}{10}

\bibitem{ezratty_perspective_2023}
Olivier Ezratty.
\newblock ``Perspective on superconducting qubit quantum computing''.
\newblock \href{https://dx.doi.org/10.1140/epja/s10050-023-01006-7}{The European Physical Journal A {\bf 59}, 94}~(2023).

\bibitem{de_leon_materials_2021}
Nathalie~P. De~Leon, Kohei~M. Itoh, Dohun Kim, Karan~K. Mehta, Tracy~E. Northup, Hanhee Paik, Benjamin~S. Palmer, Nitin Samarth, Sorawis Sangtawesin, and David~W. Steuerman.
\newblock ``Materials challenges and opportunities for quantum computing hardware''.
\newblock \href{https://dx.doi.org/10.1126/science.abb2823}{Science {\bf 372}, eabb2823}~(2021).

\bibitem{moreno-pineda_molecular_2018}
Eufemio Moreno-Pineda, Clément Godfrin, Franck Balestro, Wolfgang Wernsdorfer, and Mario Ruben.
\newblock ``Molecular {Spin} {Qudits} for {Quantum} {Algorithms}''.
\newblock \href{https://dx.doi.org/10.1039/c5cs00933b}{Chemical Society Reviews {\bf 47}, 501--513}~(2018).

\bibitem{moreno-pineda_measuring_2021}
Eufemio Moreno-Pineda and Wolfgang Wernsdorfer.
\newblock ``Measuring molecular magnets for quantum technologies''.
\newblock \href{https://dx.doi.org/10.1038/s42254-021-00340-3}{Nature Reviews Physics {\bf 3}, 645--659}~(2021).

\bibitem{chi_programmable_2022}
Yulin Chi, Jieshan Huang, Zhanchuan Zhang, Jun Mao, Zinan Zhou, Xiaojiong Chen, Chonghao Zhai, Jueming Bao, Tianxiang Dai, Huihong Yuan, Ming Zhang, Daoxin Dai, Bo~Tang, Yan Yang, Zhihua Li, Yunhong Ding, Leif~K. Oxenløwe, Mark~G. Thompson, Jeremy~L. O’Brien, Yan Li, Qihuang Gong, and Jianwei Wang.
\newblock ``A programmable qudit-based quantum processor''.
\newblock \href{https://dx.doi.org/10.1038/s41467-022-28767-x}{Nature Communications {\bf 13}, 1166}~(2022).

\bibitem{ringbauer_universal_2022}
Martin Ringbauer, Michael Meth, Lukas Postler, Roman Stricker, Rainer Blatt, Philipp Schindler, and Thomas Monz.
\newblock ``A universal qudit quantum processor with trapped ions''.
\newblock \href{https://dx.doi.org/10.1038/s41567-022-01658-0}{Nature Physics {\bf 18}, 1053--1057}~(2022).

\bibitem{wang_qudits_2020}
Yuchen Wang, Zixuan Hu, Barry~C. Sanders, and Sabre Kais.
\newblock ``Qudits and {High}-{Dimensional} {Quantum} {Computing}''.
\newblock \href{https://dx.doi.org/10.3389/fphy.2020.589504}{Frontiers in Physics {\bf 8}, 589504}~(2020).

\bibitem{chiesa_embedded_2021}
Alessandro Chiesa, Francesco Petiziol, Emilio Macaluso, Sandro Wimberger, Paolo Santini, and Stefano Carretta.
\newblock ``Embedded quantum-error correction and controlled-phase gate for molecular spin qubits''.
\newblock \href{https://dx.doi.org/10.1063/9.0000166}{AIP Advances {\bf 11}, 025134}~(2021).

\bibitem{petiziol_counteracting_2021}
Francesco Petiziol, Alessandro Chiesa, Sandro Wimberger, Paolo Santini, and Stefano Carretta.
\newblock ``Counteracting dephasing in {Molecular} {Nanomagnets} by optimized qudit encodings''.
\newblock \href{https://dx.doi.org/10.1038/s41534-021-00466-3}{npj Quantum Information {\bf 7}, 133}~(2021).

\bibitem{miyahara_decoherence_2023}
Hideyuki Miyahara, Yiyou Chen, Vwani Roychowdhury, and Louis-Serge Bouchard.
\newblock ``Decoherence mitigation by embedding a logical qubit in a qudit''.
\newblock \href{https://dx.doi.org/10.1007/s11128-023-04035-9}{Quantum Information Processing {\bf 22}, 278}~(2023).

\bibitem{gunderman_local-dimension-invariant_2020}
Lane~G. Gunderman.
\newblock ``Local-dimension-invariant qudit stabilizer codes''.
\newblock \href{https://dx.doi.org/10.1103/PhysRevA.101.052343}{Physical Review A {\bf 101}, 052343}~(2020).

\bibitem{wernsdorfer_synthetic_2019}
Wolfgang Wernsdorfer and Mario Ruben.
\newblock ``Synthetic {Hilbert} {Space} {Engineering} of {Molecular} {Qu} \textit{d} its: {Isotopologue} {Chemistry}''.
\newblock \href{https://dx.doi.org/10.1002/adma.201806687}{Advanced Materials {\bf 31}, 1806687}~(2019).

\bibitem{zheng_entanglement_2022}
Yunzhe Zheng, Hemant Sharma, and Johannes Borregaard.
\newblock ``Entanglement {Distribution} with {Minimal} {Memory} {Requirements} {Using} {Time}-{Bin} {Photonic} {Qudits}''.
\newblock \href{https://dx.doi.org/10.1103/PRXQuantum.3.040319}{PRX Quantum {\bf 3}, 040319}~(2022).

\bibitem{bouchard_high-dimensional_2017}
Frédéric Bouchard, Robert Fickler, Robert~W. Boyd, and Ebrahim Karimi.
\newblock ``High-dimensional quantum cloning and applications to quantum hacking''.
\newblock \href{https://dx.doi.org/10.1126/sciadv.1601915}{Science Advances {\bf 3}, e1601915}~(2017).

\bibitem{impagliazzo_ternary_2011}
Nikolay~Petrovich Brusentsov and José Ramil~Alvarez.
\newblock ``Ternary {Computers}: {The} {Setun} and the {Setun} 70''.
\newblock In John Impagliazzo and Eduard Proydakov, editors, Perspectives on {Soviet} and {Russian} {Computing}.
\newblock \href{https://dx.doi.org/10.1007/978-3-642-22816-2_10}{Volume 357 of {IFIP} {Advances} in {Information} and {Communication} {Technology}, pages 74--80}.
\newblock Springer Berlin Heidelberg, Berlin, Heidelberg~(2011).

\bibitem{boixo_characterizing_2018}
Sergio Boixo, Sergei~V. Isakov, Vadim~N. Smelyanskiy, Ryan Babbush, Nan Ding, Zhang Jiang, Michael~J. Bremner, John~M. Martinis, and Hartmut Neven.
\newblock ``Characterizing quantum supremacy in near-term devices''.
\newblock \href{https://dx.doi.org/10.1038/s41567-018-0124-x}{Nature Physics {\bf 14}, 595--600}~(2018).

\bibitem{arute_quantum_2019}
Frank Arute, Kunal Arya, Ryan Babbush, Dave Bacon, Joseph~C. Bardin, Rami Barends, Rupak Biswas, Sergio Boixo, Fernando G. S.~L. Brandao, David~A. Buell, Brian Burkett, Yu~Chen, Zijun Chen, Ben Chiaro, Roberto Collins, William Courtney, Andrew Dunsworth, Edward Farhi, Brooks Foxen, Austin Fowler, Craig Gidney, Marissa Giustina, Rob Graff, Keith Guerin, Steve Habegger, Matthew~P. Harrigan, Michael~J. Hartmann, Alan Ho, Markus Hoffmann, Trent Huang, Travis~S. Humble, Sergei~V. Isakov, Evan Jeffrey, Zhang Jiang, Dvir Kafri, Kostyantyn Kechedzhi, Julian Kelly, Paul~V. Klimov, Sergey Knysh, Alexander Korotkov, Fedor Kostritsa, David Landhuis, Mike Lindmark, Erik Lucero, Dmitry Lyakh, Salvatore Mandrà, Jarrod~R. McClean, Matthew McEwen, Anthony Megrant, Xiao Mi, Kristel Michielsen, Masoud Mohseni, Josh Mutus, Ofer Naaman, Matthew Neeley, Charles Neill, Murphy~Yuezhen Niu, Eric Ostby, Andre Petukhov, John~C. Platt, Chris Quintana, Eleanor~G. Rieffel, Pedram Roushan, Nicholas~C. Rubin, Daniel Sank, Kevin~J.
  Satzinger, Vadim Smelyanskiy, Kevin~J. Sung, Matthew~D. Trevithick, Amit Vainsencher, Benjamin Villalonga, Theodore White, Z.~Jamie Yao, Ping Yeh, Adam Zalcman, Hartmut Neven, and John~M. Martinis.
\newblock ``Quantum supremacy using a programmable superconducting processor''.
\newblock \href{https://dx.doi.org/10.1038/s41586-019-1666-5}{Nature {\bf 574}, 505--510}~(2019).

\bibitem{godfrin_operating_2017}
Clément Godfrin, Abdelkarim Ferhat, Rafik Ballou, Svetlana Klyatskaya, Mario Ruben, Wolfgang Wernsdorfer, and Franck Balestro.
\newblock ``Operating {Quantum} {States} in {Single} {Magnetic} {Molecules}: {Implementation} of {Grover}’s {Quantum} {Algorithm}''.
\newblock \href{https://dx.doi.org/10.1103/PhysRevLett.119.187702}{Physical Review Letters {\bf 119}, 187702}~(2017).

\bibitem{thiele_electrically_2014}
Stefan Thiele, Franck Balestro, Rafik Ballou, Svetlana Klyatskaya, Mario Ruben, and Wolfgang Wernsdorfer.
\newblock ``Electrically driven nuclear spin resonance in single-molecule magnets''.
\newblock \href{https://dx.doi.org/10.1126/science.1249802}{Science {\bf 344}, 1135--1138}~(2014).

\bibitem{otten_impacts_2021}
Matthew Otten, Keshav Kapoor, A.~Bariş Özgüler, Eric~T. Holland, James~B. Kowalkowski, Yuri Alexeev, and Adam~L. Lyon.
\newblock ``Impacts of noise and structure on quantum information encoded in a quantum memory''.
\newblock \href{https://dx.doi.org/10.1103/PhysRevA.104.012605}{Physical Review A {\bf 104}, 012605}~(2021).

\bibitem{jankovic_noisy_2024}
Denis Janković, Jean-Gabriel Hartmann, Mario Ruben, and Paul-Antoine Hervieux.
\newblock ``Noisy qudit vs multiple qubits: conditions on gate efficiency for enhancing fidelity''.
\newblock \href{https://dx.doi.org/10.1038/s41534-024-00829-6}{npj Quantum Information {\bf 10}, 59}~(2024).

\bibitem{nielsen_quantum_2000}
Michael~A. Nielsen and Isaac~L. Chuang.
\newblock ``Quantum computation and quantum information''.
\newblock \href{https://dx.doi.org/10.1017/CBO9780511976667}{Cambridge University Press}. Cambridge, NY~(2000).

\bibitem{nielsen_simple_2002}
Michael~A. Nielsen.
\newblock ``A simple formula for the average gate fidelity of a quantum dynamical operation''.
\newblock \href{https://dx.doi.org/10.1016/S0375-9601(02)01272-0}{Physics Letters A {\bf 303}, 249--252}~(2002).

\bibitem{lindblad_generators_1976}
Göran Lindblad.
\newblock ``On the generators of quantum dynamical semigroups''.
\newblock \href{https://dx.doi.org/10.1007/BF01608499}{Communications in Mathematical Physics {\bf 48}, 119--130}~(1976).

\bibitem{gorini_completely_1976}
Vittorio Gorini, Andrzej Kossakowski, and E.~C.~George Sudarshan.
\newblock ``Completely positive dynamical semigroups of \textit{{N}} -level systems''.
\newblock \href{https://dx.doi.org/10.1063/1.522979}{Journal of Mathematical Physics {\bf 17}, 821--825}~(1976).

\bibitem{breuer_theory_2007}
Heinz-Peter Breuer and Francesco Petruccione.
\newblock ``The {Theory} of {Open} {Quantum} {Systems}''.
\newblock \href{https://dx.doi.org/10.1093/acprof:oso/9780199213900.001.0001}{Oxford University Press}. Oxford~(2007).
\newblock First edition.

\bibitem{manzano_short_2020}
Daniel Manzano.
\newblock ``A short introduction to the {Lindblad} master equation''.
\newblock \href{https://dx.doi.org/10.1063/1.5115323}{AIP Advances {\bf 10}, 025106}~(2020).

\bibitem{watrous_theory_2018}
John Watrous.
\newblock ``The {Theory} of {Quantum} {Information}''.
\newblock \href{https://dx.doi.org/10.1017/9781316848142}{Cambridge University Press}. Cambridge~(2018).
\newblock First edition.

\bibitem{bengtsson_geometry_2006}
Ingemar Bengtsson and Karol Zyczkowski.
\newblock ``Geometry of {Quantum} {States}: {An} {Introduction} to {Quantum} {Entanglement}''.
\newblock \href{https://dx.doi.org/10.1017/CBO9780511535048}{Cambridge University Press}. Cambridge~(2006).
\newblock First edition.

\bibitem{havel_robust_2003}
Timothy~F. Havel.
\newblock ``Robust procedures for converting among {Lindblad}, {Kraus} and matrix representations of quantum dynamical semigroups''.
\newblock \href{https://dx.doi.org/10.1063/1.1518555}{Journal of Mathematical Physics {\bf 44}, 534--557}~(2003).

\bibitem{loring_computing_2014}
Terry~A. Loring.
\newblock ``Computing a logarithm of a unitary matrix with general spectrum''.
\newblock \href{https://dx.doi.org/10.1002/nla.1927}{Numerical Linear Algebra with Applications {\bf 21}, 744--760}~(2014).

\bibitem{aifer_quantum_2022}
Maxwell Aifer and Sebastian Deffner.
\newblock ``From quantum speed limits to energy-efficient quantum gates''.
\newblock \href{https://dx.doi.org/10.1088/1367-2630/ac6821}{New Journal of Physics {\bf 24}, 055002}~(2022).

\bibitem{vourdas_quantum_2004}
Apostolos Vourdas.
\newblock ``Quantum systems with finite {Hilbert} space''.
\newblock \href{https://dx.doi.org/10.1088/0034-4885/67/3/R03}{Reports on Progress in Physics {\bf 67}, 267--320}~(2004).

\bibitem{howard_qudit_2012}
Mark Howard and Jiri Vala.
\newblock ``Qudit versions of the qubit π / 8 gate''.
\newblock \href{https://dx.doi.org/10.1103/PhysRevA.86.022316}{Physical Review A {\bf 86}, 022316}~(2012).

\bibitem{martinez-carranza_alternative_2012}
Jose Martínez-Carranza, Francisco Soto-Eguibar, and Héctor~M. Moya-Cessa.
\newblock ``Alternative analysis to perturbation theory in quantum mechanics: {Dyson} series in matrix form''.
\newblock \href{https://dx.doi.org/10.1140/epjd/e2011-20654-5}{The European Physical Journal D {\bf 66}, 22}~(2012).

\bibitem{villegas-martinez_application_2016}
Braulio~M. Villegas-Martínez, Francisco Soto-Eguibar, and Héctor~M. Moya-Cessa.
\newblock ``Application of {Perturbation} {Theory} to a {Master} {Equation}''.
\newblock \href{https://dx.doi.org/10.1155/2016/9265039}{Advances in Mathematical Physics {\bf 2016}, 1--7}~(2016).

\bibitem{campbell_law_1896}
John~E. Campbell.
\newblock ``On a {Law} of {Combination} of {Operators} bearing on the {Theory} of {Continuous} {Transformation} {Groups}''.
\newblock \href{https://dx.doi.org/10.1112/plms/s1-28.1.381}{Proceedings of the London Mathematical Society {\bf s1-28}, 381--390}~(1896).

\bibitem{hall_lie_2015}
Brian~C. Hall.
\newblock ``Lie groups, {Lie} algebras, and representations: an elementary introduction''.
\newblock \href{https://dx.doi.org/10.1007/978-3-319-13467-3}{Number 222 in Graduate texts in mathematics}. Springer. Cambridge, NY~(2015).
\newblock Second edition.

\bibitem{collins_weingarten_2022}
Benoit Collins, Sho Matsumoto, and Jonathan Novak.
\newblock ``The {Weingarten} {Calculus}''.
\newblock \href{https://dx.doi.org/10.1090/noti2474}{Notices of the American Mathematical Society {\bf 69}, 1}~(2022).

\bibitem{chiesa_molecular_2020}
Alessandro Chiesa, Emilio Macaluso, Francesco Petiziol, Sandro Wimberger, Paolo Santini, and Stefano Carretta.
\newblock ``Molecular {Nanomagnets} as {Qubits} with {Embedded} {Quantum}-{Error} {Correction}''.
\newblock \href{https://dx.doi.org/10.1021/acs.jpclett.0c02213}{The Journal of Physical Chemistry Letters {\bf 11}, 8610--8615}~(2020).

\bibitem{horodecki_general_1999}
Michał Horodecki, Paweł Horodecki, and Ryszard Horodecki.
\newblock ``General teleportation channel, singlet fraction, and quasidistillation''.
\newblock \href{https://dx.doi.org/10.1103/PhysRevA.60.1888}{Physical Review A {\bf 60}, 1888--1898}~(1999).

\bibitem{volkin1968iterated}
Howard~C. Volkin, United States.~National Aeronautics, Space Administration, and Lewis~Research Center.
\newblock ``Iterated commutators and functions of operators''.
\newblock {NASA} technical note. National Aeronautics and Space Administration. ~(1968).
\newblock  url:~\url{https://ntrs.nasa.gov/citations/19680027053}.

\bibitem{hartmann_code_2025}
Jean-Gabriel Hartmann, Denis Janković, Rémi Pasquier, Mario Ruben, and Paul-Antoine Hervieux.
\newblock ``Code and {Data} for ``{Nonlinearity} of the {Fidelity} in {Open} {Qudit} {Systems}: {Gate} and {Noise} {Dependence} in {High}-dimensional {Quantum} {Computing}"''.
\newblock \href{https://dx.doi.org/10.5281/ZENODO.13618284}{Zenodo}. ~(2025).

\bibitem{mezzadri_how_2007}
Francesco Mezzadri.
\newblock ``How to generate random matrices from the classical compact groups''.
\newblock \href{https://dx.doi.org/10.48550/arXiv.math-ph/0609050}{Notices of the American Mathematical Society {\bf 54}, 592--604}~(2007).

\bibitem{dyson_threefold_1962}
Freeman~J. Dyson.
\newblock ``The {Threefold} {Way}. {Algebraic} {Structure} of {Symmetry} {Groups} and {Ensembles} in {Quantum} {Mechanics}''.
\newblock \href{https://dx.doi.org/10.1063/1.1703863}{Journal of Mathematical Physics {\bf 3}, 1199--1215}~(1962).

\bibitem{mehta_random_2004}
Madan~L. Mehta.
\newblock ``Random matrices''.
\newblock \href{https://dx.doi.org/10.1016/S0079-8169(04)80088-6}{Number 142 in Pure and applied mathematics series}. Elsevier. Amsterdam~(2004).
\newblock Third edition.

\bibitem{wigner_characteristic_1955}
Eugene~P. Wigner.
\newblock ``Characteristic {Vectors} of {Bordered} {Matrices} {With} {Infinite} {Dimensions}''.
\newblock \href{https://dx.doi.org/10.2307/1970079}{The Annals of Mathematics {\bf 62}, 548}~(1955).

\bibitem{johansson_qutip_2012}
Jesper~R. Johansson, Paul~D. Nation, and Franco Nori.
\newblock ``{QuTiP}: {An} open-source {Python} framework for the dynamics of open quantum systems''.
\newblock \href{https://dx.doi.org/10.1016/j.cpc.2012.02.021}{Computer Physics Communications {\bf 183}, 1760--1772}~(2012).

\bibitem{johansson_qutip_2013}
Jesper~R. Johansson, Paul~D. Nation, and Franco Nori.
\newblock ``{QuTiP} 2: {A} {Python} framework for the dynamics of open quantum systems''.
\newblock \href{https://dx.doi.org/10.1016/j.cpc.2012.11.019}{Computer Physics Communications {\bf 184}, 1234--1240}~(2013).

\bibitem{cirq_developers_cirq_2023}
Cirq Developers.
\newblock ``Cirq''.
\newblock \href{https://dx.doi.org/10.5281/ZENODO.4062499}{Google Quantum AI}. ~(2023).

\bibitem{suzen_spectral_2017}
Mehmet Süzen, Cornelius Weber, and Joan~J. Cerdà.
\newblock ``Spectral {Ergodicity} {In} {Deep} {Learning} {Architectures} {Via} {Surrogate} {Random} {Matrices}''~(2017).
\newblock  \href{http://arxiv.org/abs/1704.08303}{arXiv:1704.08303}.

\bibitem{giannopoulos_concentration_2000}
Apostolos~A. Giannopoulos and Vitali~D. Milman.
\newblock ``Concentration {Property} on {Probability} {Spaces}''.
\newblock \href{https://dx.doi.org/10.1006/aima.2000.1949}{Advances in Mathematics {\bf 156}, 77--106}~(2000).

\bibitem{ledoux_concentration_2001}
Michel Ledoux.
\newblock ``The concentration of measure phenomenon''.
\newblock \href{https://dx.doi.org/10.1090/surv/089}{Number Issue 89 in Mathematical surveys and monographs}. American Mathematical Society. Providence, RI~(2001).

\bibitem{meckes_random_2019}
Elizabeth~S. Meckes.
\newblock ``The {Random} {Matrix} {Theory} of the {Classical} {Compact} {Groups}''.
\newblock \href{https://dx.doi.org/10.1017/9781108303453}{Cambridge University Press}. Cambridge~(2019).
\newblock First edition.

\end{thebibliography}
